\documentclass[aps,prd,preprint,eqsecnum,nofootinbib,groupedaddress,showpacs,floatfix]{revtex4}

\usepackage[dvips]{graphicx}
\usepackage{multirow}
\usepackage{placeins}
\usepackage{wasysym}
\newcommand{\BlackHat}{{\sc BlackHat}}

\newcommand{\SHERPA}{{\sc SHERPA}}

\newcommand{\AMEGIC}{{\sc AMEGIC++}}
\newcommand{\COMIX}{{\sc COMIX}}
\newcommand{\SISCone}{{\sc SISCone}}
\newcommand{\ntuple}{{$n$-tuple}}
\newcommand{\ntuples}{{$n$-tuples}}

\newif\ifdraft
\draftfalse
\newif\ifpreprint
\preprinttrue

\def\fig#1{fig.~{\ref{#1}}}

\def\figs#1#2{figs.~{\ref{#1}} and {\ref{#2}}}
\def\Figs#1#2{Figs.~{\ref{#1}} and {\ref{#2}}}
\def\sect#1{section~{\ref{#1}}}
\def\eqn#1{eq.~(\ref{#1})}

\def\eqns#1#2{eqs.~(\ref{#1}) and~(\ref{#2})}

\def\tab#1{table~{\ref{#1}}}
\def\Tab#1{Table~{\ref{#1}}}
\def\tabs#1#2{tables~{\ref{#1}} and~{\ref{#2}}}

\def\nn{\nonumber}

\def\Wj{$W\,\!+\,1$}
\def\Wjj{$W\,\!+\,2$}
\def\Wjjj{$W\,\!+\,3$}
\def\Wjjjj{$W\,\!+\,4$}
\def\Wjjjjj{$W\,\!+\,5$}
\def\Wjjjjjj{$W\,\!+\,6$}
\def\Wjn{$W\,\!+\,n$}
\def\Wmj{$W^-\,\!+\,1$}
\def\Wmjj{$W^-\,\!+\,2$}
\def\Wmjjj{$W^-\,\!+\,3$}
\def\Wmjjjj{$W^-\,\!+\,4$}
\def\Wmjjjjj{$W^-\,\!+\,5$}
\def\Wmjjjjjj{$W^-\,\!+\,6$}
\def\Wpj{$W^+\,\!+\,1$}
\def\Wpjj{$W^+\,\!+\,2$}
\def\Wpjjj{$W^+\,\!+\,3$}
\def\Wpjjjj{$W^+\,\!+\,4$}
\def\Wpjjjjj{$W^+\,\!+\,5$}
\def\Wpjjjjjj{$W^+\,\!+\,6$}

\def\Wjjjjx{$W\,\!+\,4,5$}

\def\Wjn{$W\,\!+\,n$}
\def\Wjnp1{$W\,\!+\,(n\!+\!1)$}
\def\Wmjn{$W^-\,\!+\,n$}
\def\Wmjnm{$W^-\,\!+\,(n\!-\!1)$}
\def\Wpjn{$W^+\,\!+\,n$}
\def\Wpjnm{$W^+\,\!+\,(n\!-\!1)$}

\def\Wjnm{$W\!+\,(n\!-\!1)$}

\def\Zj{$Z\,\!+\,1$}
\def\Zjj{$Z\,\!+\,2$}
\def\Zjjj{$Z\,\!+\,3$}

\def\Zjn{$Z\,\!+\,n$}

\def\jet{{\rm jet}}
\def\kT{k_{\rm T}}
\def\pT{p_{\rm T}}
\def\pTn#1{p_{{\rm T}#1}}
\def\pTmin{\pT^{\text{min}}}
\def\pTV{\pT^V}

\def\root{{\sc root}}
\def\Ord{{\cal O}}

\def\ETsl{{\s E}_{\rm T}}
\def\ET{E_{\rm T}}

\def\HTpartonicp{{\hat H}_{\rm T}'}
\def\HT{H_{\rm T}}
\def\HTjets{H_{\rm T}^{\rm jets}}

\def\MT{M_{\rm T}}

\newbox\charbox
\newbox\slabox
\def\s#1{{      
        \setbox\charbox=\hbox{$#1$}
        \setbox\slabox=\hbox{$/$}
        \dimen\charbox=\ht\slabox
        \advance\dimen\charbox by -\dp\slabox
        \advance\dimen\charbox by -\ht\charbox
        \advance\dimen\charbox by \dp\charbox
        \divide\dimen\charbox by 2
        \raise-\dimen\charbox\hbox to \wd\charbox{\hss/\hss}
        \llap{$#1$}
}}

\begin{document}

\title{
\ifpreprint
\hbox{\rm\small
UCLA/14/TEP/108$\null\hskip 3.2cm \null$
SLAC--PUB--16144$\null\hskip 4.4cm \null$
IPPP/11/82\break}
\vskip -.35 cm 
\hbox{\rm\small
SB/F/442-14$\null\hskip 4.3cm \null$ 
Saclay--IPhT--T14/185$\hskip 2.3cm \null$
FR-PHENO-2014-002 \break}
\hbox{$\null$\break}
\fi
Extrapolating $W$-Associated Jet-Production Ratios at the LHC}

\author{Z.~Bern${}^a$, L.~J.~Dixon${}^b$, F.~Febres Cordero${}^{c,d}$, 
S.~H{\" o}che${}^b$, D.~A.~Kosower${}^e$, H.~Ita${}^{d}$ and 
D.~Ma\^{\i}tre${}^{f}$
\\
$\null$
\\
${}^a$Department of Physics and Astronomy, UCLA, Los Angeles, CA
90095-1547, USA \\
${}^b$SLAC National Accelerator Laboratory, Stanford University,
             Stanford, CA 94309, USA \\
${}^c${Physikalisches Institut, Albert-Ludwigs-Universit\"at Freiburg, D--79104 Freiburg, Germany}\\
${}^d$Departamento de F\'{\i}sica, Universidad Sim\'on Bol\'{\i}var, 
 Caracas 1080A, Venezuela\\
${}^e$Institut de Physique Th\'eorique, CEA--Saclay,
          F--91191 Gif-sur-Yvette cedex, France\\
${}^f$Department of Physics, University of Durham, Durham DH1 3LE, UK\\
}

\begin{abstract}
Electroweak vector-boson production, accompanied
by multiple jets, is an important background to 
searches for physics beyond the Standard Model. 
A precise and quantitative understanding of this process
is helpful in constraining deviations from known physics.
We study four key ratios in \Wjn-jet production at the LHC.
We compute the ratio of cross sections for \Wjn- to \Wjnm-jet
production as a function of the minimum jet transverse momentum.
We also study the ratio differentially, as a function of the
$W$-boson transverse momentum; as a function of the scalar sum of the jet
transverse energy, $\HTjets$; and as a function of certain jet transverse
momenta. We show how to use such ratios to extrapolate differential
cross sections to \Wjjjjjj-jet production at next-to-leading order, 
and we cross-check the method against a direct calculation at leading order. 
We predict the differential distribution in $\HTjets${} for
\Wjjjjjj~jets at next-to-leading order using such an extrapolation.
We use the \BlackHat{} software library together with \SHERPA{} to perform the 
computations.
\end{abstract}

\pacs{12.38.-t, 12.38.Bx, 13.87.-a, 14.70.Hp \hspace{1cm}}

\maketitle

\section{Introduction}

Searches for new physics beyond the Standard Model rely on
quantitative theoretical predictions for known-physics backgrounds.
Such predictions are also important to the emerging precision studies
of the Higgs-like boson~\cite{AtlasHiggs,CMSHiggs} discovered last
year, of the top quark, and of the self-interactions of electroweak
vector bosons.  Signals of new physics typically hide beneath
Standard-Model backgrounds in a broad range of search strategies.
Sniffing out the signals requires a good quantitative understanding of
the backgrounds as well as the corresponding theoretical
uncertainties.  The challenge of obtaining such an understanding
increases with the increasing jet multiplicities used in cutting-edge
search strategies.  For some search strategies, the uncertainty
surrounding predictions of Standard-Model background rates can be
lessened by using data-driven estimates; this approach still requires
theoretical input to predict the ratios of signal to control processes
or regions.

Predictions for Standard-Model rates at the Large Hadron Collider (LHC)
require calculations in perturbative QCD, which enters all aspects of
short-distance collisions at a hadron collider.  Leading-order (LO)
predictions in QCD suffer from a strong dependence on the unphysical
renormalization and factorization scales.  This dependence gets
increasingly strong with growing jet multiplicity.  Next-to-leading (NLO)
calculations reduce this dependence, typically to a 10--15\% residual
sensitivity, and offer the first quantitatively reliable order in
perturbation theory.

Basic measurements of cross sections or differential distributions
suffer from a number of experimental and theoretical uncertainties.
Ratios of cross sections should be subject to greatly reduced
uncertainties, in particular those due to the jet energy scale,
lepton efficiency or acceptance, or the proton-proton luminosity.
We may also expect ratios to suffer less from
theoretical uncertainties due to uncalculated higher-order
corrections, though quantifying this reduction is not necessarily
easy.  In this paper, we study a variety of ratios based on
NLO results for
\Wpjn-jet and \Wmjn-jet production with $n\le 5$.  We study the
so-called jet-production ratio~\cite{JetProductionRatio}: the ratio of
\Wjn-jet production to \Wjnm-jet production.  (This ratio is also
sometimes called the ``Berends'', ``Berends--Giele'' or ``staircase''
ratio.)  We study the jet-production ratio for inclusive total cross
sections as a function of
the minimum jet transverse momentum $\pTmin$, and find a remarkable
universality for $n> 2$.  We also study several differential
cross sections: with respect to the vector-boson transverse momentum;
with respect to certain jet transverse momenta; and with respect
to the total jet transverse energy $\HTjets$.  Such ratios are also
central to data-driven estimates of backgrounds, in which a measurement of
one process is used to estimate another.  
The jet-production ratio is
useful for making estimates of backgrounds with additional jets. 
As an example, we predict the differential distribution in
the total jet transverse energy in \Wjjjjjj-jet production to NLO
accuracy.  
Englert {\it et al.{}}~\cite{Englert} and
Gerwick {\it et al.{}}~\cite{Gerwick} have
studied jet-production ratios in vector-boson production and
pure-jet production.
They found that in QCD one expects a constant ratio for jet production at
fixed $\pTmin$, when all jets are subject to the same cut.  Our results
are in agreement with these expectations for a broad range of $\pTmin$ values.

There have been many experimental measurements of vector production
in association with jets at both the Tevatron and the LHC.  Here
we focus specifically on measurements of various ratios at the LHC.
The CMS collaboration has
measured~\cite{ExptJetProductionRatioLHC} the jet-production ratios at
the LHC for production in association with a $W$ boson, as well as the
ratio of $W\,+\,$jets to $Z\,+\,$jets and the $W$ charge asymmetry
as a function of the number of jets.  (The usefulness of
the charge asymmetry, or $W^+/W^-$ ratio, in reducing uncertainties
was emphasized by Kom and Stirling~\cite{KomStirling};
it has been computed to NLO for up to five associated jets~\cite{W4j,W5j}.)
More recently, CMS measured the $Z/\gamma$ ratio as a function
of jets~\cite{CMSZgamma}; this ratio has also played a role
in CMS's determination of the $Z(\to\nu\nu)\,+\,$jets background
to supersymmetry searches~\cite{CMSSUSY} (see also ref.~\cite{AskEtal}).
The ATLAS collaboration has recently presented studies of
vector-boson production in association with 
jets~\cite{ATLASRatios}, as well as measuring
the ratio $W/Z$ in association with up to four jets and
comparing it with NLO predictions~\cite{Aad2014rta}.

NLO QCD predictions for production of vector bosons with a lower
multiplicity of jets (one or two jets) have been available for many
years~\cite{MCFM,FernandoWjetpapers}. 
In recent years, the advent of new
on-shell techniques (see refs.~\cite{RecentOnShellReviews} for recent
reviews) for computing one-loop amplitudes at larger multiplicity has
also made possible NLO results for
three~\cite{W3jPRL,W3jDistributions,EMZW3j,BlackHatZ3jet},
four~\cite{W4j,Z4j} and even five associated jets~\cite{W5j}.
High-multiplicity NLO results have been used to study the 
jet-production ratios for $W$-boson production in association with up to
four jets~\cite{W3jDistributions,W4j}, but at a single value of $\pTmin$.  

Here we use on-shell techniques and
employ the same computational setup as in ref.~\cite{W5j}.
An NLO QCD result is comprised of virtual, Born, and real-emission
contributions, along with appropriate infrared subtraction terms.  We
compute the virtual corrections numerically using the \BlackHat{}
code~\cite{BlackHatI}.  For
processes with up to three associated jets, we use the \AMEGIC{}
package~\cite{Amegic} to compute Born and real-emission contributions
along with their Catani--Seymour~\cite{CS} infrared subtraction terms.
For four or five associated jets, we use the \COMIX{}
package~\cite{Comix}.  We use \SHERPA{}~\cite{Sherpa} to manage the
overall calculation, including the integration over phase
space~\cite{AntennaIntegrator}.  To facilitate new studies with
different (tighter) cuts, different scale choices or PDF sets, we
store intermediate results: we record the momenta for all partons along
with matrix-element information, including the coefficients of various
scale- or PDF-dependent functions, in
\root{}-format~\cite{ROOT} files~\cite{NtuplesNote}.  This
makes it possible to evaluate cross sections and distributions for
different scales and PDF error sets without re-running from scratch.
Experiments have made use of this ability in applying \BlackHat{}
predictions. (See, for example, ref.~\cite{NTupleUse}.)

In carrying out the computations for this study, we neglect a number
of small contributions.  For four and five associated
jets, we make use of a leading-color approximation for the virtual
contributions that has been validated for processes with up to four
jets, with corrections that are under
3\%~\cite{W3jDistributions,ItaOzeren}.  We also neglected small
contributions from virtual top quarks.  In the five-jet case, all four-quark
pair terms in the real-emission contributions were also neglected.

This paper is organized as follows.  In \sect{BasicSetUpSection} we 
summarize the basic setup used in the computation.  In \sect{ResultsSection}
we present our results for cross sections, ratios and distributions. 
We give our summary and conclusions in \sect{ConclusionSection}.
Tables with numerical results are shown in Appendices A--D.


\section{Basic Setup}
\label{BasicSetUpSection}

In this paper we study ratios involving the \Wjn-jet processes with
$n\le 5$ to NLO in QCD.  We include the decay of the vector boson
($W^{\pm}$) into a charged lepton and neutrino at the amplitude level,
with no on-shell approximations made for the $W$ boson.

We use several standard kinematic variables to characterize scattering
events; for completeness we give their definitions here.  The
pseudorapidity $\eta$ is given by $\eta= -\ln(\tan {\theta}/{2})$,
where $\theta$ is the polar angle with respect to the beam axis.  We
denote the angular separation of two objects (partons, jets or
leptons) by $\Delta R= \sqrt{(\Delta \phi)^2+ (\Delta y)^2}$ with
$\Delta \phi$ the difference in the azimuthal angles, and $\Delta y$
the difference in the rapidities.  
 
For use in scale choices, we define the quantity
\begin{equation}
\HTpartonicp \equiv \sum_j \pT^j + \ET^W\,,
\label{HThatpscale}
\end{equation}
where the sum runs over all final-state partons $j$ and $\ET^W \equiv
\sqrt{M_W^2+(\pT^W)^2}$.  This transverse energy is a modified version
of the simple sum of transverse energies of all massless outgoing
partons and leptons.  (The particular choice~(\ref{HThatpscale})
does not bias the leptonic angular distributions
in the $W$ rest frame~\cite{Wpolarization}.)
For observable distributions, we use the jet-based quantity
\begin{equation}
\HTjets \equiv \sum_{j\in{\rm jets}} \ET^j\,,
\label{HTjetsdef}
\end{equation}
the total transverse energy of jets passing all cuts.  This quantity,
unlike $\HTpartonicp$, excludes the transverse energy of the vector
boson.

In our study, we consider the inclusive processes $p p \rightarrow$
\Wjn{} jets at an LHC center-of-mass energy of $\sqrt{s} = 7$ TeV 
for $n\le 5$, with
the following basic set of cuts:
\begin{equation}
\begin{array}{lll}
\ET^{e} > 20 \hbox{ GeV} \,, \hskip 1.5 cm 
&|\eta_e| < 2.5\,, \hskip 1.5 cm 
&\ET^{\nu} > 20 \hbox{ GeV}\,,  \hskip 1.5 cm \nn\\
\pT^\jet > 25 \hbox{ GeV}\,, \hskip 1.5 cm 
&|\eta_\jet|<3\,,  \hskip 1.5 cm &\MT^W > 20  \hbox{ GeV}\,. 
\end{array}
\label{Cuts}
\end{equation}
We define jets using the anti-$\kT$ algorithm~\cite{antikT} with
parameter $R = 0.5$.  The jets are ordered in $\pT$, and are labeled
$i,j=1,2,3,\ldots$ in order of decreasing transverse momentum $\pT$,
with $1$ being the leading (hardest) jet.  
The transverse mass of the $W$-boson is computed from the kinematics
of its decay products, $W\rightarrow e\nu_e$,
\begin{equation}
\MT^W=\sqrt{2\ET^e \ET^\nu (1- \cos(\Delta \phi_{e\nu}))}\,.
\end{equation}
At LO, the missing transverse energy, $\ETsl$, is just the neutrino
transverse energy, $\ET^\nu$.  Beyond LO, there are small difference,
because of the effect of hadronic energy falling outside of the
detector.  We leave the assessment of this difference to the
experimenters.  Accordingly, we perform our calculation for a detector
with complete coverage for hadrons, so that again $\ETsl$ is the same
as $\ET^\nu$ even at NLO.

As described in ref.~\cite{NtuplesNote}, we save intermediate results
in publicly available \root{}-format~\cite{ROOT} \ntuple{}
files~\cite{NtuplesNote} in order to facilitate new studies with
different (tighter) cuts, different scale choices or PDF sets. These
files contain momenta for all partons, along with the coefficients of
various scale- or PDF-dependent functions associated with the squared
matrix elements.  Using these files we can
generate cross sections and distributions for different scales and PDF
error sets without re-running from scratch.  We impose a looser set of
cuts when generating the
underlying \ntuples, restricting only the minimum jet transverse
momentum, to $\pT^\jet > 25$~GeV. The \ntuples{} also contain the needed
information for
anti-$k_T$, $k_T$ and \SISCone{} algorithms~\cite{JetAlgorithms} for
$R=0.4, 0.5, 0.6, 0.7$, as implemented in the FASTJET
package~\cite{FastJet}.  In the \SISCone{} case the merging parameter
$f$ is chosen to be $0.75$.  This allows the \ntuples{} to be used for
studying the effects of varying the jet algorithm.  The
cuts~(\ref{Cuts}) are imposed only in analyzing the \ntuples{}.

We use the MSTW2008 LO and NLO parton distribution functions
(PDFs)~\cite{MSTW2008} at the respective orders.  We use a five-flavor
running $\alpha_s(\mu)$ and the value of $\alpha_s(M_Z)$ supplied with
the parton distribution functions.  The lepton-pair invariant mass
follows a relativistic Breit-Wigner distribution with width given by
$\Gamma_W = 2.06$ GeV and mass $M_W = 80.419$ GeV.  We take the leptonic
decay products to be massless.  In this approximation the results for
muon final states are of course identical to those for the electron.
The other electroweak parameters are also chosen as in
ref.~\cite{W3jDistributions}.

As our calculation is a parton-level one, we do not apply corrections
due to non-perturbative effects such as those induced by the
underlying event or hadronization.  For comparisons to experiment it
is of course important to account for these, although for the cross-section
ratios studied in this paper we do not expect substantial effects.

The light quarks ($u,d,c,s,b$) are all treated as massless.  As mentioned
in the introduction, we do not
include contributions to the amplitudes from a real or virtual top
quark; its omission should have a percent-level effect on the cross
sections~\cite{W4j,Z4j}, and presumably even less in ratios.
We use a leading-color
approximation for the \Wjjjj- and \Wjjjjj-jet calculations.
We also
approximate the Cabibbo-Kobayashi-Maskawa matrix by the unit matrix.
For the cuts we impose, this approximation results in a change of
under 1\% in total cross sections for \Wjjj-jet production, and should
likewise be completely negligible in our study.  We also neglected
the tiny processes involving eight quarks or anti-quarks in
real-emission contributions to $W+5$ jet production. 

There have been recent advances in matching parton showers to NNLO
predictions and in merging NLO results for different jet 
multiplicities with parton showers. While matching is suitable for
describing the production of a particular final state (say $W\,+\,1$~jet
production) at particle level, merging allows the prediction of signatures
involving different numbers of final-state jets (like $W\,+\,1$~jet,
$W\,+\,2$~jet, etc.) from the same sample of events and with the same
formal accuracy.  In our study this method will be used to obtain an
alternative prediction.

Matching was pioneered by the MC@NLO and POWHEG methods~\cite{MCNLO,POWHEG}.
The multi-scale improved NLO method (MINLO)~\cite{MINLO}, which augments
matched NLO calculations with a natural scale choice and Sudakov form factors
later evolved into a full matching technique at NNLO~\cite{NNLOPS}.
An independent procedure was introduced in ref.~\cite{UNLOPS}.

The first merging methods at LO accuracy were devised a decade 
ago~\cite{MLM,CKKW,CKKWL}. Using MC@NLO and POWHEG matching they were recently
extended to NLO accuracy~\cite{MEPSNLO,UNLOPS,FXFX}, with a substantial 
gain in theoretical precision. For the predictions included here we use
the method presented in ref.~\cite{MEPSNLO}. We utilize NLO matrix
elements up to $W\,+\,2$~jets and LO matrix elements up to $W\,+\,4$~jets.

We use the same computational setup as in ref.~\cite{W5j}.  The
virtual contributions are provided by the \BlackHat{} package which
is based on on-shell methods.  The \BlackHat{} library has previously
been used to generate the virtual contributions to NLO predictions for
a variety of processes, including those of pure jets, or jets in
association with a vector boson, a 
photon or a pair of photons~\cite{W3jDistributions,BlackHatZ3jet,W4j,Z4j,PhotonZ,FourJets,W5j}. This
library uses on-shell methods which are based on the underlying
properties of factorization and unitarity obeyed by any amplitude.
(See refs.~\cite{Primitive,Zqqgg,ItaOzeren,W5j} for references to the 
underlying methods.)

\begin{table}[t]
\vskip .4 cm
\centering
\begin{tabular}{||c||c|c||c|c|}
\hline
Jets &  $W^-$ LO  & $W^-$ NLO & $W^+$ LO & $W^+$ NLO   \\  \hline

1 & $284.0(0.1)^{+26.2}_{-24.6}$ & $351.2 (0.9)^{+16.8}_{-14.0}$ &
  $416.8(0.6)^{+38.0}_{-35.5}$ & $516(3)^{+29.}_{-23}$\\ \hline
2 & $83.76(0.09)^{+25.45}_{-18.20}$ & $83.5(0.3)^{+1.6}_{-5.2}$ & 
$130.0(0.1)^{+39.3}_{-28.1}$ & $125.1(0.8)^{+1.8}_{-7.4}$\\ \hline
3 & $21.03(0.03)^{+10.66}_{-6.55}$ & $18.3(0.1)^{+0.3}_{-1.8}$ &
 $34.72(0.05)^{+17.44}_{-10.75}$ & $29.5(0.2)^{+0.4}_{-2.8}$\\ \hline
4 & $4.93(0.02)^{+3.49}_{-1.90}$ & $3.87(0.06)^{+0.14}_{-0.62}$ &
 $8.65(0.01)^{+6.06}_{-3.31}$ & $6.63(0.07)^{+0.21}_{-1.03}$\\ \hline
5 & $1.076(0.003)^{+0.985}_{-0.480}$ & 
$0.77(0.02)^{+0.07}_{-0.19}$ & $2.005(0.006)^{+1.815}_{-0.888}$ &
 $1.45(0.04)^{+0.12}_{-0.34}$\\ \hline
\end{tabular}
\caption{Total cross sections in picobarns, as reported in ref.~\cite{W5j},
for \Wjn{} jet production at the LHC at $\sqrt{s}=7$~TeV, using the 
anti-$k_T$ jet algorithm with $R=0.5$, the cuts~(\ref{Cuts}),
and the central scale choice $\mu_R = \mu_F = \HTpartonicp/2$.
The NLO results for \Wjjjjx{}-jet production use the leading-color
approximation discussed in the text.  The numerical
  integration uncertainty is given in parentheses, and the scale dependence
  is quoted in superscripts and subscripts.
}
\label{CrossSectionAnti-kt-R5Table}
\end{table}

\section{Results}
\label{ResultsSection}

The present study, which will illustrate the principles of 
extrapolating to higher multiplicities, is based on our NLO results for
\Wjn-jet production at the LHC with $\sqrt{s}=7$~TeV,
as reported in ref.~\cite{W5j}.  The results reported there,
reproduced in \tab{CrossSectionAnti-kt-R5Table},
are for the total cross sections for \Wjn-jet production with the
standard minimum jet transverse momentum, $\pT^\jet = 25$~GeV.
The central value of the renormalization scale $\mu_R$ and
factorization scale $\mu_F$ has been set to $\HTpartonicp/2$,
and the upper and lower uncertainties come from varying $\mu_R = \mu_F$
by a factor of two in either direction around the central value.
The same ratios could be studied in the future at 8, 13, or 14~TeV.

Ratios of cross sections and of distributions are expected to provide
a cleaner comparison between experiment and theory
than the underlying absolute cross sections from which they are formed.
Experimentally, ratios should have reduced dependence on various
systematic uncertainties, most notably uncertainty in the jet-energy
scale.  The theoretical predictions of typical ratios are either
independent of $\alpha_s$ at leading order, or else behave as
$\Ord(\alpha_s)$; accordingly, they usually have a
much smaller dependence on the renormalization and factorization
scales than the underlying quantities\footnote{In a quantity of
 $\Ord(\alpha_s^0)$, the scale variation should not be expected to
 decrease in going from LO to NLO, and it is not useful
  even as a proxy for remaining theoretical uncertainties. 
  For ratios of $\Ord(\alpha_s)$, the scale
  variation is expected to decrease, and we find that it indeed does.
  Its precise value depends, however, on additional arbitrary choices:
  what central scales should be chosen in the numerator and
  denominator? (In \Tab{CrossSectionAnti-kt-R5Table}, the scales are
  chosen differently for different multiplicities.)  Should the
  variation be taken in correlated or uncorrelated fashion in the
  numerator and denominator? ({In either case, the information
    presented in \Tab{CrossSectionAnti-kt-R5Table} does not suffice to
    evaluate the variation in ratios.})  These questions also cast
  doubt on the utility of scale variation as a proxy for theoretical
  uncertainties for the ratios we consider, and therefore
  we omit such variation.}.  Ratios also reduce some
of the limitations of fixed-order results compared to parton-shower
simulations, as well uncertainties from parton-distribution
functions~\cite{KomStirling}.  We will display two explicit examples
of reduced uncertainty with respect to the parton-distribution
functions in this section.

\begin{figure}[tb]
\begin{center}
\begin{minipage}[b]{1.03\linewidth}
\null\hskip -5mm
\includegraphics[clip,scale=0.44]{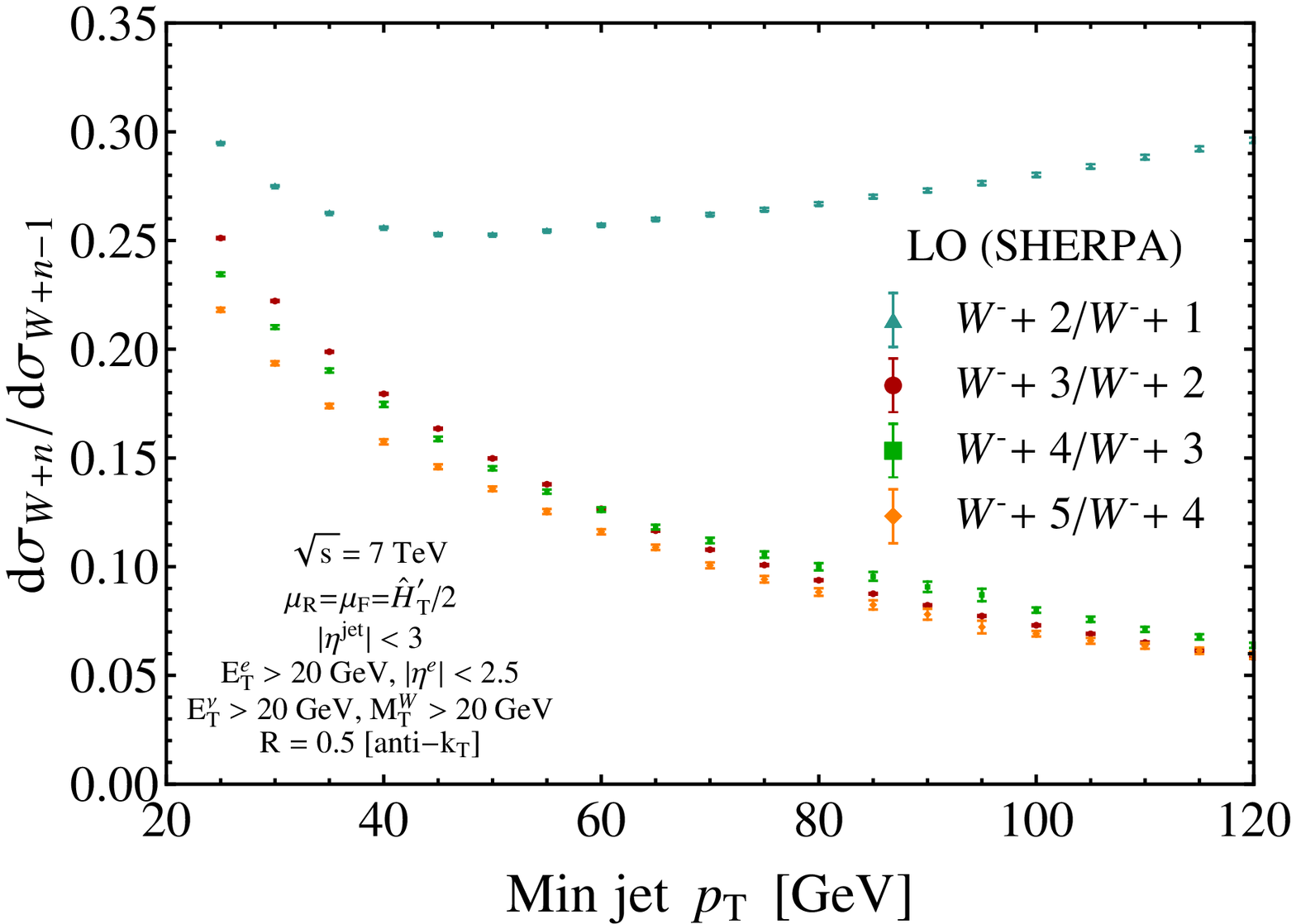}
\includegraphics[clip,scale=0.44]{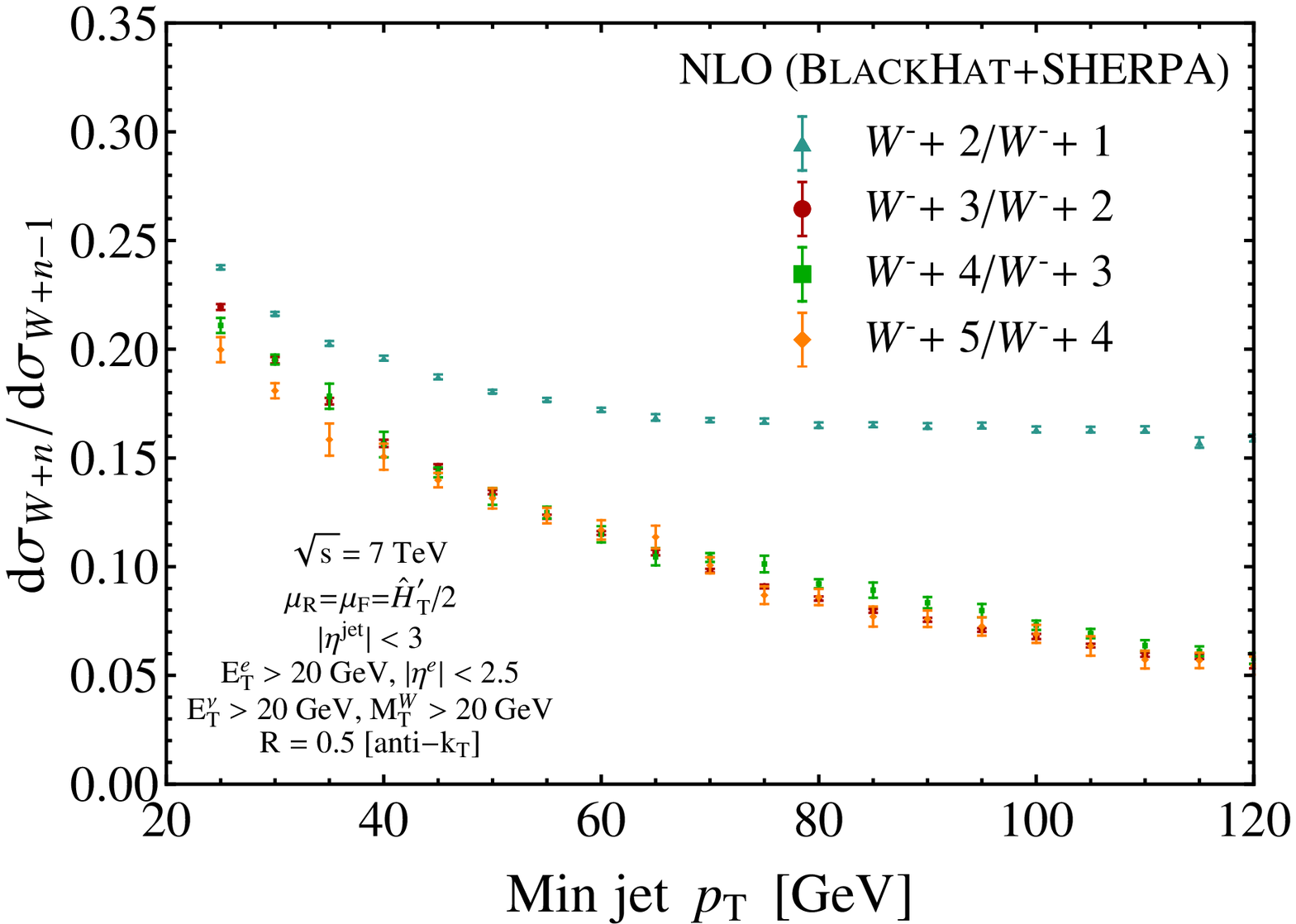}
\end{minipage}
\end{center}
\caption{The ratio of the $W^-+n$-jet to the $W^-+(n-1)$-jet cross section 
as a function of 
the minimum jet transverse momentum, $\pTmin$.
The left plot shows the ratio at LO, and the right plot at NLO.
The error bars represent numerical integration errors.
}
\label{WmMinJetPtJetProductionRatioFigure}
\end{figure}

\begin{figure}[tb]
\begin{center}
\begin{minipage}[b]{1.03\linewidth}
\null\hskip -5mm
\includegraphics[clip,scale=0.44]{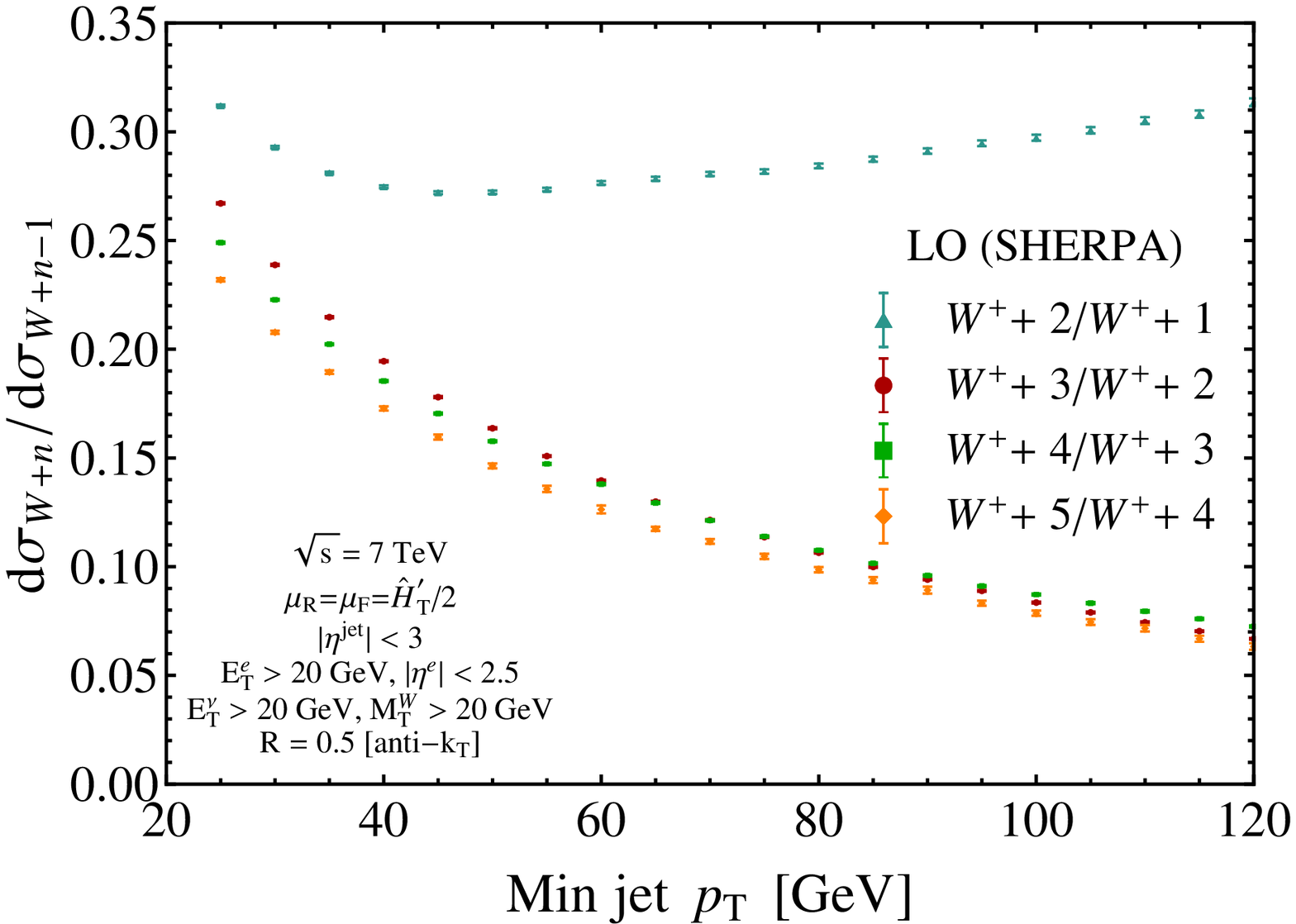}
\includegraphics[clip,scale=0.44]{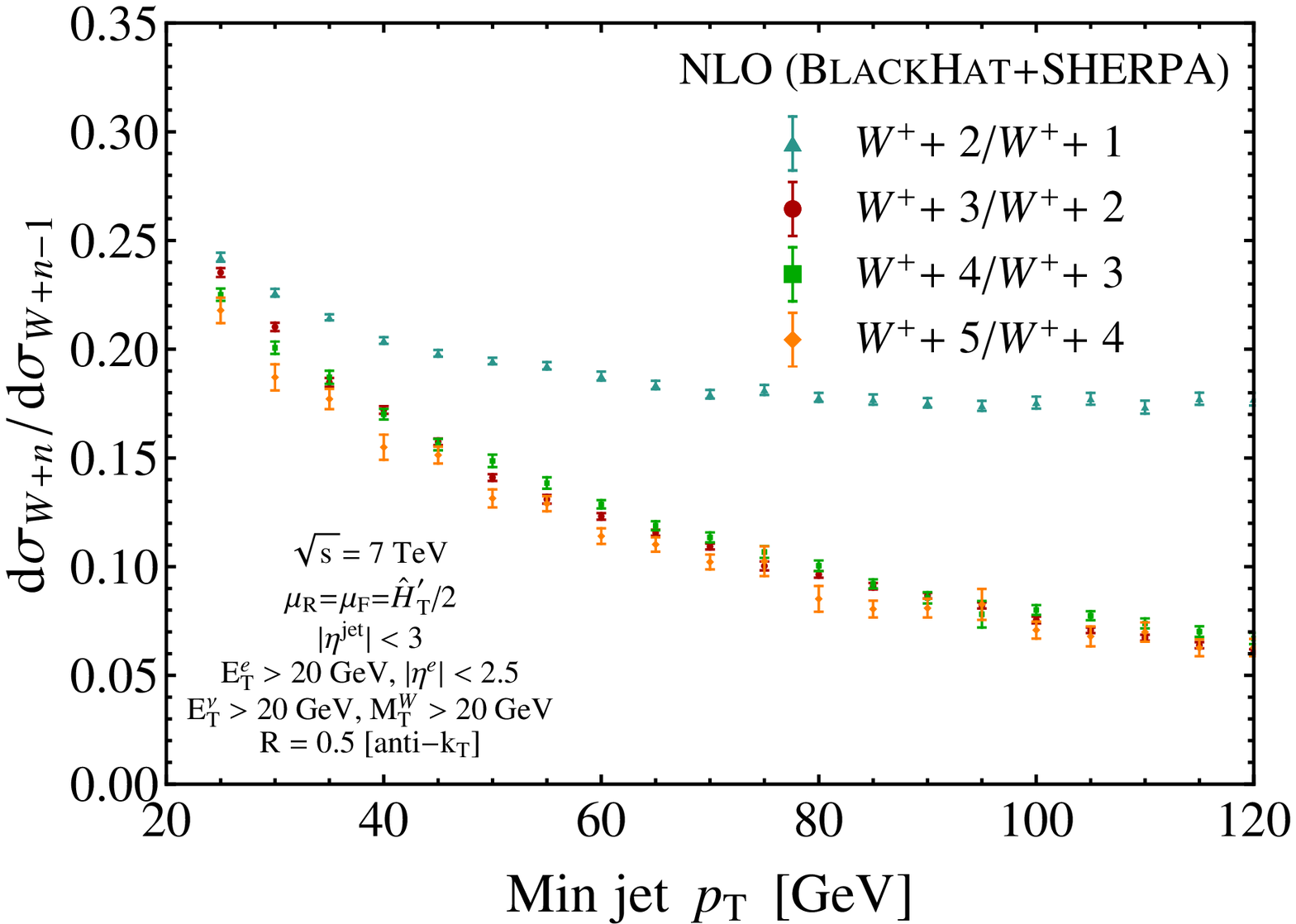}
\end{minipage}
\end{center}
\caption{The ratio of the $W^++n$-jet to the $W^++(n-1)$-jet cross section 
as a function of 
the minumum jet $\pT$.
The left plot shows the ratio at LO, and the right plot at NLO.
}
\label{WpMinJetPtJetProductionRatioFigure}
\end{figure}

\begin{figure}[tb]
\begin{center}
\begin{minipage}[b]{1.03\linewidth}
\null\hskip -5mm
\includegraphics[clip,scale=0.44]{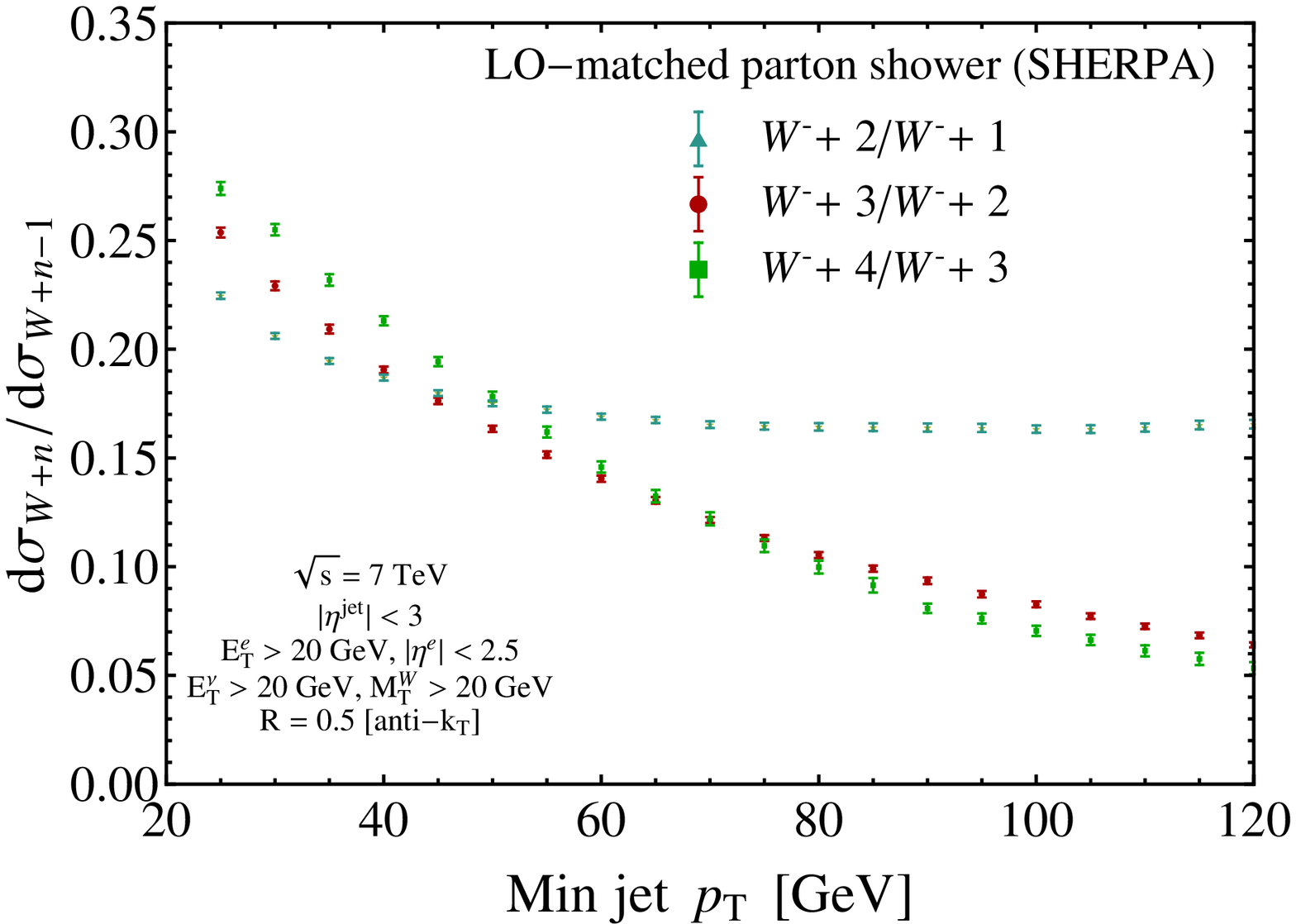}
\includegraphics[clip,scale=0.44]{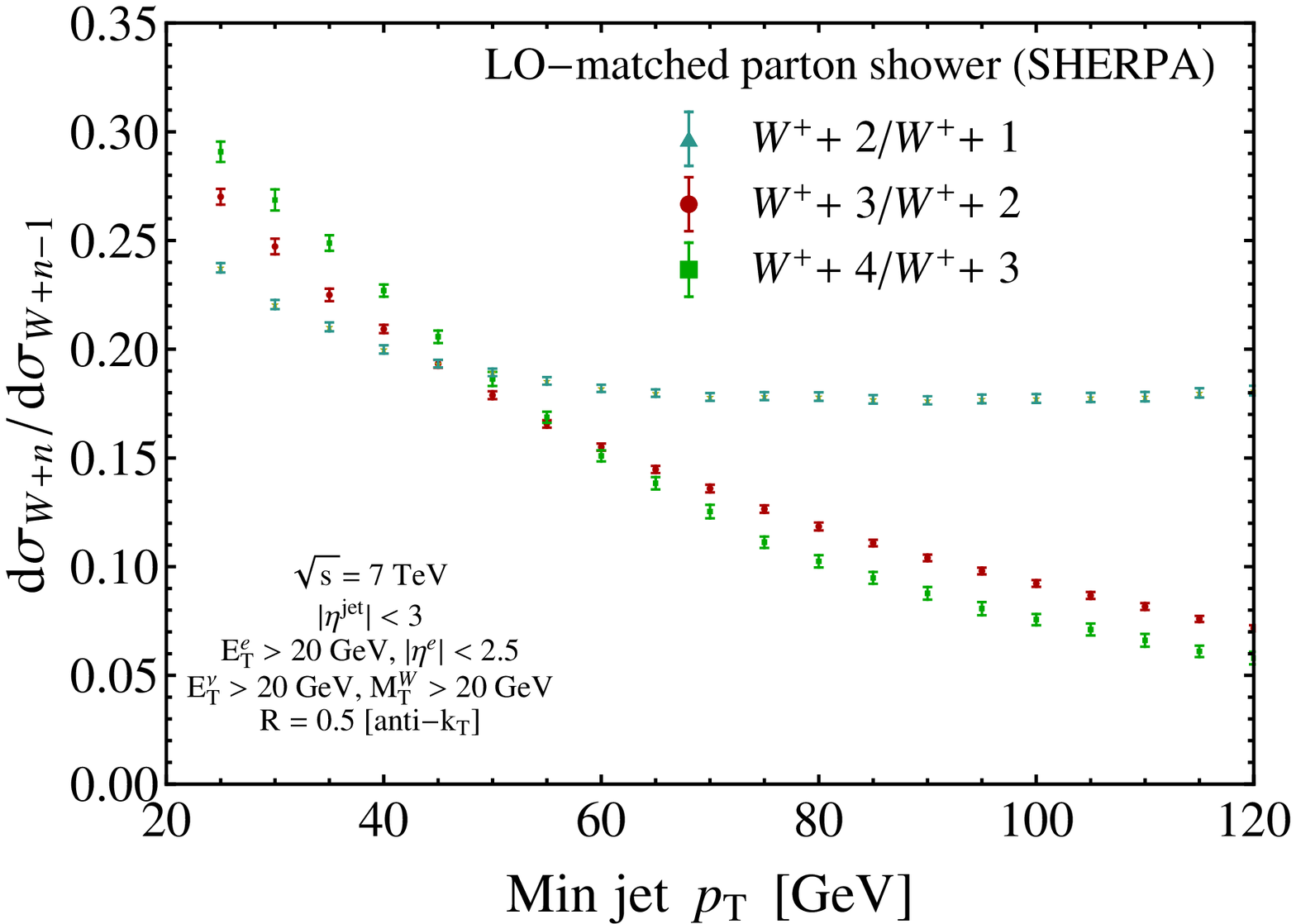}
\end{minipage}
\end{center}
\caption{The ratio of the $W+n$-jet to the $W+(n-1)$-jet cross section 
as a function of the minimum jet transverse momentum, $\pTmin$,
for a parton shower matched to LO.  The left plot shows the ratio for $W^-$,
and the right plot for $W^+$.
}
\label{WMinJetPtJetProductionRatioPSFigure}
\end{figure}

\subsection{Dependence on Minimum Jet Transverse Momentum}
\label{MinJetPTSection}

We first examine the dependence of the jet-production ratio on the minimum
jet $\pT$\footnote{We thank Maria Spiropulu for pointing out to us the
  importance of this quantity.}.  We have studied the ratio of
\Wjn-jet to \Wjnm-jet production at both LO and NLO for a range of
minimum jet $\pT$s from 25 to 120~GeV.  We provide tables of our
detailed results in appendix~\ref{MinJetpTAppendix}.  We show these
data graphically in
\figs{WmMinJetPtJetProductionRatioFigure}{WpMinJetPtJetProductionRatioFigure}
as a function of the same cut.  We see that the \Wjj-jet to \Wj-jet
ratio is very different from the other ratios.  It suffers from large
NLO corrections, is numerically quite different, and is flat with
increasing jet $\pTmin$ (at NLO).  In contrast, the ratios of total
cross sections for \Wjn-jet to \Wjnm-jet production with $n\ge3$ are
similar numerically, fall in a similar manner, and have only modest
NLO corrections.  Only at the lowest jet $\pTmin$ values does the
\Wjj-jet to \Wj-jet ratio take on a value comparable to the other ratios,
a similarity which is likely accidental.

The dissimilarity of the \Wjj-jet to \Wj-jet ratio and the large NLO
corrections to it are not surprising.  At LO, \Wj-jet production does
not include contributions from the $gg$ initial-state; these arise
only at NLO.  In contrast, at LO \Wjn-jet production with $n\ge2$ already
includes contributions from all initial-state partons.  Furthermore, both
\Wj-jet and \Wjj-jet production suffer from kinematic constraints
which are relaxed at NLO.  For example, at LO the leading jet in
\Wjj-jet production must be opposite in azimuthal angle
to the vector boson; at NLO, configurations with the leading jet
near the vector boson are possible.

Ratios for $n\ge 3$ are largely independent of $n$
for the range of $\pTmin$ values that we have studied.  The slight differences
between different ratios at LO narrow considerably at NLO.
The similarity of ratios suggests a certain universality.  The narrowing
of differences when going from an LO to an NLO prediction
suggests that some of the residual differences at LO are a result of using
different renormalization and factorization scales in the numerator and
denominator of the ratios, together with the strong
scale dependence of the LO cross sections.
The similarity is in agreement with the results of
refs.~\cite{Englert,Gerwick} arguing for constant ratios when all jets
are subject to an identical $\pTmin$ cut, as is the case here.

For comparison, in \fig{WMinJetPtJetProductionRatioPSFigure}
we show the same ratios computed using the \SHERPA{}
parton shower matched to LO.  The \Wjj-jet to \Wj-jet ratio is similar
in its $\pTmin$ dependence to the NLO result, suggesting 
that it should not suffer large corrections beyond NLO.  The
\Wjjj-jet to \Wjj-jet and \Wjjjj-jet to \Wjjj-jet ratios are similar
in shape to the NLO results, although they are about 20\% higher
at smaller values of $\pTmin$.
The LO-matched results are broadly consistent with universality 
for $n\ge3$ seen in the NLO prediction.   We do not provide ratios for
\Wjjjjj-jet production, because of computational limitations with
current technology (more specifically, the parton-shower based clustering).

In ref.~\cite{ExptJetProductionRatioLHC}, the dependence of the
jet-production ratio on $n$ was studied.  This ratio
(or rather its inverse) was parametrized as
\begin{equation}
\sigma(\text{\Wjn\ jets})/\sigma(\text{\Wjnp1\ jets}) = \alpha+\beta n\,.
\end{equation}
The universality of jet-production ratios at different values of $\pTmin$
suggests that we try a parametrization that allows us to study
their $p_{\rm T}$-dependence.
We consider the following parametrization,
\begin{equation}
\sigma(\text{\Wjn\ jets})/\sigma(\text{\Wjnm\ jets}) = 
r_n(\pTmin)\,,
\end{equation}
where we take the following form for $r_n(p)$,
\begin{equation}
r_n(p) = b_n p^{-\eta_n} e^{-d_n p}\,.
\label{FitForm}
\end{equation}
As explained above, we consider only values of
$n$ greater than 1; the $n=1$ case should be treated separately.  
Our main interest is in the overall power behavior described by the
second factor on the right-hand side.  We fit $b_n$, $\eta_n$ and $d_n$
to the results for the \Wjjj-jet to \Wjj-jet, 
\Wjjjj-jet to \Wjjj-jet,
and \Wjjjjj-jet to \Wjjjj-jet production 
ratios (corresponding to the last three columns of
the tables in appendix~\ref{MinJetpTAppendix}).  A power of
$\alpha_s$ is absorbed into parameter $b_n$.

The form~(\ref{FitForm}) is purely a fit. While it captures the overall
features of the curves, it should not be expected to reflect the exact
underlying physics.  As we increase the statistical precision of the results,
the quality of the fit should be expected to decrease.
Nonetheless, we can distinguish between fits that
do not work at all and those that are reasonable. For example,
omitting the last factor in \eqn{FitForm} gives very poor fits,
with $\chi^2$/dof of order $150$ for the \Wmjjj-jet to \Wmjj-jet
production ratio, for
example, whereas the fits of the form in \eqn{FitForm} to the LO
results give $\chi^2$/dof around $2.2$ (or equivalently a likelihood
of $3\cdot 10^{-3}$; that is, the probability that the $\chi^2$ 
will exceed $2.2$).  
While this fit is thus marginal, it is not terrible.  The fits to the
LO \Wmjjjj-jet to \Wmjjj-jet and \Wmjjjjj-jet to \Wmjjjj-jet ratios are better.
The fits to the NLO results are also better, simply because the statistical
uncertainties are larger.  (The fits to \Wpjn-jet ratios are all acceptable.)
It is remarkable that we obtain such good
fits with only three parameters and a simple functional form.  Given
expected experimental uncertainties, the form in \eqn{FitForm} should
give a very good fit to experimental data as well.  

We perform a logarithmic fit to the form in \eqn{FitForm}.  That is,
we perform a linear (least-squares) fit of the log of ratios of
\Wjn-jet production to the log of the right-hand side of
\eqn{FitForm}.  The results of our fits are shown in
\tab{WmMinJetPTFit} for \Wmjn-jet production, and in
\tab{WpMinJetPTFit} for \Wpjn-jet production.  We compute the error
estimates using an ensemble of 10,000 fits to synthetic data.  Each
synthetic data point is taken from a Gaussian distribution with
central value given by the computed \Wjn-jet cross section (at the
appropriate $\pTmin$) and width given by the computed statistical
uncertainty in the values underlying the ratios in tables
\ref{WmJetProductionRatioJetPtLOTable}--\ref{WpJetProductionRatioJetPtNLOTable}. The
cross sections shown in tables
\ref{WmJetProductionRatioJetPtLOTable}--\ref{WpJetProductionRatioJetPtNLOTable}
for different values of $\pTmin$ are not independent, because the same
underlying samples of events are used to compute them.  Thus, for example, all
events that contribute to the cross section with $\pTmin=80$~GeV also
contribute to the cross section with $\pTmin=50$~GeV.  For the LO
fits, we nonetheless treat the statistical uncertainties as
independent, as including the correlations yields only insignificant
differences.  For the NLO fits, we include the full correlation matrix
in generating the synthetic data; this makes a noticeable difference
for the \Wjjjj-jet to \Wjjj-jet and \Wjjjjj-jet to \Wjjjj-jet ratios.  (In
performing each of the 10,000 fits, the different data points are
weighted, in a least-squares procedure, by the diagonal statistical
uncertainties, which we treat as independent; this makes only a small
difference to the final parameters.)  The quoted uncertainties are for
each parameter taken independently, and do not take into account any
correlations between fit parameters.  Two of the fit parameters,
$\eta_n$ and $b_n$, are dimensionless, while $d_n$ has units of
GeV${}^{-1}$.

\def\fitWmLOa{~$ 0.480 \pm 0.008$~ &~$0.0077 \pm 0.0002$~ &~$1.43 \pm 0.03$~}
\def\fitWmNLOa{~$ 0.46 \pm 0.03$~ &~$0.0074 \pm 0.0005$~ &~$1.15 {}^{+0.10}_{-0.09}$~}
\def\fitWmLOb{~$ 0.48 \pm 0.02$~ &~$0.0057 \pm 0.0003$~ &~$1.26 \pm 0.06$~}
\def\fitWmNLOb{~$ 0.44 \pm 0.07$~ &~$0.006 \pm 0.001$~ &~$1.0 \pm 0.2$~}
\def\fitWmLOc{~$ 0.47 \pm 0.02$~ &~$0.0064 \pm 0.0004$~ &~$1.15 {}^{+0.08}_{-0.07}$~}
\def\fitWmNLOc{~$ 0.26 \pm 0.12$~ &~$0.009 \pm 0.002$~ &~$0.6 \pm 0.2$~}

\begin{table}
\begin{tabular}{||cc|l|l|l||}
\hline
\multicolumn{2}{||c}{\multirow{2}{*}{Ratio}} & \multicolumn{3}{|c||}{Fit Values}\\
\cline{3-5}
&& \multicolumn{1}{|c|}{$\eta_n$} & \multicolumn{1}{|c|}{$d_n$} & \multicolumn{1}{|c||}{$b_n$} \\
\hline
\multirow{2}{*}{$\displaystyle\quad\frac{W^-+ 3}{W^-+2}\quad$}
& \multicolumn{1}{|c|}{LO} &\fitWmLOa \\
\cline{2-5}
& \multicolumn{1}{|c|}{~NLO~~} &\fitWmNLOa \\
\hline
\multirow{2}{*}{$\displaystyle\frac{W^-+ 4}{W^-+3}$}
& \multicolumn{1}{|c|}{LO} &\fitWmLOb \\
\cline{2-5}
& \multicolumn{1}{|c|}{~NLO~~} &\fitWmNLOb \\
\hline
\multirow{2}{*}{$\displaystyle\frac{W^-+ 5}{W^-+4}$}
& \multicolumn{1}{|c|}{LO} &\fitWmLOc \\
\cline{2-5}
& \multicolumn{1}{|c|}{~NLO~~} &\fitWmNLOc \\
\hline
\end{tabular}
\caption{Fit parameters for the jet-production ratio in \Wmjn{} jets as a function of the minimum jet $\pT$, using the form in \eqn{FitForm}.
}
\label{WmMinJetPTFit}
\end{table}

\def\fitWpLOa{~$ 0.457 \pm 0.006$~ &~$0.0071 \pm 0.0001$~ &~$1.39 \pm 0.03$~}
\def\fitWpNLOa{~$ 0.49 \pm 0.04$~ &~$0.0060 \pm 0.0007$~ &~$1.3 {}^{+0.2}_{-0.1}$~}
\def\fitWpLOb{~$ 0.459 \pm 0.007$~ &~$0.0055 \pm 0.0001$~ &~$1.25 \pm 0.03$~}
\def\fitWpNLOb{~$ 0.38 \pm 0.06$~ &~$0.007 \pm 0.001$~ &~$0.9 {}^{+0.2}_{-0.1}$~}
\def\fitWpLOc{~$ 0.42 \pm 0.02$~ &~$0.0066 \pm 0.0004$~ &~$1.06 {}^{+0.06}_{-0.05}$~}
\def\fitWpNLOc{~$ 0.50 \pm 0.11$~ &~$0.005 \pm 0.002$~ &~$1.3 {}^{+0.5}_{-0.4}$~}

\begin{table}
\begin{tabular}{||cc|l|l|l||}
\hline
\multicolumn{2}{||c}{\multirow{2}{*}{Ratio}} & \multicolumn{3}{|c||}{Fit Values}\\
\cline{3-5}
&& \multicolumn{1}{|c|}{$\eta_n$} & \multicolumn{1}{|c|}{$d_n$} & \multicolumn{1}{|c||}{$b_n$} \\
\hline
\multirow{2}{*}{$\displaystyle\quad\frac{W^++ 3}{W^++2}\quad$}
& \multicolumn{1}{|c|}{LO} &\fitWpLOa \\
\cline{2-5}
& \multicolumn{1}{|c|}{~NLO~} &\fitWpNLOa \\
\hline
\multirow{2}{*}{$\displaystyle\frac{W^++ 4}{W^++3}$}
& \multicolumn{1}{|c|}{LO} &\fitWpLOb \\
\cline{2-5}
& \multicolumn{1}{|c|}{~NLO~} &\fitWpNLOb \\
\hline
\multirow{2}{*}{$\displaystyle\frac{W^++ 5}{W^++4}$}
& \multicolumn{1}{|c|}{LO} &\fitWpLOc \\
\cline{2-5}
& \multicolumn{1}{|c|}{~NLO~} &\fitWpNLOc \\
\hline
\end{tabular}
\caption{Fit parameters for the jet-production ratio in \Wpjn{} jets as a function of the minimum jet $\pT$, using the form in \eqn{FitForm}.
}
\label{WpMinJetPTFit}
\end{table}

\def\tighten{\vspace{-6mm}}
\def\tightcaption{\baselineskip=15pt}

\begin{figure}[tb]
\begin{center}
\begin{minipage}[b]{1.\linewidth}
\begin{tabular}{cc}
\hskip -3mm
\includegraphics[clip,scale=0.54]{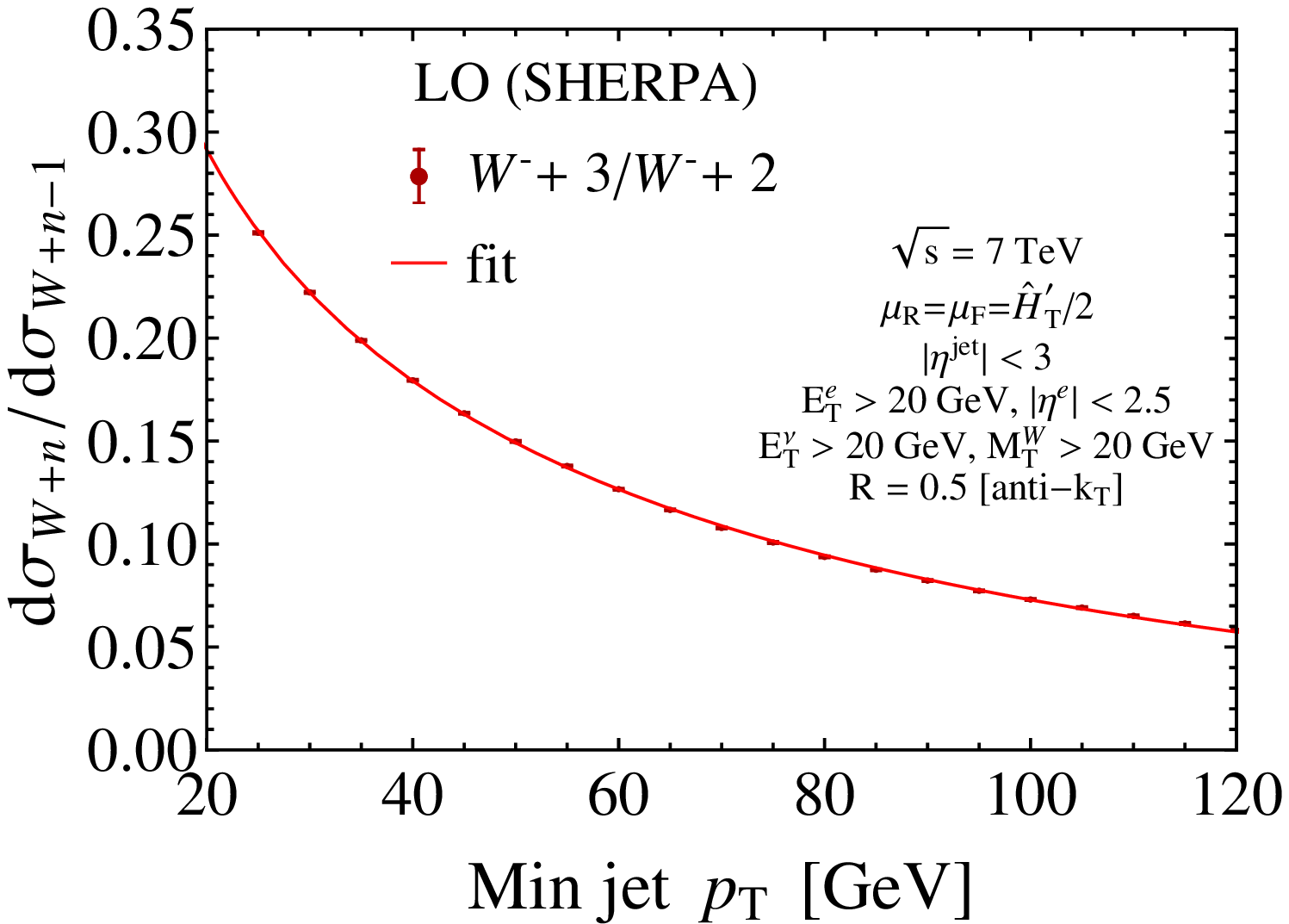}
&\includegraphics[clip,scale=0.54]{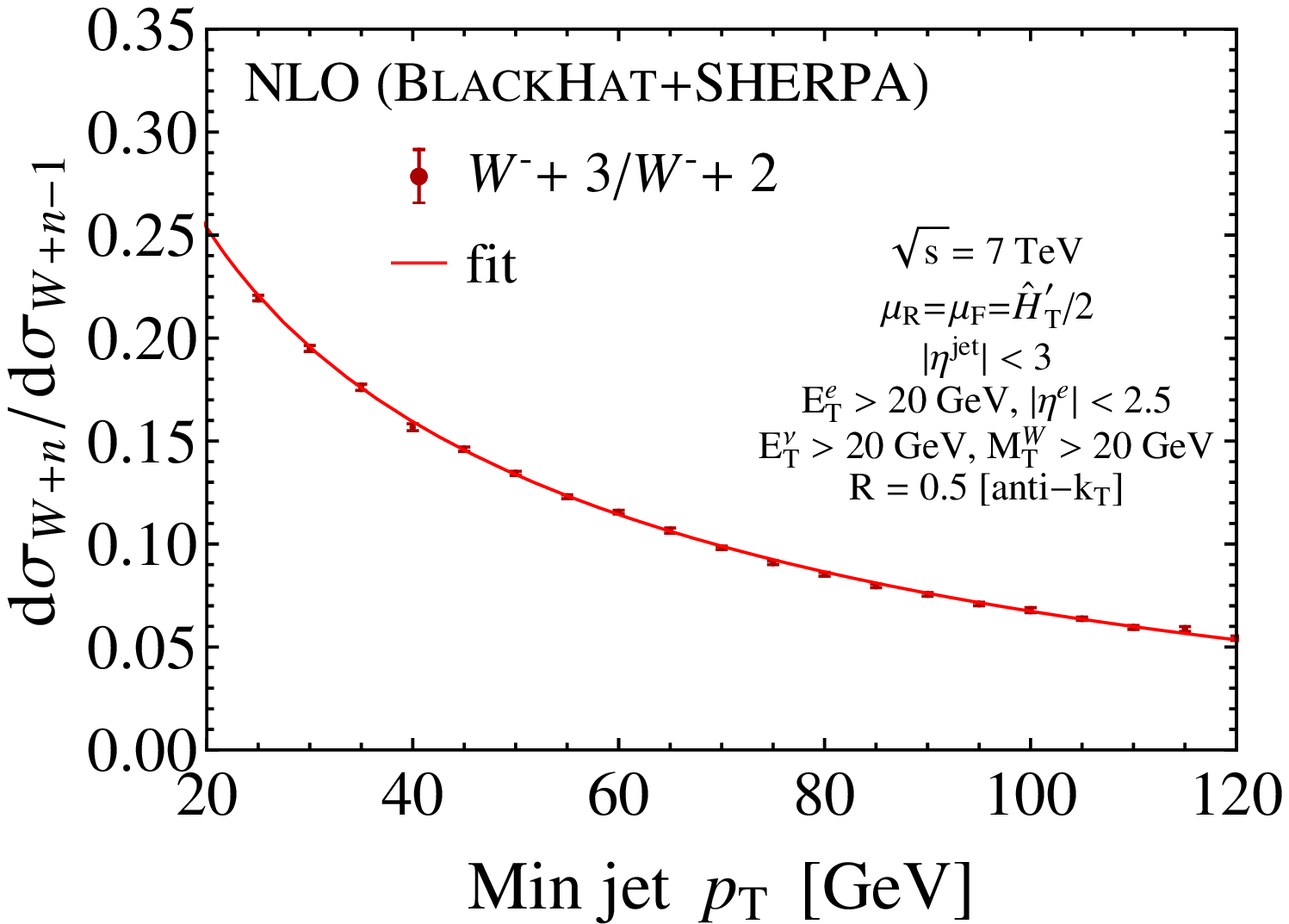}\\[-3mm]
(a)&(b)\\[3mm]
\includegraphics[clip,scale=0.54]{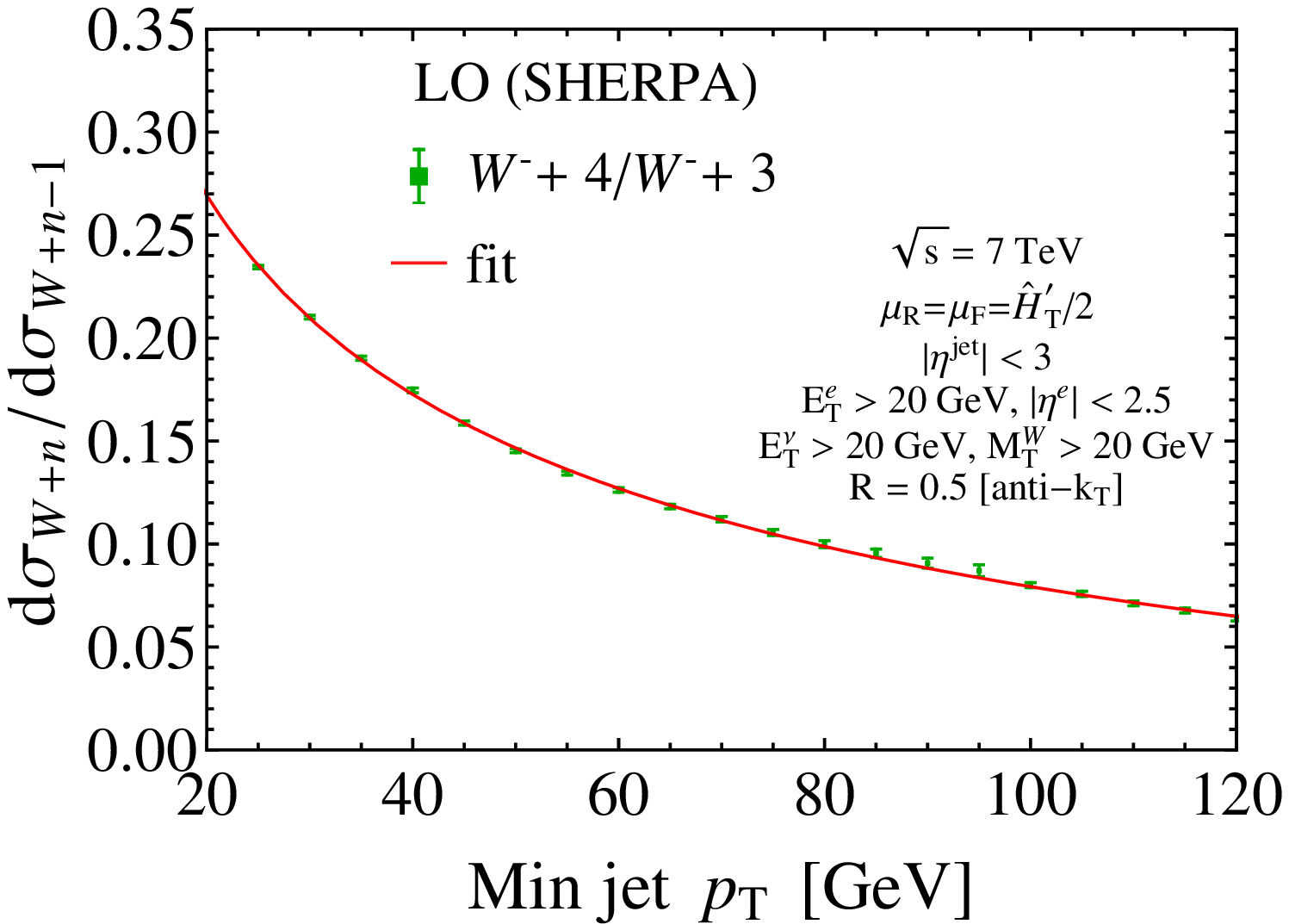}
&\includegraphics[clip,scale=0.54]{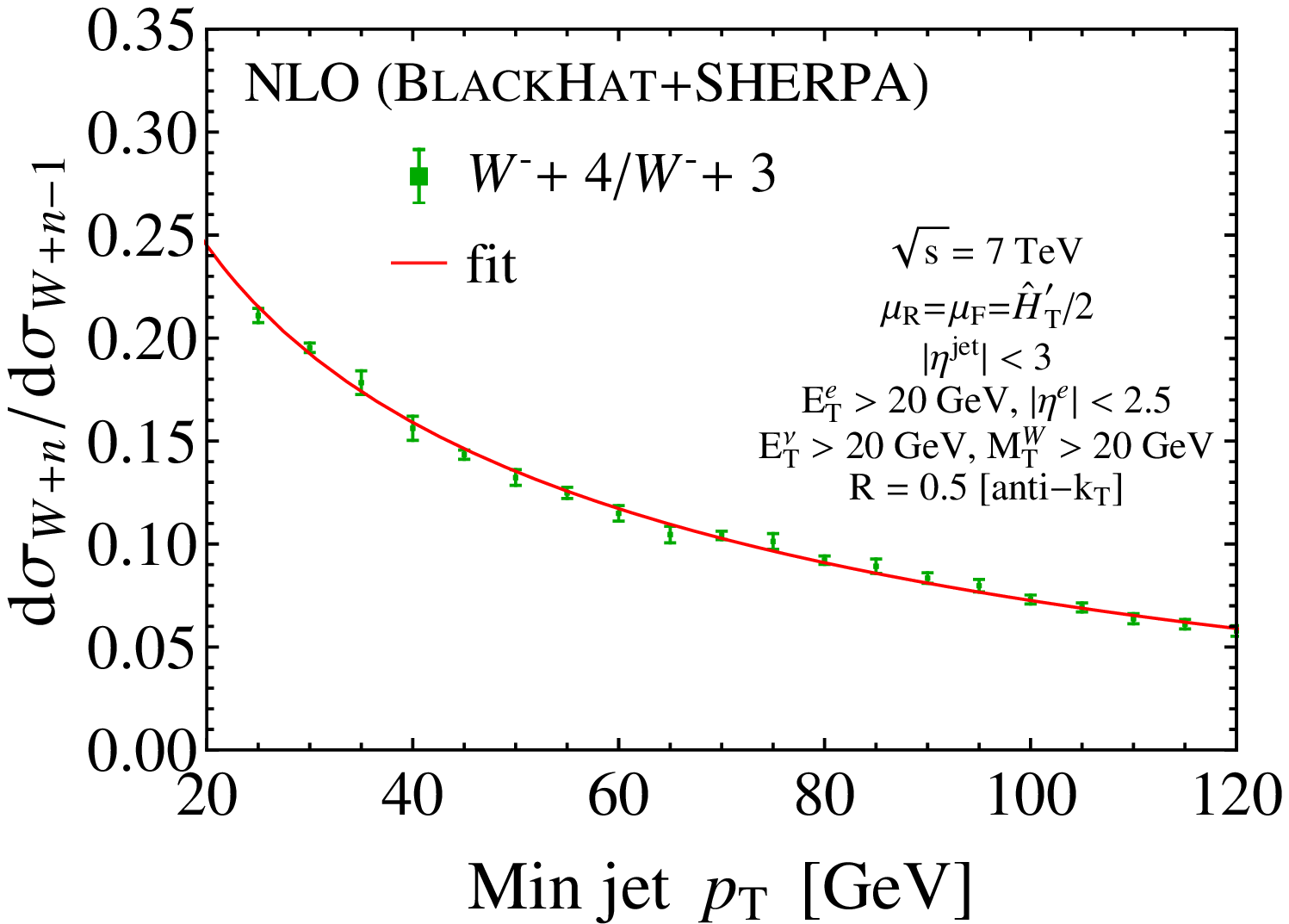}\\[-3mm]
(c)&(d)\\[3mm]
\includegraphics[clip,scale=0.54]{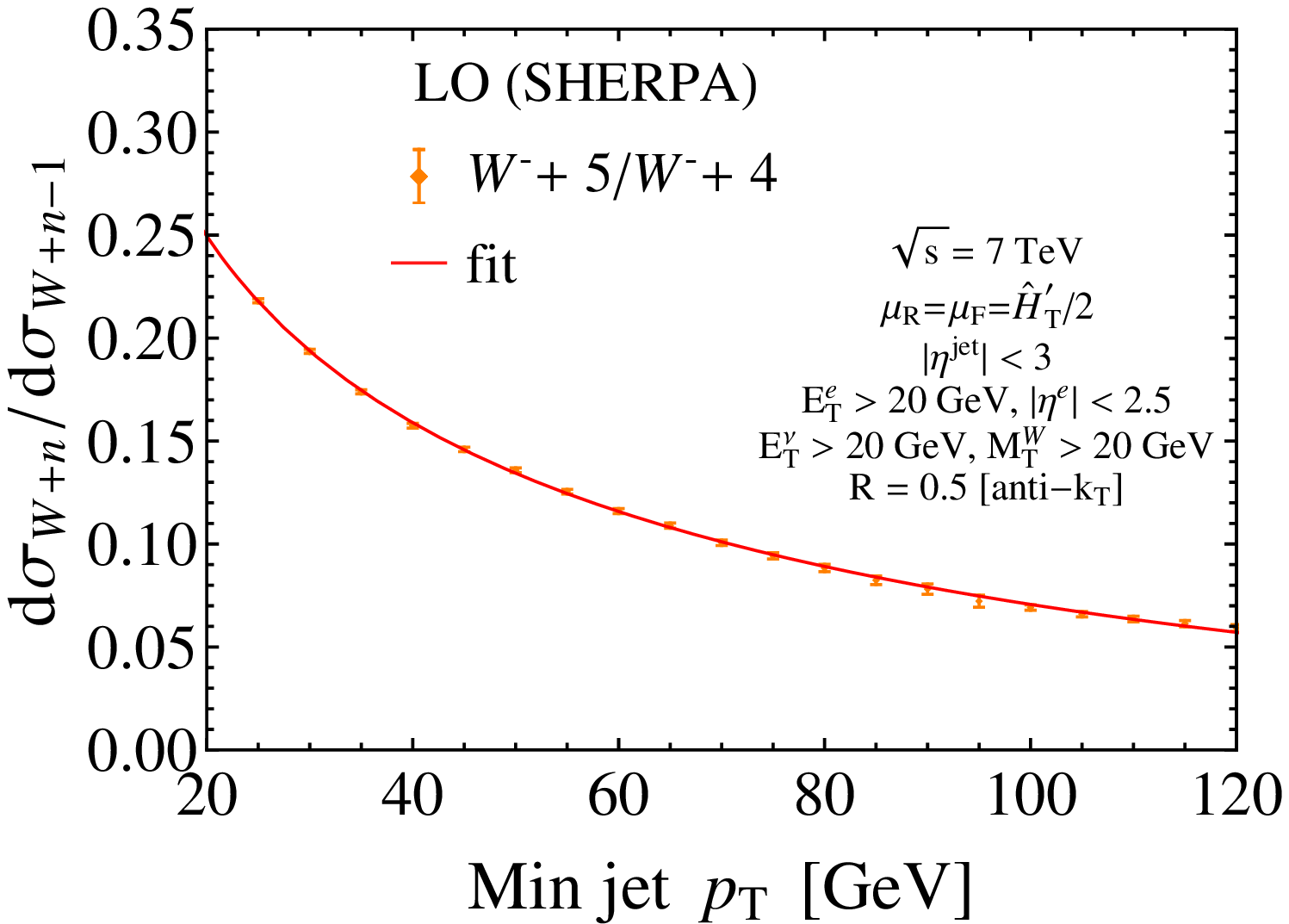}
&\includegraphics[clip,scale=0.54]{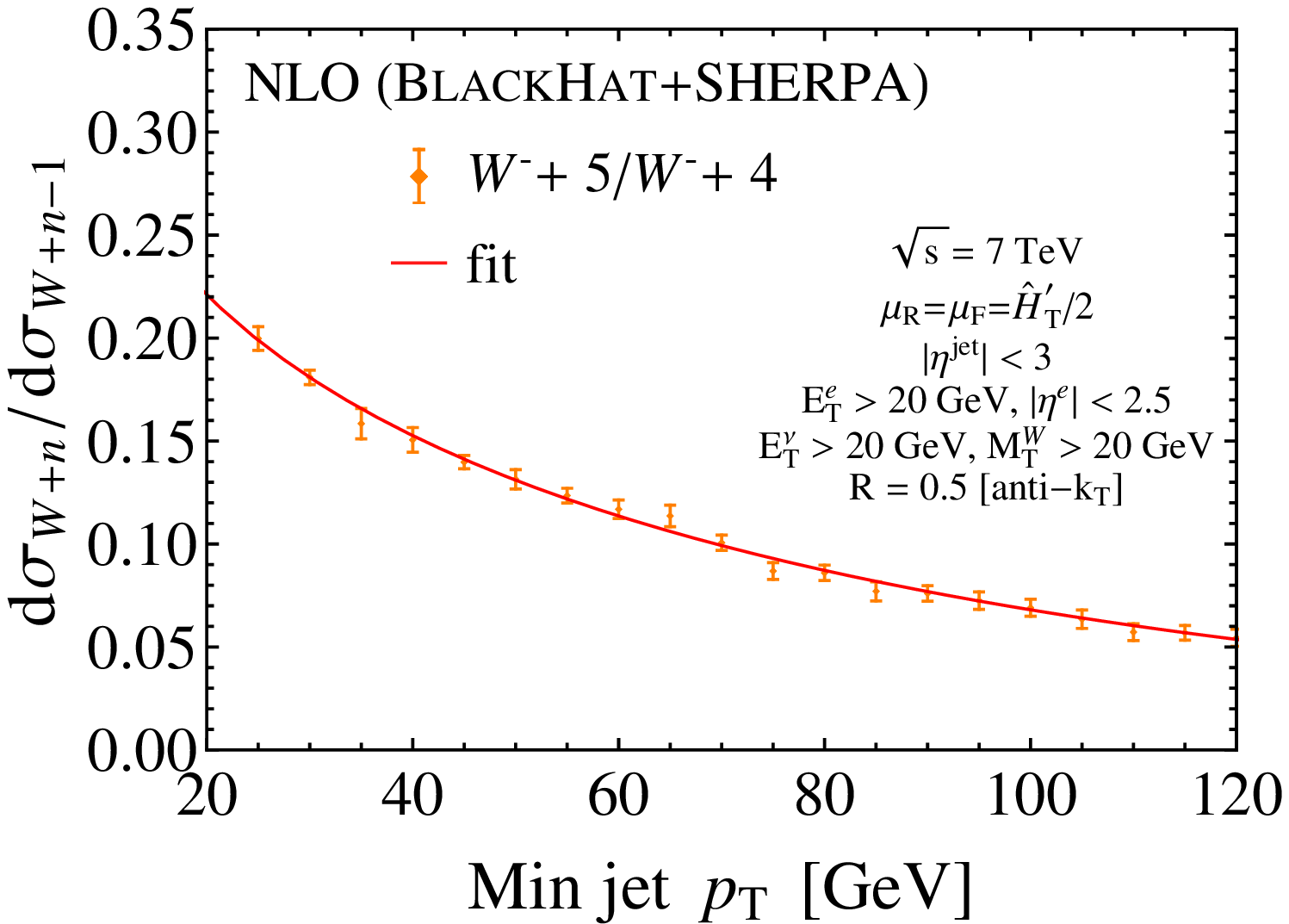}\\[-3mm]
(e)&(f)
\end{tabular}
\end{minipage}
\end{center}
\tighten
\caption{\tightcaption
Fits to the ratio of the \Wmjn-jet to \Wmjnm-jet cross sections
as a function of the jet $\pTmin$ cut.  
In the left column we show the ratios at LO and in the right column, at NLO.
From top to bottom, $n$ goes from 3 to 5.}
\vskip .5 cm 
\label{WmMinJetPtJetProductionRatioFitFigure}
\end{figure}

\begin{figure}[tb]
\begin{center}
\begin{minipage}[b]{1.\linewidth}
\hskip -3mm\begin{tabular}{cc}
\includegraphics[clip,scale=0.54]{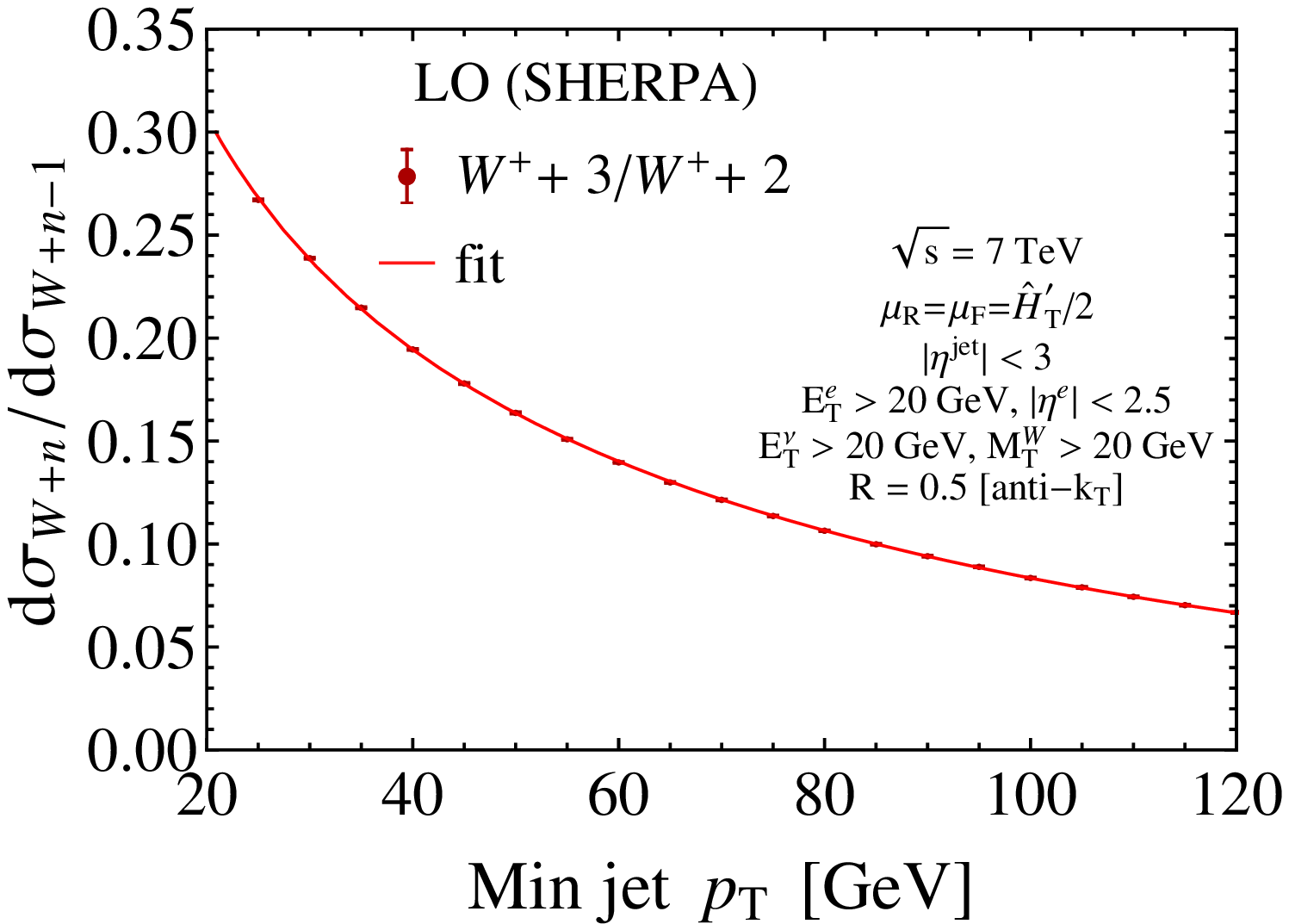}
&\includegraphics[clip,scale=0.54]{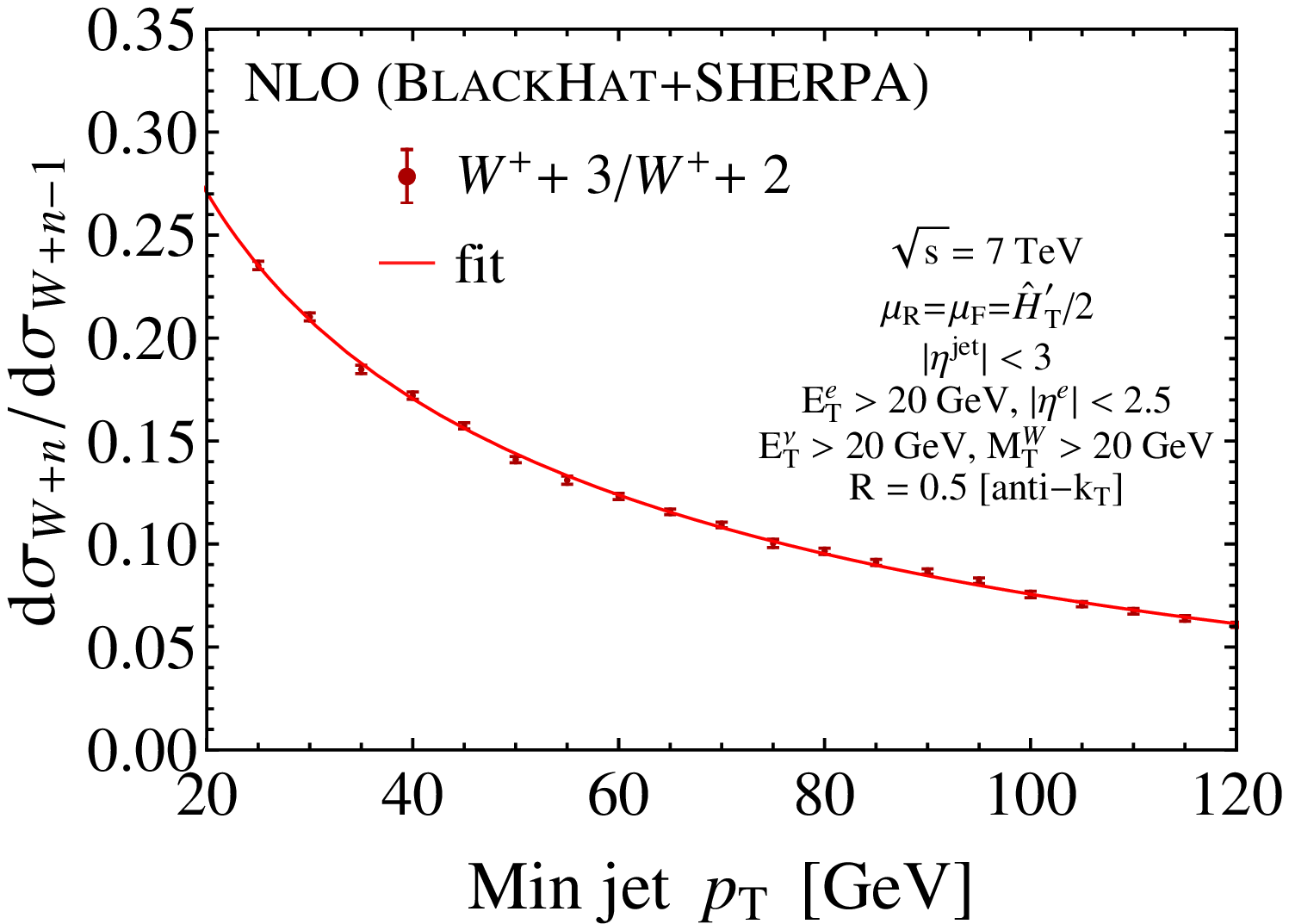}\\[-3mm]
(a)&(b)\\[3mm]
\includegraphics[clip,scale=0.54]{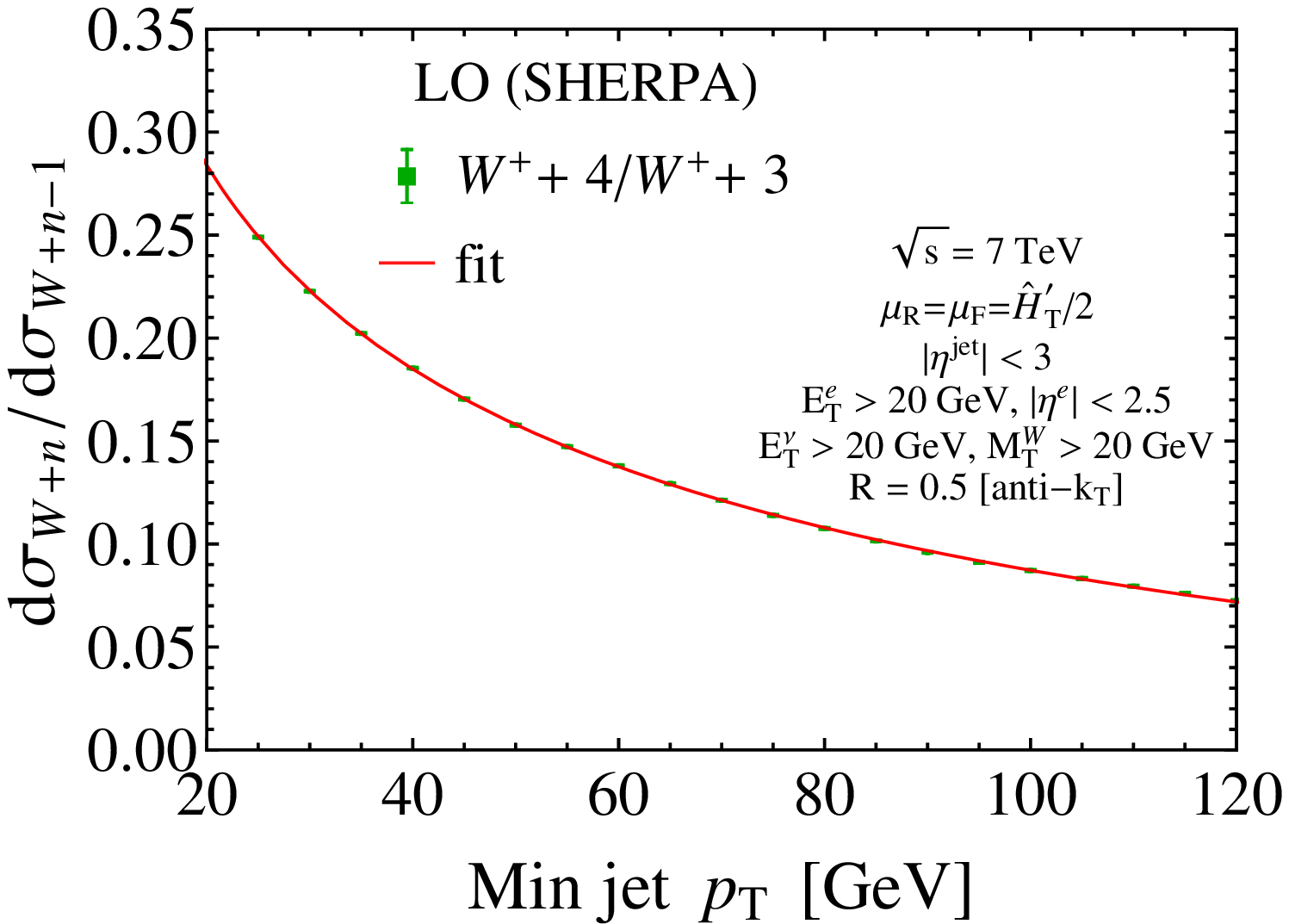}
&\includegraphics[clip,scale=0.54]{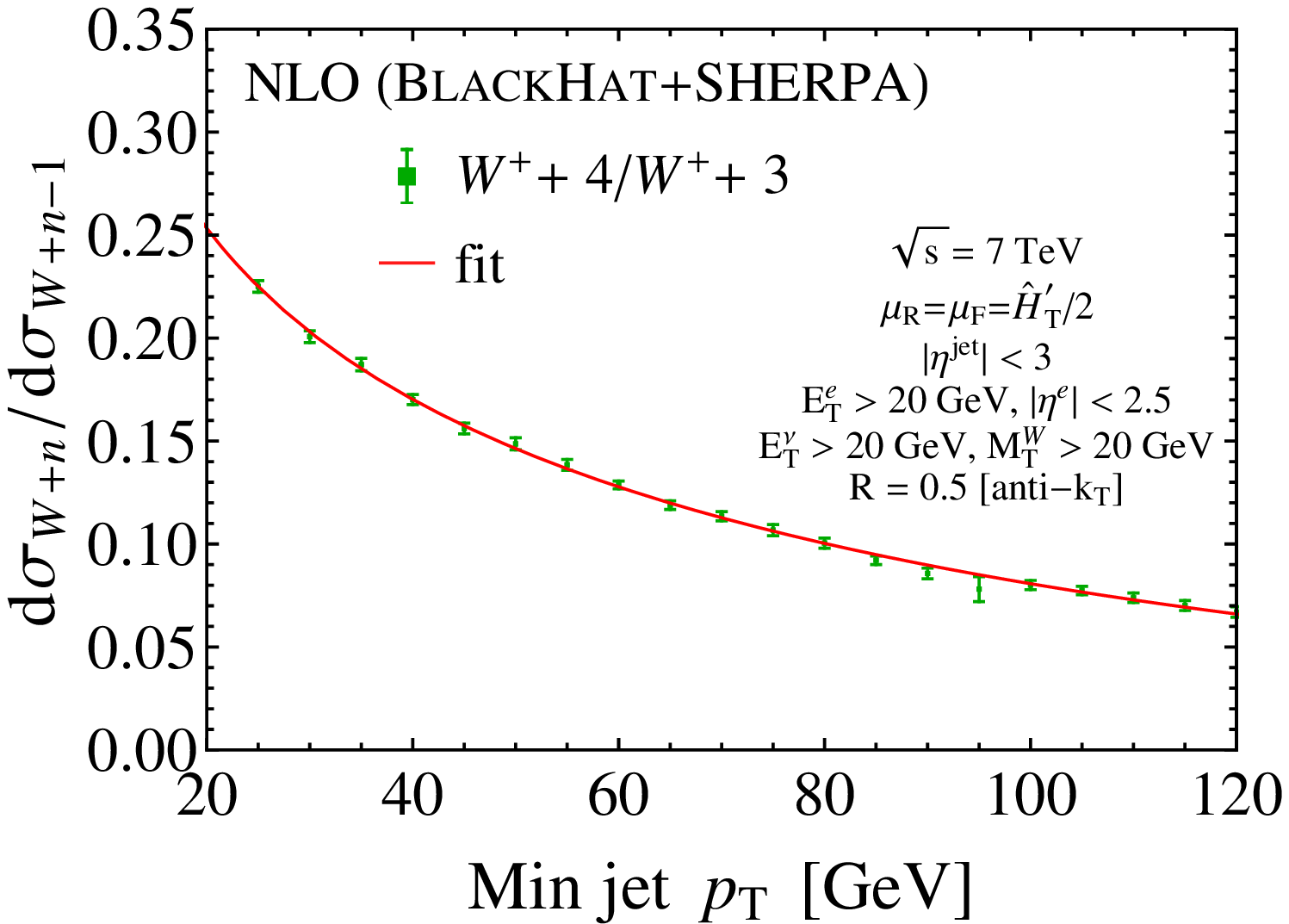}\\[-3mm]
(c)&(d)\\[3mm]
\includegraphics[clip,scale=0.54]{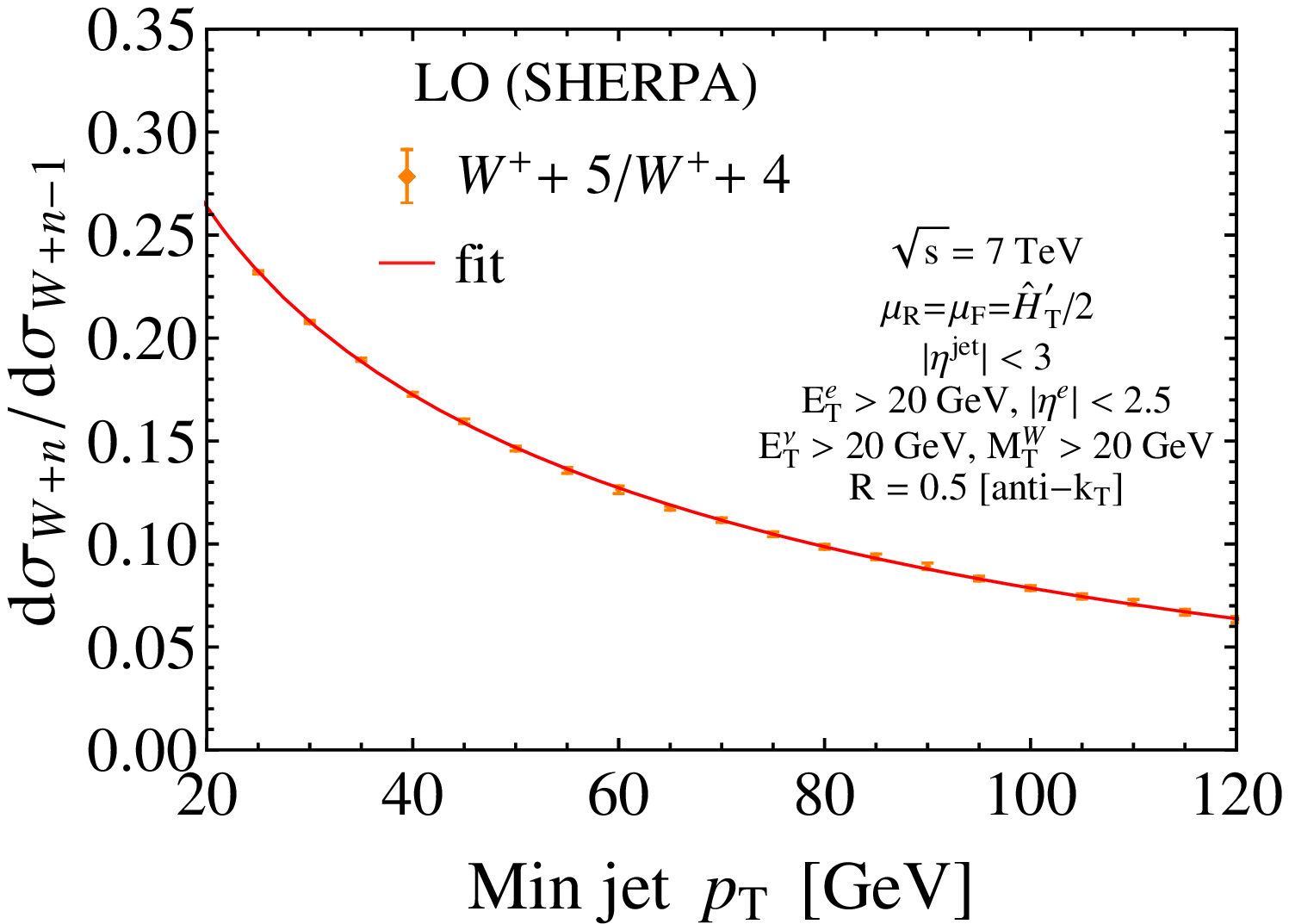}
&\includegraphics[clip,scale=0.54]{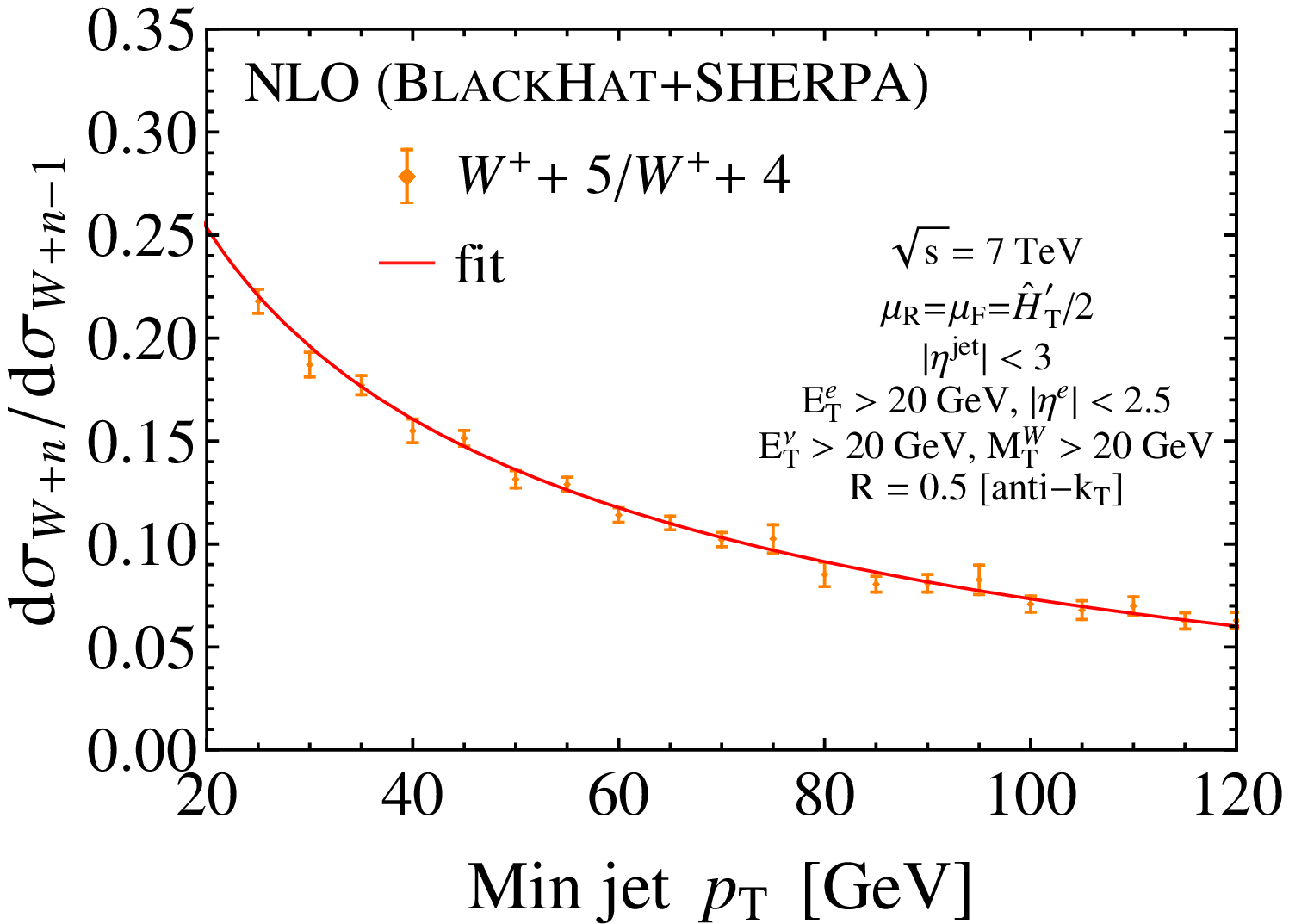}\\[-3mm]
(e)&(f)
\end{tabular}
\end{minipage}
\end{center}
\tighten
\caption{\tightcaption
Fits to the ratio of the \Wpjn-jet to \Wpjnm-jet cross sections
as a function of the minimum jet $\pT$ cut.
In the left column we show the ratios at LO and in the right column, at NLO.
From top to bottom, $n$ goes from 3 to 5.
}
\label{WpMinJetPtJetProductionRatioFitFigure}
\vskip .5 cm 
\end{figure}

The values of $d_n$ are not universal, but those for the primary
exponent of interest $\eta_n$ are nearly independent of the number of
jets, and also change very little in going from LO to NLO.  We show
the fits to the \Wmjjj-jet to \Wmjj-jet, \Wmjjjj-jet to \Wmjjj-jet,
and \Wmjjjjj-jet to \Wmjjjj-jet ratios in
\fig{WmMinJetPtJetProductionRatioFitFigure}, with the LO ratios in the
left-hand column, and the NLO ratios in the right-hand column.  The
central value for the exponent in the \Wmjjjjj-jet to \Wmjjjj-jet
ratio is different, but the statistical uncertainty is large, and the
result for $\eta_n$ remains marginally consistent with the other
$\eta_n$s.  We have not studied the sensitivity of this fit to the
various cuts defining the sample, such as the cut on the lepton
rapidity.  The exponents are similar for \Wpjn-jet production, for
which we show the fits to the ratios in
\fig{WpMinJetPtJetProductionRatioFitFigure}.

\begin{figure}[tb]
\begin{center}
\begin{minipage}[b]{1.03\linewidth}
\null\hskip -5mm
\includegraphics[clip,scale=0.44]{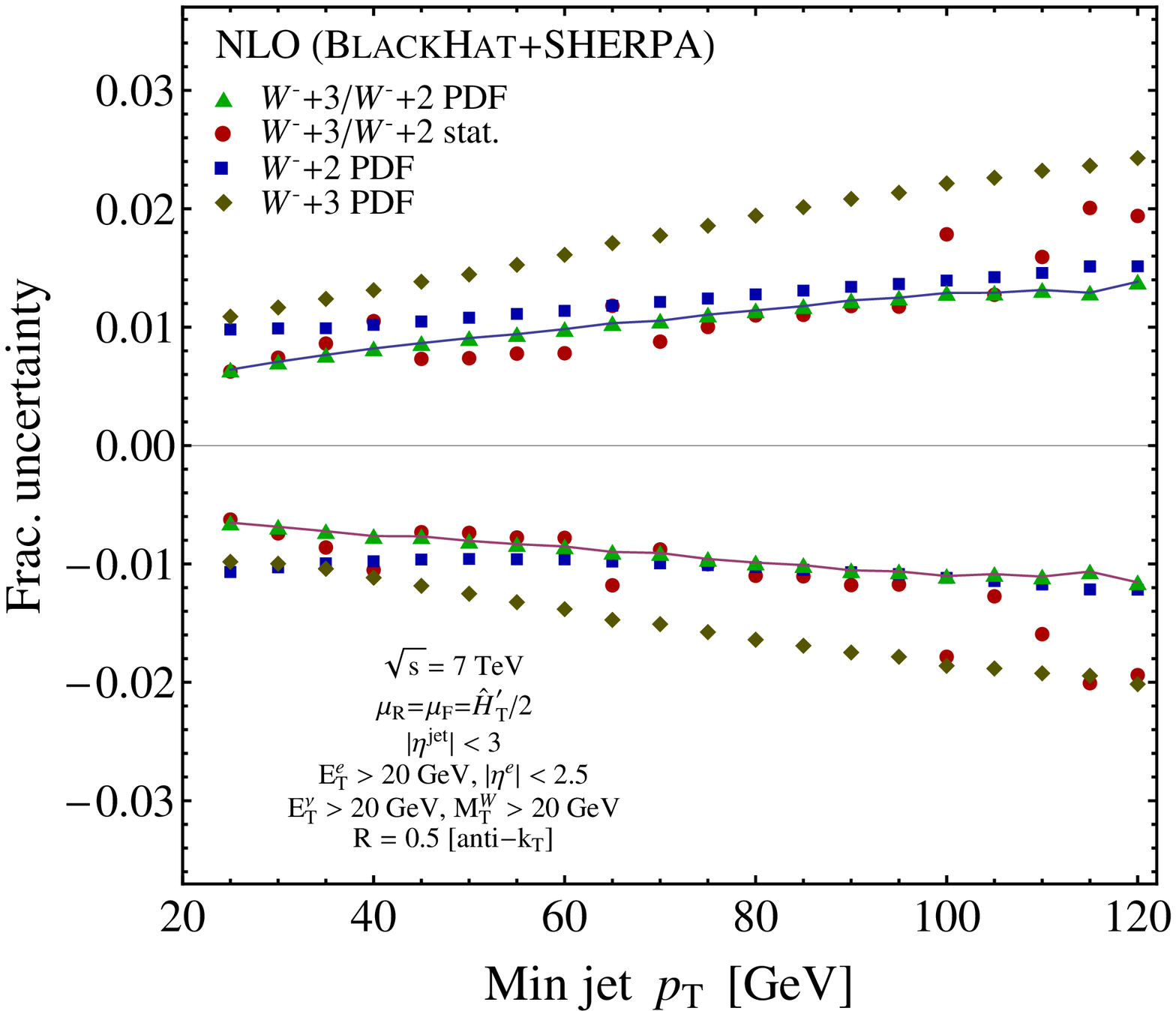}
\includegraphics[clip,scale=0.44]{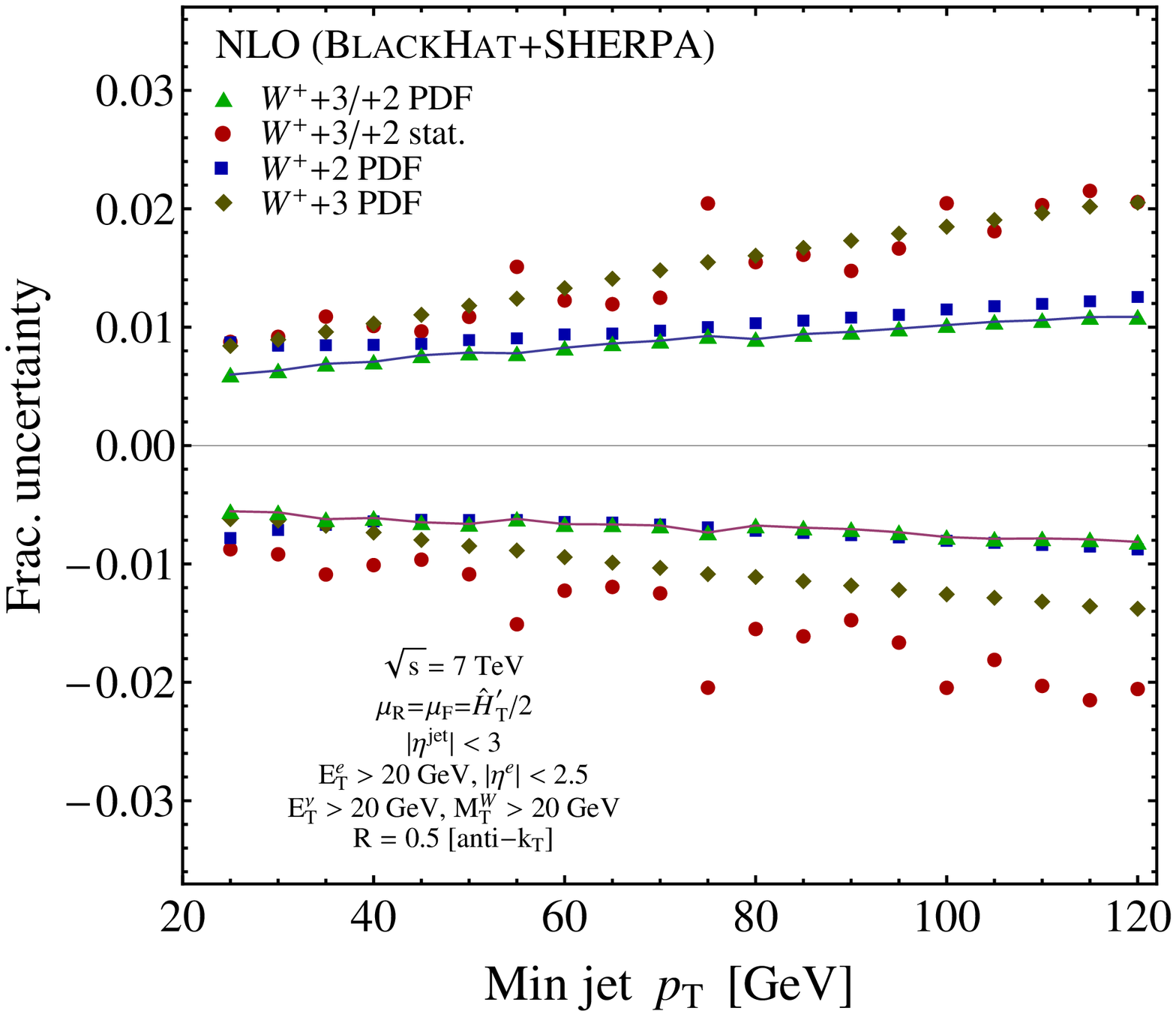}
\end{minipage}
\end{center}
\caption{PDF uncertainties in \Wjj-jet and \Wjjj-jet production at NLO
  as a function of the minimum jet transverse momentum cut $\pTmin$.
  The plots show the PDF uncertainties on the \Wjjj/\Wjj-jet ratio
  (green triangles joined by lines), the separate uncertainties on the
  \Wjjj- (olive diamonds) and \Wjj-jet (dark blue squares) cross
  sections.  The statistical uncertainty on the \Wjjj/\Wjj-jet ratio
  (dark red circles) is also shown for comparison.  The left plot
  shows the $W^{-}$ cross sections, and the right plot the $W^+$ ones.
}
\label{pTminPDFUncertaintyFigure}
\end{figure}

As explained in the introduction, we save our results in an
intermediate format which makes it straightforward and efficient to evaluate
cross sections and distributions for PDF error sets~\cite{NtuplesNote}.  
We have made use of the \ntuples{} to evaluate the PDF uncertainties
on the cross sections and on their ratios as a function of $\pTmin$.  To do so,
we calculate the different \Wjn-jet cross sections, as well as their ratios,
for each element of an MSTW PDF set, and use the standard weighting
procedure to obtain 68\% upper and lower confidence intervals.
We display the results for \Wjj- and \Wjjj-jet production, along with
their ratio, in \fig{pTminPDFUncertaintyFigure}.  The figure shows that
the PDF uncertainties are small, ranging from 0.5\%{} for smaller values
of $\pTmin$ to just below 1\% for the range of cuts we have studied.
The PDF uncertainties on the ratio are slightly smaller than those on the
\Wjj-jet cross section, and a factor of two smaller than those on the
\Wjjj-jet cross section.  The PDF uncertainties on the ratio are comparable
to the statistical uncertainties for \Wmjn-jet production, and smaller
than the statistical uncertainties for \Wpjn-jet production, for the
samples used in this study.
\FloatBarrier

\subsection{Dependence on the Vector Boson Transverse Momentum}
\label{WpTSection}

\begin{figure}[tb]
\begin{center}
\null\hskip -5mm\begin{minipage}[b]{1.03\linewidth}
\includegraphics[clip,scale=0.45]{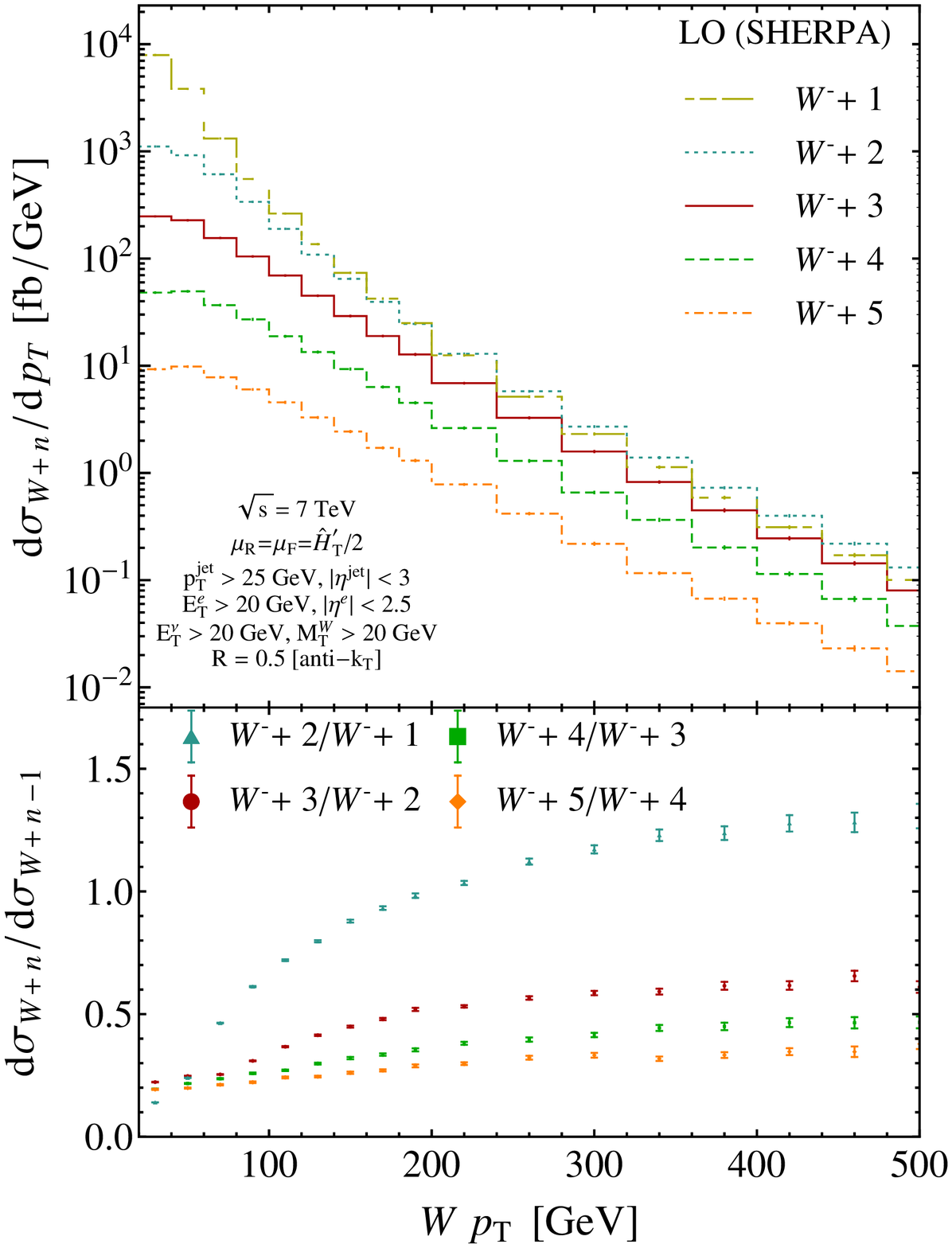}
\hspace*{-5mm}
\includegraphics[clip,scale=0.45]{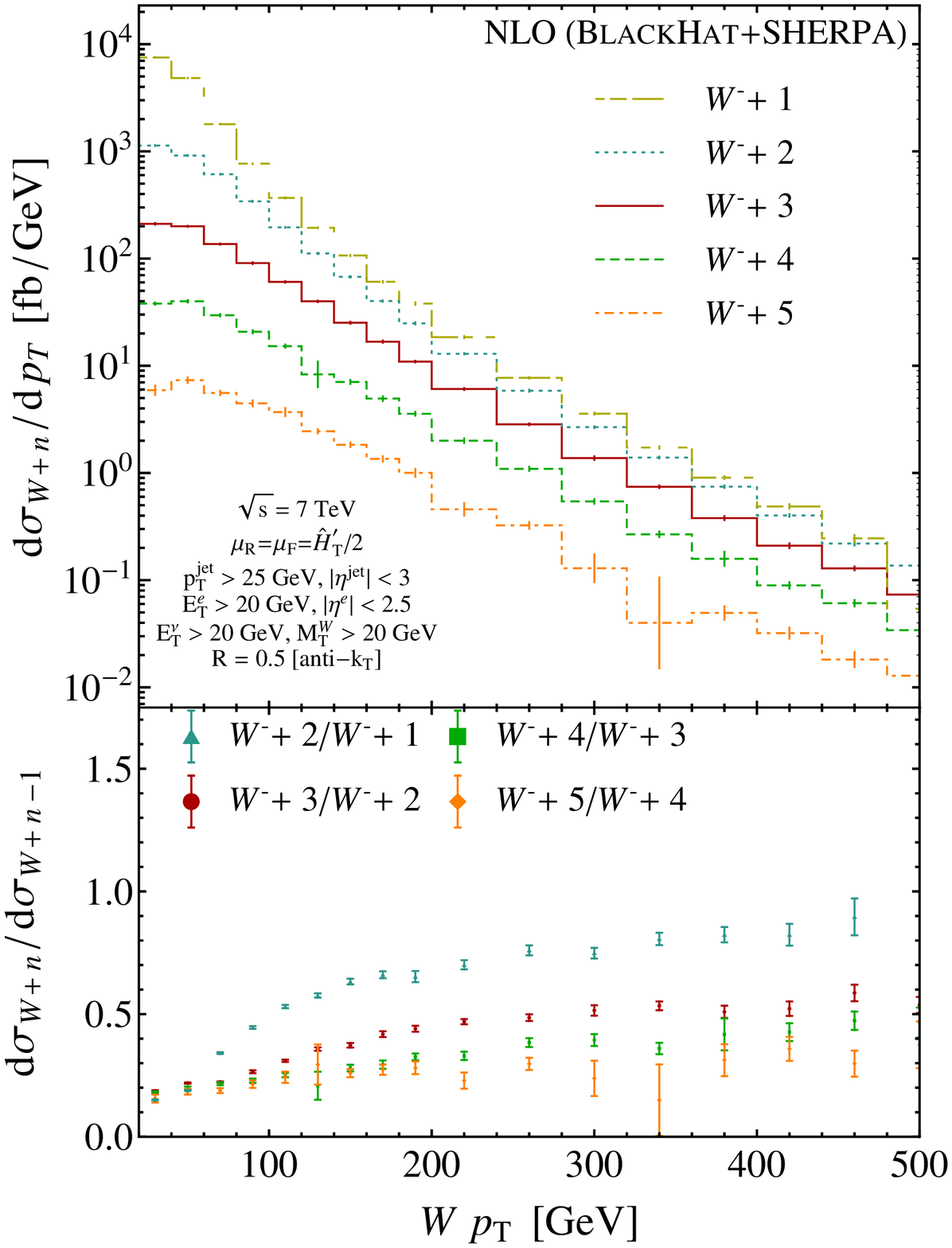}
\end{minipage}
\end{center}
\caption{The LO and NLO vector-boson $\pT$ distributions for \Wmjn-jet
  production at the LHC. The upper panels show the distributions in
  fb/GeV: \Wmj-jet production through \Wmjjjjj-jet production are
  ordered from top to bottom.  The thin vertical lines, where visible,
  indicate the statistical uncertainties.  The lower panels show the
  jet-production ratios.  The left plot are the distributions and
  ratios at LO, the right plot at NLO.  }
\label{WmPtJetProductionRatioFigure}
\end{figure}

\begin{figure}[tb]
\begin{center}
\null\hskip -5mm\begin{minipage}[b]{1.03\linewidth}
\includegraphics[clip,scale=0.45]{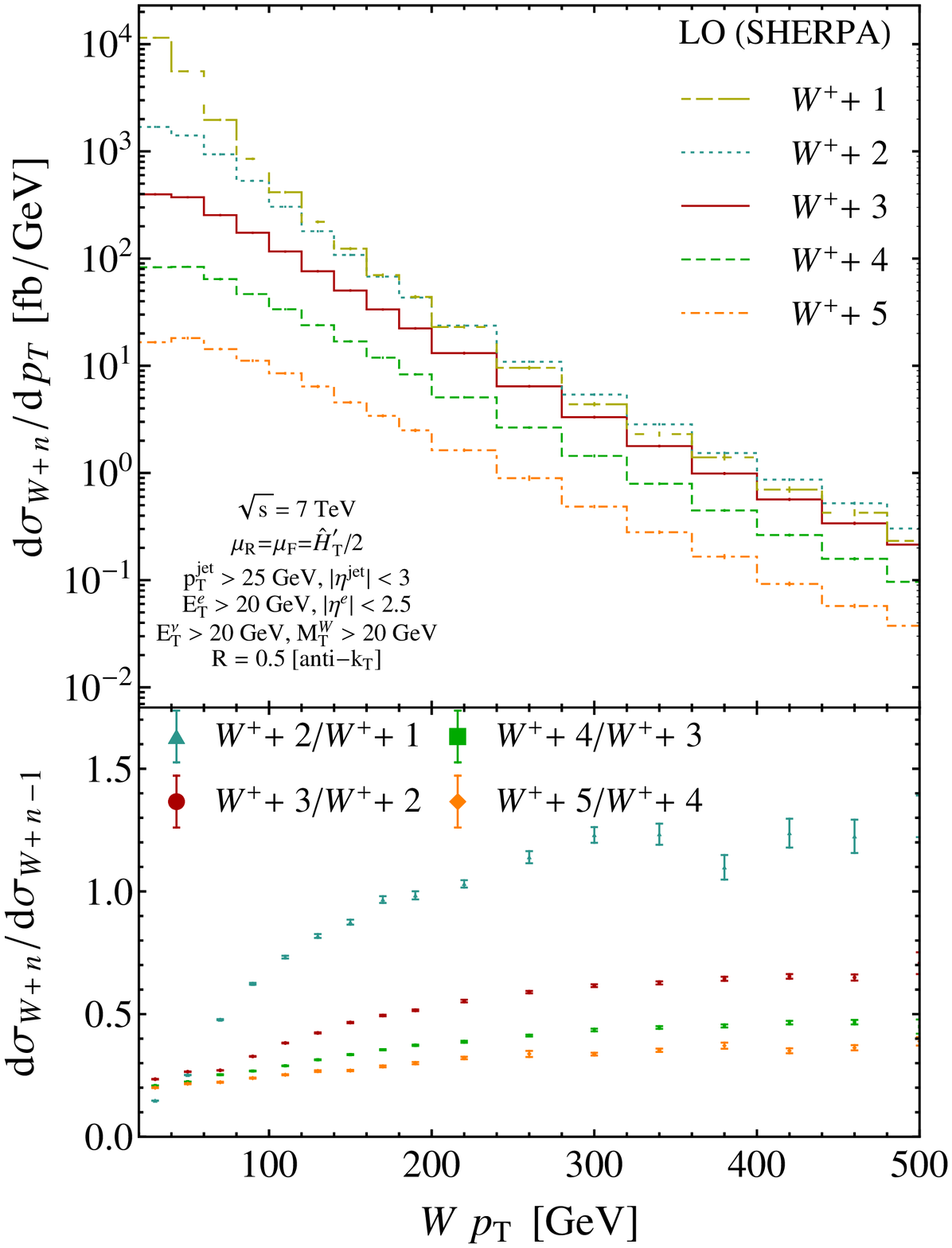}
\hspace*{-5mm}
\includegraphics[clip,scale=0.45]{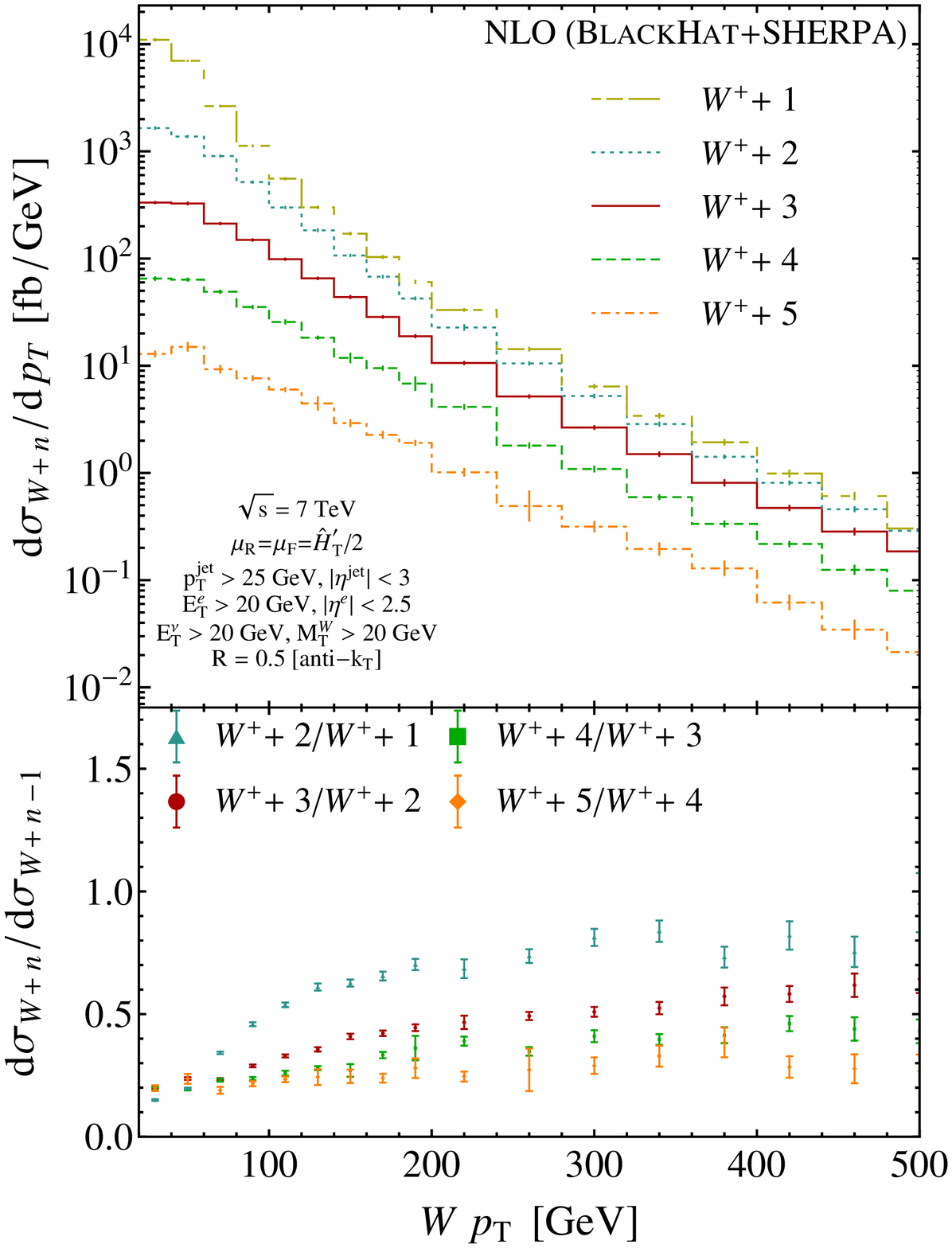}
\end{minipage}
\end{center}
\caption{The LO and NLO vector-boson $\pT$ distributions
for \Wpjn-jet production at the LHC. The upper panels
show the distributions in fb/GeV: \Wpj-jet production
through \Wpjjjjj-jet production are ordered from top to bottom.  
The lower panels show the jet-production ratios.
The left plot shows the distributions and ratios at LO,
the right plot at NLO.
}
\label{WpPtJetProductionRatioFigure}
\end{figure}

We turn next to the dependence of the jet-production ratio on the 
$W$-boson transverse momentum. The dependence of jet-production ratios
on the vector boson $\pT$ was studied previously for jet production
in association with $Z$ bosons at the Tevatron~\cite{BlackHatZ3jet}. 
We provide tables of the LO and NLO
differential cross sections at the LHC as a function of the
$W$ $\pT$ in \Wj-jet through \Wjjjjj-jet production in
appendix~\ref{WpTAppendix}.  We display these
differential cross sections in the upper panels of
\fig{WmPtJetProductionRatioFigure} for the $W^-$, and of
\fig{WpPtJetProductionRatioFigure} for the $W^+$.

The corresponding jet-production ratios are shown differentially in
the $W$ $\pT$ in the lower panels of these figures.  In the lowest
$\pT$ bins, up to a $\pT$ of order the $W$ mass, the ratio takes on a
value near $0.25$, roughly independent of the number of jets, and the
NLO corrections are modest.  This is in agreement with the ratios
of total cross sections that can be obtained from
\Tab{CrossSectionAnti-kt-R5Table}.  In order to get a feeling for how well
these ratios will continue to hold when cuts are tightened,
{\it e.g.} a cut on $\ETsl$ in searches for new physics at
ever-higher energy scales, we can examine their dependence
on the vector-boson transverse momentum.  Ref.~\cite{BlackHatZ3jet}
already noted strong sensitivity of the jet-production ratios to the
$Z$ boson's $\pT$ for \Zjn-jet production at the Tevatron (at
$\sqrt{s}=1.96$~TeV).  This was especially true for the \Zjj-jet to
\Zj-jet ratio, but held for the \Zjjj-jet to \Zjj-jet ratio as well.

\Figs{WmPtJetProductionRatioFigure}{WpPtJetProductionRatioFigure} reveal that
the seeming independence of the jet-production ratio from the number of jets is
also misleading at the LHC.  Once again, it holds only for vector-boson
$\pT$s of less than 60~GeV.  At higher $\pT$, the ratios change noticeably as
the number of jets changes.  Of course, if no cut is placed on the
vector-boson $\pT$, the bulk of the cross-section arises from lower $\pT$,
and the ratio of total cross sections will be insensitive to the number of
jets.  This insensitivity cannot be extrapolated safely to measurements with
a large value of the cut on the $W^\pm$ $\pT$.  Furthermore, the ratios for
\Wjj-jet/\Wj-jet and \Wjjj-jet/\Wjj-jet production show substantial NLO
corrections.  This is as expected, following the discussion in the previous
subsection.  The sensitivity of the ratios to the $W$ $\pT$ decreases
slowly with increasing number of jets, although it still remains
noticeable for \Wjjjjj-jet production.  The NLO corrections are smaller
beyond \Wjj{} jets.

An approximate fit to these differential jet-production ratios was provided
in ref.~\cite{BlackHatZ3jet}.  The fit's functional form was motivated by
the expectation that at very large vector-boson transverse momentum $\pTV$,
the matrix element would be maximized for an asymmetric configuration of jets,
corresponding to a near-singular configuration of the partons. A typical
configuration, for example, would have one hard jet recoiling against the
vector boson, and additional
jets (if any) with small transverse momenta just above the minimum jet
transverse momentum.  In these configurations, the short-distance matrix
element will factorize into a matrix element for production of one hard
gluon, and a singular factor (a splitting function in collinear limits, or an
eikonal one in soft limits).  The phase-space integrals over these
near-singular configurations give rise to potentially large
logarithms. Because the minimum jet--jet distance $R$ is relatively large,
collinear logarithms should not play an important role; on the other hand,
$\pTV/\pTmin$ can become large (where $\pTmin$ is the minimum jet $\pT$),
so its logarithm will play a role.

\def\fitWmptLOa{~$ 2.50 \pm 0.06$~ &~$-0.36 \pm 0.03$~ &~$0.1 \pm 0.2$~}
\def\fitWmptNLOa{~$ 1.75 \pm 0.05$~ &~$-0.15 \pm 0.04$~&~\hfil---~}
\def\fitWmptLOb{~$ 1.32 \pm 0.03$~ &~$-0.64 \pm 0.03$~ &~$1.0 \pm 0.2$~}
\def\fitWmptNLOb{~$ 1.36 \pm 0.03$~ &~$-0.52 \pm 0.04$~&~\hfil---~}
\def\fitWmptLOc{~$ 0.80 \pm 0.01$~ &~$-0.74 \pm 0.03$~&~\hfil---~}
\def\fitWmptNLOc{~$ 0.83 \pm 0.04$~ &~$-0.6 \pm 0.1$~&~\hfil---~}
\def\fitWmptLOd{~$ 0.56 \pm 0.01$~ &~$-0.66 \pm 0.04$~&~\hfil---~}
\def\fitWmptNLOd{~$ 0.50 \pm 0.08$~ &~$-0.2 \pm 0.3$~&~\hfil---~}

\begin{table}
\begin{tabular}{||cc|l|l|l||}
\hline
\multicolumn{2}{||c}{\multirow{2}{*}{Process}} & \multicolumn{3}{|c||}{Fit Values}\\
\cline{3-5}
&& \multicolumn{1}{|c|}{$N_n/N_{n-1}$} & \multicolumn{1}{|c|}{$c_n$} & \multicolumn{1}{|c||}{$\gamma_n$} \\
\hline
\multirow{2}{*}{$\displaystyle\quad{W^-+ 2}\quad$}
& \multicolumn{1}{|c|}{LO} &\fitWmptLOa \\
\cline{2-5}
& \multicolumn{1}{|c|}{~NLO~~} &\fitWmptNLOa \\
\hline
\multirow{2}{*}{$\displaystyle\quad{W^-+ 3}\quad$}
& \multicolumn{1}{|c|}{LO} &\fitWmptLOb \\
\cline{2-5}
& \multicolumn{1}{|c|}{~NLO~~} &\fitWmptNLOb \\
\hline
\multirow{2}{*}{$\displaystyle{W^-+ 4}$}
& \multicolumn{1}{|c|}{LO} &\fitWmptLOc \\
\cline{2-5}
& \multicolumn{1}{|c|}{~NLO~~} &\fitWmptNLOc \\
\hline
\multirow{2}{*}{$\displaystyle{W^-+ 5}$}
& \multicolumn{1}{|c|}{LO} &\fitWmptLOd \\
\cline{2-5}
& \multicolumn{1}{|c|}{~NLO~~} &\fitWmptNLOd \\
\hline
\end{tabular}
\caption{Fit parameters for the jet-production ratio in \Wmjn{} jets as a function of the $W$ $\pT$, using the form in \eqn{WpTModel}.  Dashes indicate parameters that are fixed as described in the text rather than fitted.
}
\label{WmPTFit}
\end{table}

\def\fitWpptLOa{~$ 2.6 \pm 0.1$~ &~$-0.38 \pm 0.05$~ &~$0.7 \pm 0.3$~}
\def\fitWpptNLOa{~$ 1.84 \pm 0.07$~ &~$-0.15 \pm 0.06$~&~\hfil---~}
\def\fitWpptLOb{~$ 1.25 \pm 0.02$~ &~$-0.59 \pm 0.02$~ &~$0.9 \pm 0.1$~}
\def\fitWpptNLOb{~$ 1.22 \pm 0.04$~ &~$-0.38 \pm 0.05$~&~\hfil---~}
\def\fitWpptLOc{~$ 0.799 \pm 0.007$~ &~$-0.64 \pm 0.02$~&~\hfil---~}
\def\fitWpptNLOc{~$ 0.87 \pm 0.05$~ &~$-0.5 \pm 0.1$~&~\hfil---~}
\def\fitWpptLOd{~$ 0.58 \pm 0.01$~ &~$-0.53 \pm 0.03$~&~\hfil---~}
\def\fitWpptNLOd{~$ 0.46 \pm 0.06$~ &~$0.0 \pm 0.3$~&~\hfil---~}

\begin{table}
\begin{tabular}{||cc|l|l|l||}
\hline
\multicolumn{2}{||c}{\multirow{2}{*}{Process}} & \multicolumn{3}{|c||}{Fit Values}\\
\cline{3-5}
&& \multicolumn{1}{|c|}{$N_n/N_{n-1}$} & \multicolumn{1}{|c|}{$c_n$} & \multicolumn{1}{|c||}{$\gamma_n$} \\
\hline
\multirow{2}{*}{$\displaystyle\quad{W^++ 2}\quad$}
& \multicolumn{1}{|c|}{LO} &\fitWpptLOa \\
\cline{2-5}
& \multicolumn{1}{|c|}{~NLO~~} &\fitWpptNLOa \\
\hline
\multirow{2}{*}{$\displaystyle\quad{W^++ 3}\quad$}
& \multicolumn{1}{|c|}{LO} &\fitWpptLOb \\
\cline{2-5}
& \multicolumn{1}{|c|}{~NLO~~} &\fitWpptNLOb \\
\hline
\multirow{2}{*}{$\displaystyle{W^++ 4}$}
& \multicolumn{1}{|c|}{LO} &\fitWpptLOc \\
\cline{2-5}
& \multicolumn{1}{|c|}{~NLO~~} &\fitWpptNLOc \\
\hline
\multirow{2}{*}{$\displaystyle{W^++ 5}$}
& \multicolumn{1}{|c|}{LO} &\fitWpptLOd \\
\cline{2-5}
& \multicolumn{1}{|c|}{~NLO~~} &\fitWpptNLOd \\
\hline
\end{tabular}
\caption{Fit parameters for the jet-production ratio in \Wpjn{} jets
  as a function of the $W$ $\pT$, using the form in
  \eqn{WpTModel}.  Dashes indicate parameters that are fixed as
  described in the text rather than fitted.  }
\label{WpPTFit}
\end{table}

\def\pTmax{\pT^{{\rm max}}} 

The approximate factorization suggested
the following model for differential cross sections,
\begin{equation}
\frac{d\sigma_{V+n}}{d\pTV} = (a_s(\pTV))^n f(\pTV) 
  \biggl(\sum_{j=0}^{n-1} {\bar c}^{(n)}_j \ln^j\rho\biggr)
                       \bigl(1-\pTV/\pTmax\bigr)^{\gamma_n}\,,
\label{WpTModel0}
\end{equation}
where 
$\rho = \pTV/\pTmin$, $\pTmax = 3.5$~TeV is the maximum
transverse momentum at $\sqrt{s}=7$~TeV,
and where
\begin{equation}
a_s(\pT)\equiv \alpha_s(\pT) N_c/(2\pi)\,.
\label{asDefinition}
\end{equation}
The last factor in \eqn{WpTModel0} takes into account the different phase-space limits
and suppression due to
parton distribution functions as a function of the number of jets
$n$.  It is of course much less important at the LHC than at the Tevatron.
The function $f(\pTV)$, which describes the overall, rapidly falling behavior
of the distribution, will cancel in the ratios, leaving us with the
parameters ${\bar c}^{(n)}_j$ and $\gamma_n$.  The calculations we have
performed, and especially their statistical errors, do not allow us to fit
all the parameters in \eqn{WpTModel0} in a stable manner.  Accordingly, we
simplify the model, retaining only the two leading logarithms for each value
of $n$; and retaining distinct exponents $\gamma_n$ only for $n=2,3$ at LO.  We
set the NLO $\gamma_n$s to be equal to their LO counterparts, and also set
$\gamma_5=\gamma_4=\gamma_3$.  We adopt a slightly different parametrization,
\begin{equation}
\frac{d\sigma_{V+n}}{d\pTV} = \big(4 a_s(\pTV)\big)^n f(\pTV) N_n 
   \bigl(\ln^{n-1}\rho + c_n \ln^{n-2}\rho\bigr)
                 \bigl(1-\pTV/\pTmax\bigr)^{\gamma_n}\,,
\label{WpTModel}
\end{equation}
where we omit the $\ln^{n-2}\rho$ term for $n=1$, and also set $N_1=1$
and $\gamma_1=0$.  (Because the form of the factor in which $\gamma_n$
enters, the value of $\gamma_1$ has no significance for the ratio fits we perform; 
it will merely shift $\gamma_{n>1}$ by
whatever amount to which it is set.)  The fit quantities $N_n/N_{n-1}$, $c_n$, and $\gamma_n$
are dimensionless.

The distributions in
\figs{WmPtJetProductionRatioFigure}{WpPtJetProductionRatioFigure} have
structure at $\pT\lesssim M_W$, transitioning to a more uniform
`scaling' region at higher $\pT$.  The model in \eqn{WpTModel} might
be expected to provide adequate fits only far into this latter region,
where $\pTV \gg M_W$ (so that mass effects are negligible), and where
$\pTV \gg \pTmin$ so that the logarithms will dominate over any finite
terms.  In practice, we find that the fits turn out to work well as
far down as $\pTV$ of $\Ord(80$--100~GeV$)$ where $\pTV/\pTmin \sim
3$~to~4.

Accordingly, we include essentially all calculated points in the
expected scaling region with reasonable statistical errors: all bins
with $80\le \pTV\le 500$ for $n=2,3$ and all bins with
$100\le\pTV\le500$ for $n=4,5$.  We perform a nonlinear fit,
numerically minimizing a goodness-of-fit function.  We obtain stable
fits for this simplified model, ranging from good to acceptable.  In
performing these fits, we first fit the form~(\ref{WpTModel}) to the
computed \Wjj-jet to \Wj-jet ratio, and then use the resulting fits
for $N_2$ and $c_2$ in fitting \eqn{WpTModel} to the \Wjjj-jet to
\Wjj-jet ratio, and so on.  We obtain the fit parameters directly; the
uncertainties we obtain using the Monte-Carlo procedure described in
\sect{MinJetPTSection}.  As in the case of the $\pTmin$-dependence of
the total cross section discussed in the previous subsection, this
model is not exact.  Hence, as we increase the statistics in our
calculation, we should expect the quality of the fit to deteriorate.
For uncertainties of the magnitude of typical experimental errors, we
expect the fits to describe the data very well.  Although these
distributions have fewer computed points to fit than the
$\pTmin$-dependent total cross section discussed in the previous
section, it is still striking that the results can be parametrized
so simply.

\begin{figure}[tb]
\begin{center}
\null\hskip -3mm\begin{minipage}[b]{1.\linewidth}
\begin{tabular}{cc}
\includegraphics[clip,scale=0.43]{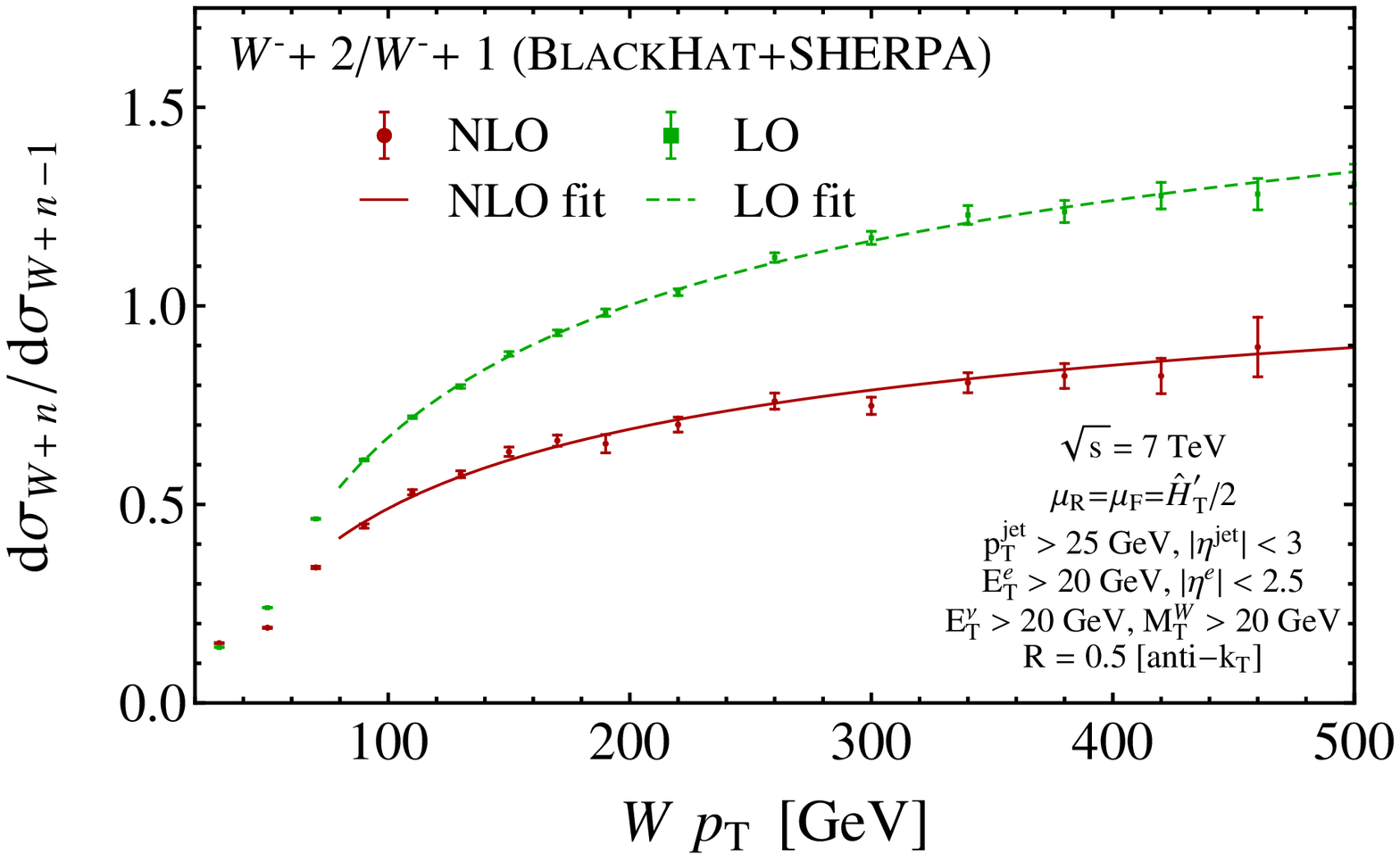}
&\includegraphics[clip,scale=0.43]{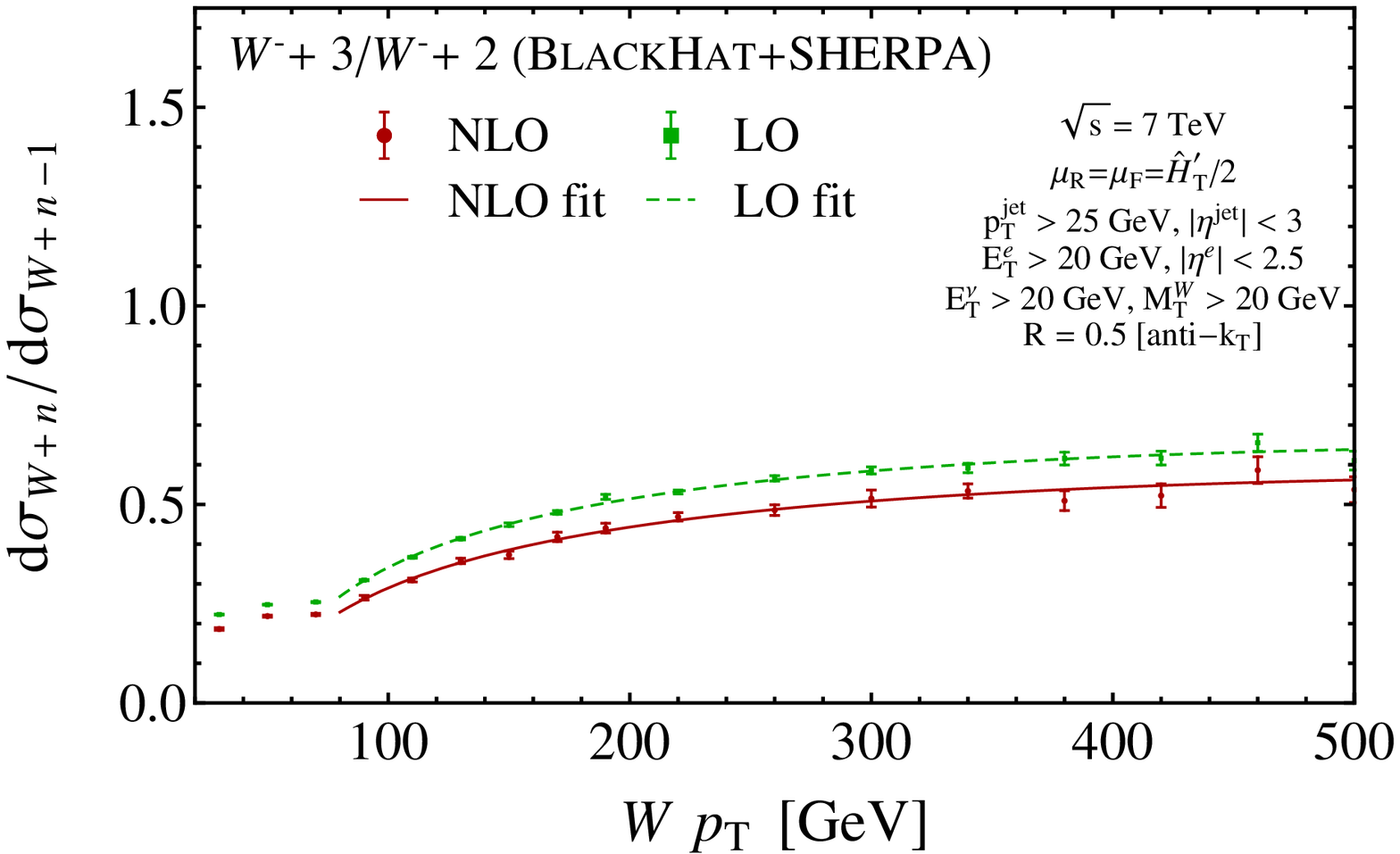}\\
(a)&(b)\\
\includegraphics[clip,scale=0.43]{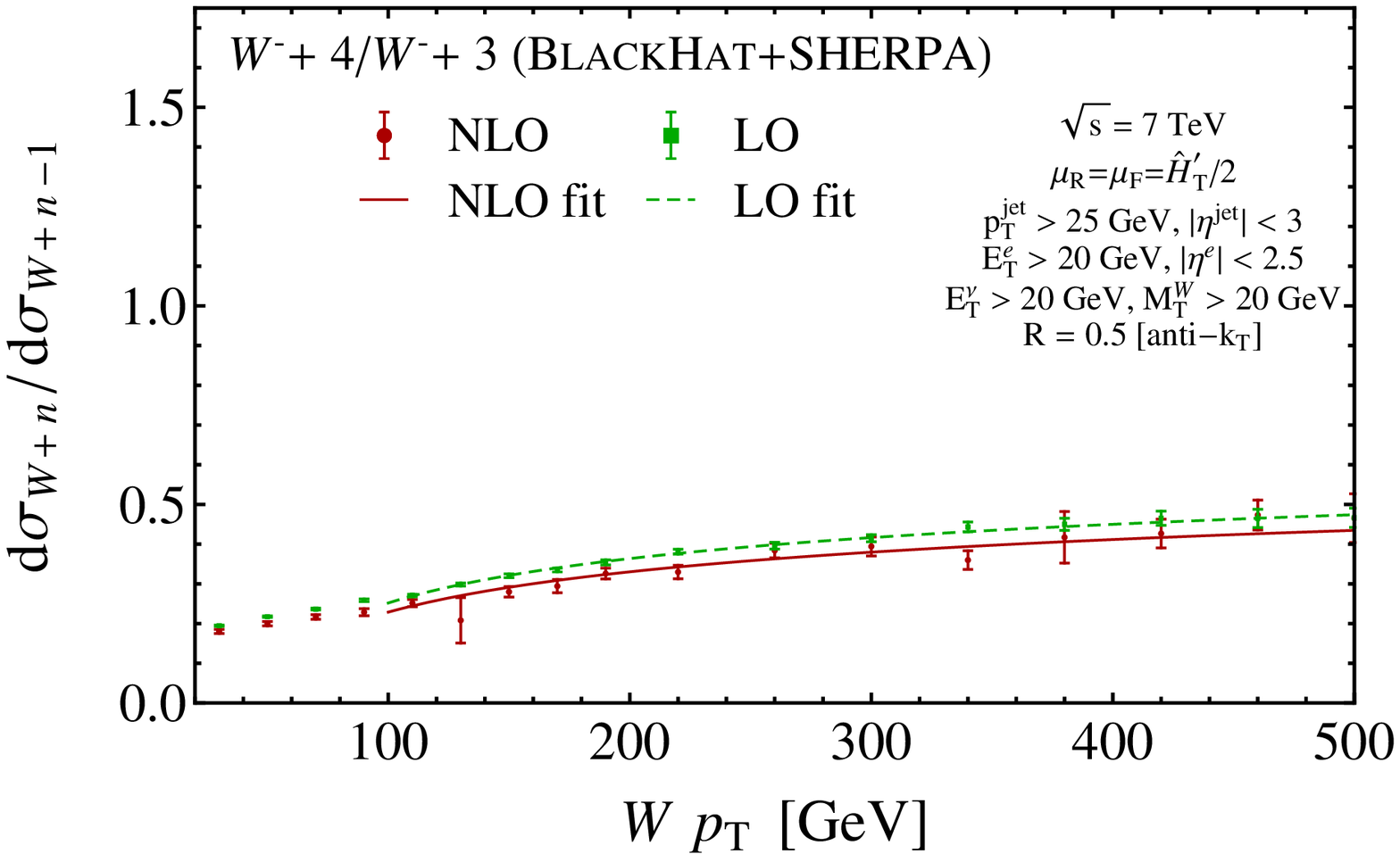}
&\includegraphics[clip,scale=0.43]{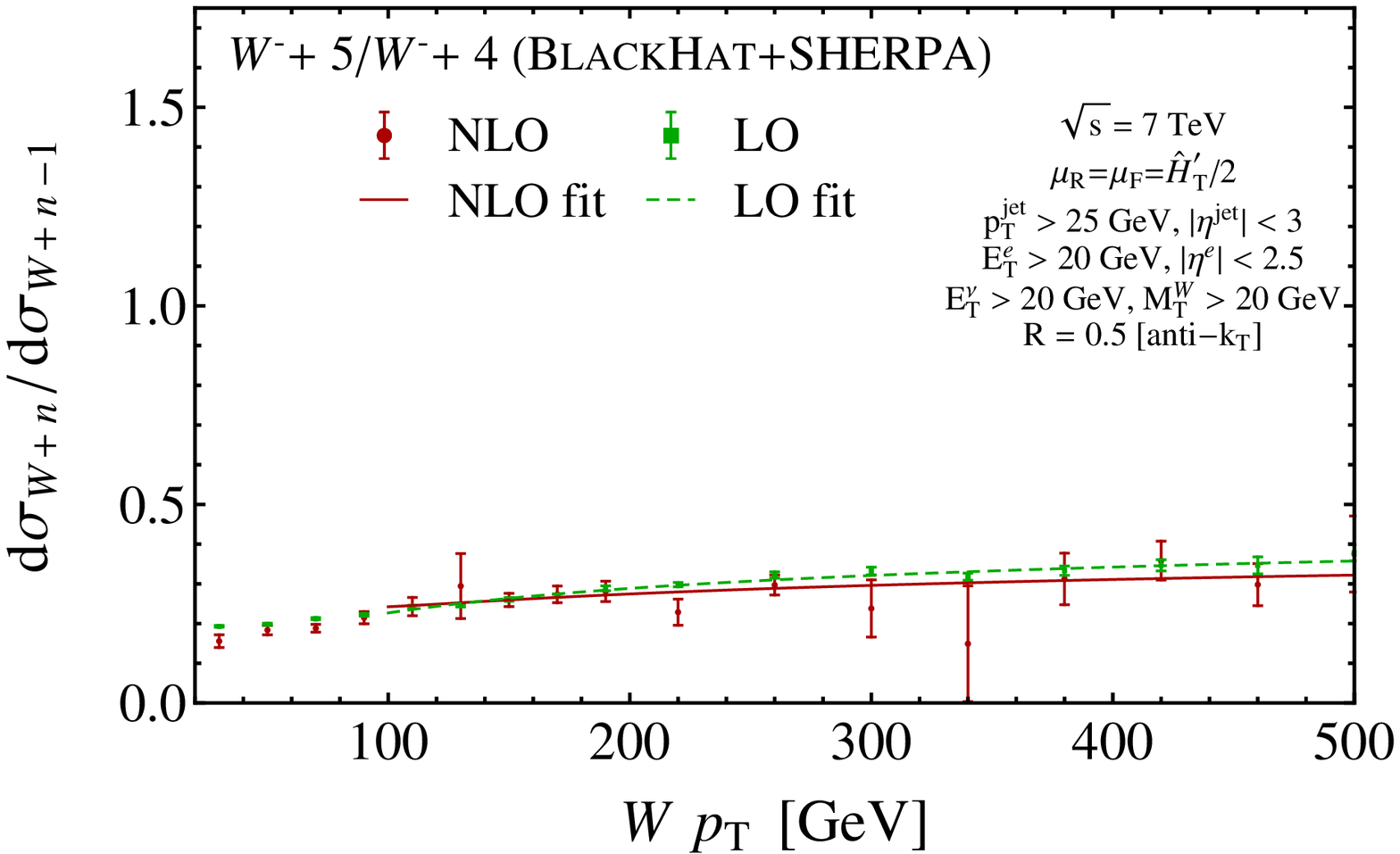}
\\
(c)&(d)
\end{tabular}
\end{minipage}
\end{center}
\caption{Fits to the ratios of the \Wmjn-jet to \Wmjnm-jet
  differential cross sections as a function of the $W$ $\pT$.  Each
  plot shows the computed LO and NLO ratios, as well as fits with the
  parameters in \tab{WmPTFit}.  
In (a), (b), (c) and (d) the cases with $n=2,3,4,5$ are shown,
respectively.}
\label{WmPtJetProductionRatioFitFigure}
\end{figure}

\begin{figure}[tb]
\begin{center}
\null\hskip -3mm\begin{minipage}[b]{1.\linewidth}
\begin{tabular}{cc}
\includegraphics[clip,scale=0.43]{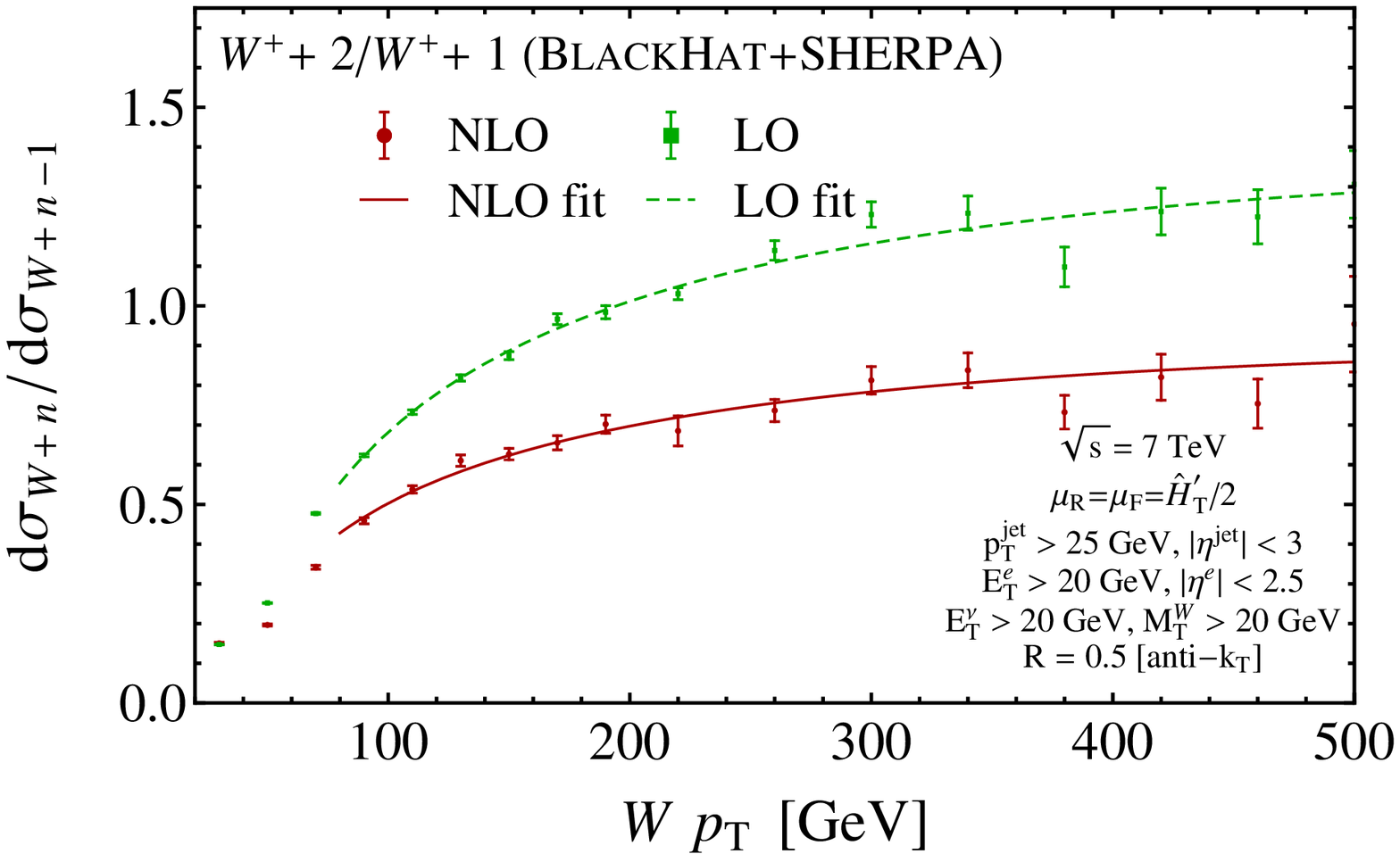}
&\includegraphics[clip,scale=0.43]{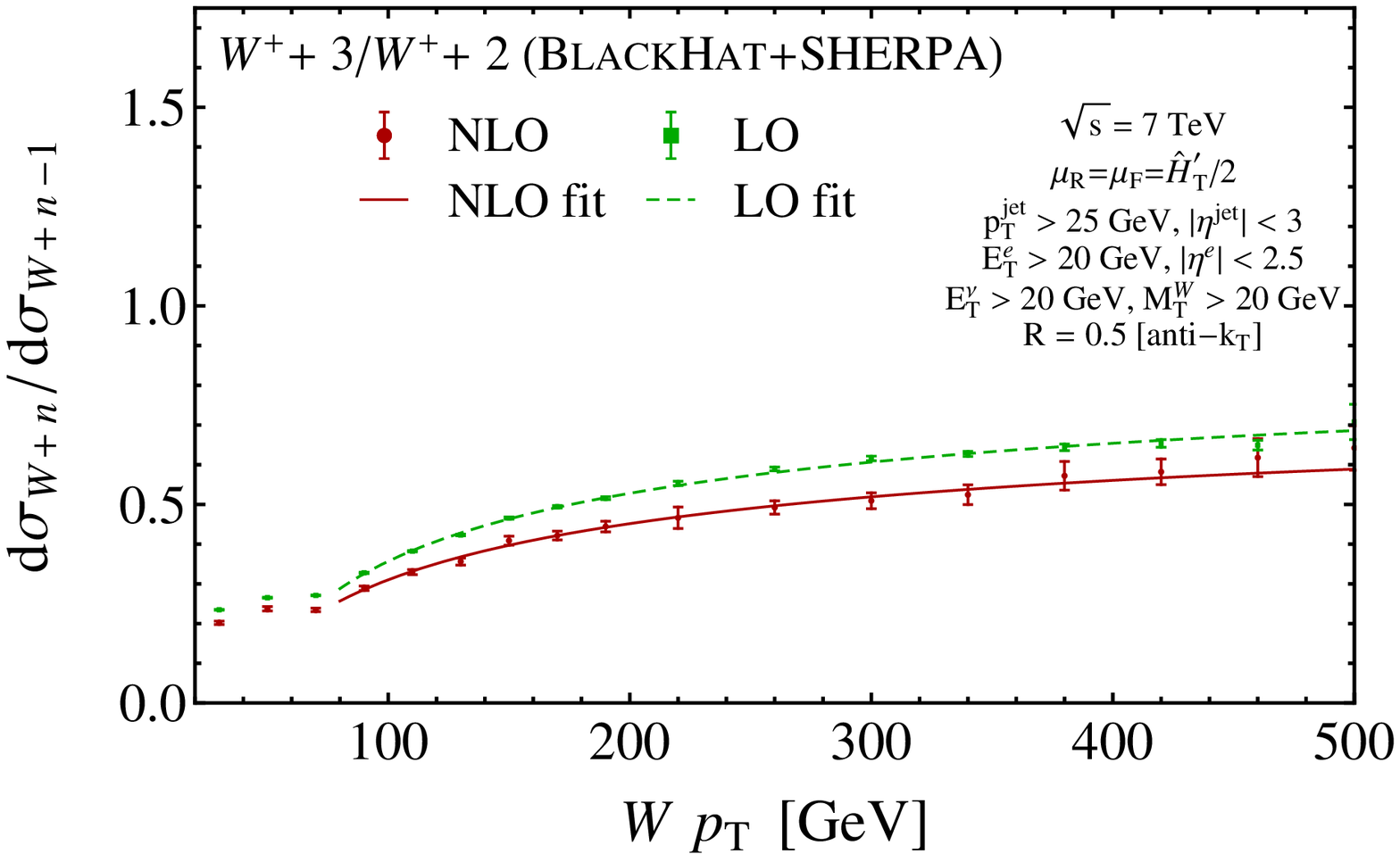}
\\
(a)&(b)\\
\includegraphics[clip,scale=0.43]{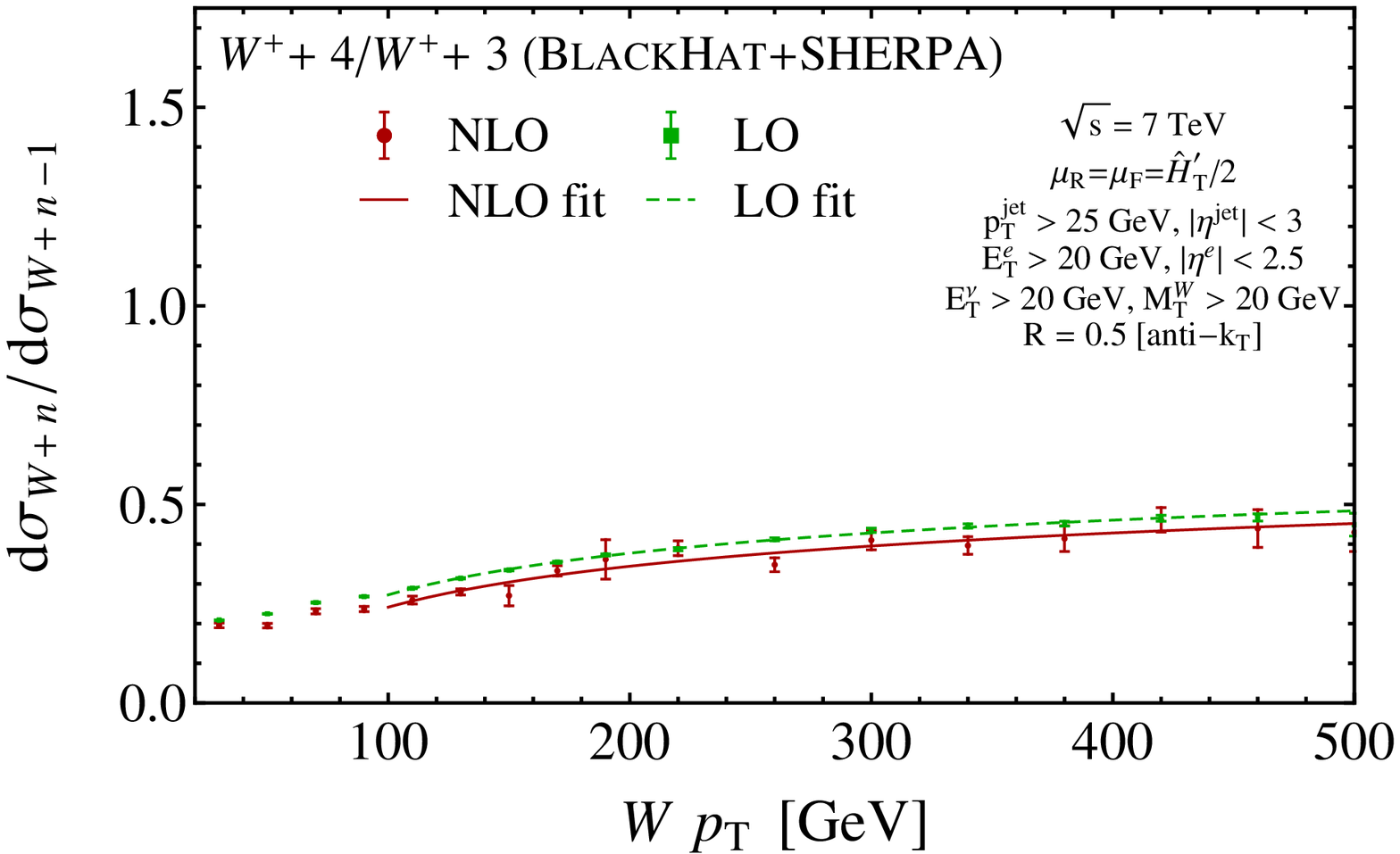}
&\includegraphics[clip,scale=0.43]{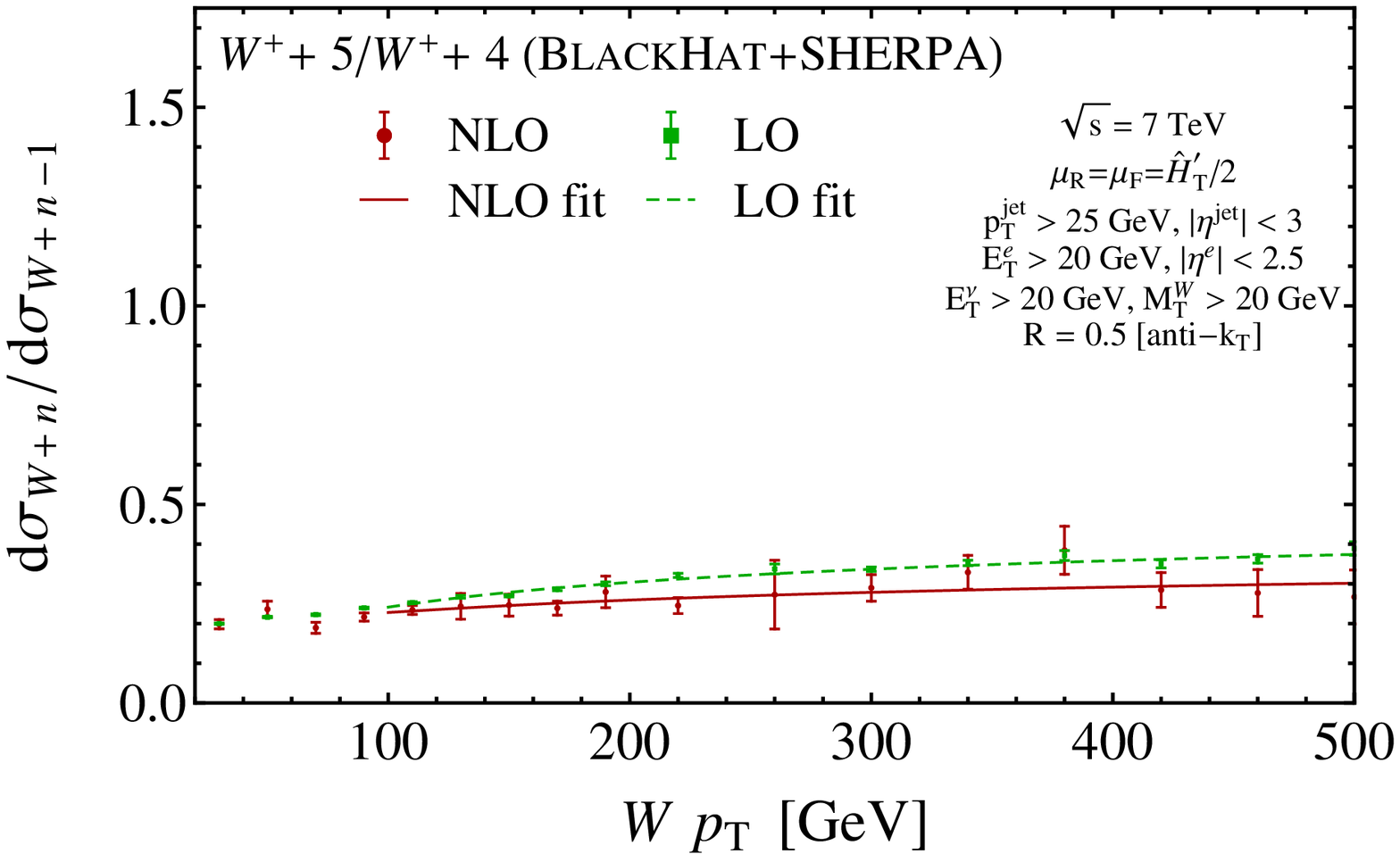}
\\
(c)&(d)
\end{tabular}
\end{minipage}
\end{center}
\caption{Fits to the ratios of the \Wpjn-jet to \Wpjnm-jet
  differential cross sections as a function of the $W$ $\pT$.  Each
  plot shows the computed LO and NLO ratios, as well as fits with the
  parameters in \tab{WpPTFit}. 
In (a), (b), (c) and (d) the cases with $n=2,3,4,5$ are shown,
respectively.}
\label{WpPtJetProductionRatioFitFigure}
\end{figure}
Our results for the $W^-$ fits are described in \tab{WmPTFit}, and
those for $W^+$ in \tab{WpPTFit}.  We display the $W^-$ fits in
\fig{WmPtJetProductionRatioFitFigure}, and the $W^+$ fits in
\fig{WpPtJetProductionRatioFitFigure}.

\FloatBarrier

\begin{figure}[tb]
\begin{center}
\begin{minipage}[b]{1.03\linewidth}
\null\hskip -8mm%
\includegraphics[clip,scale=0.45]{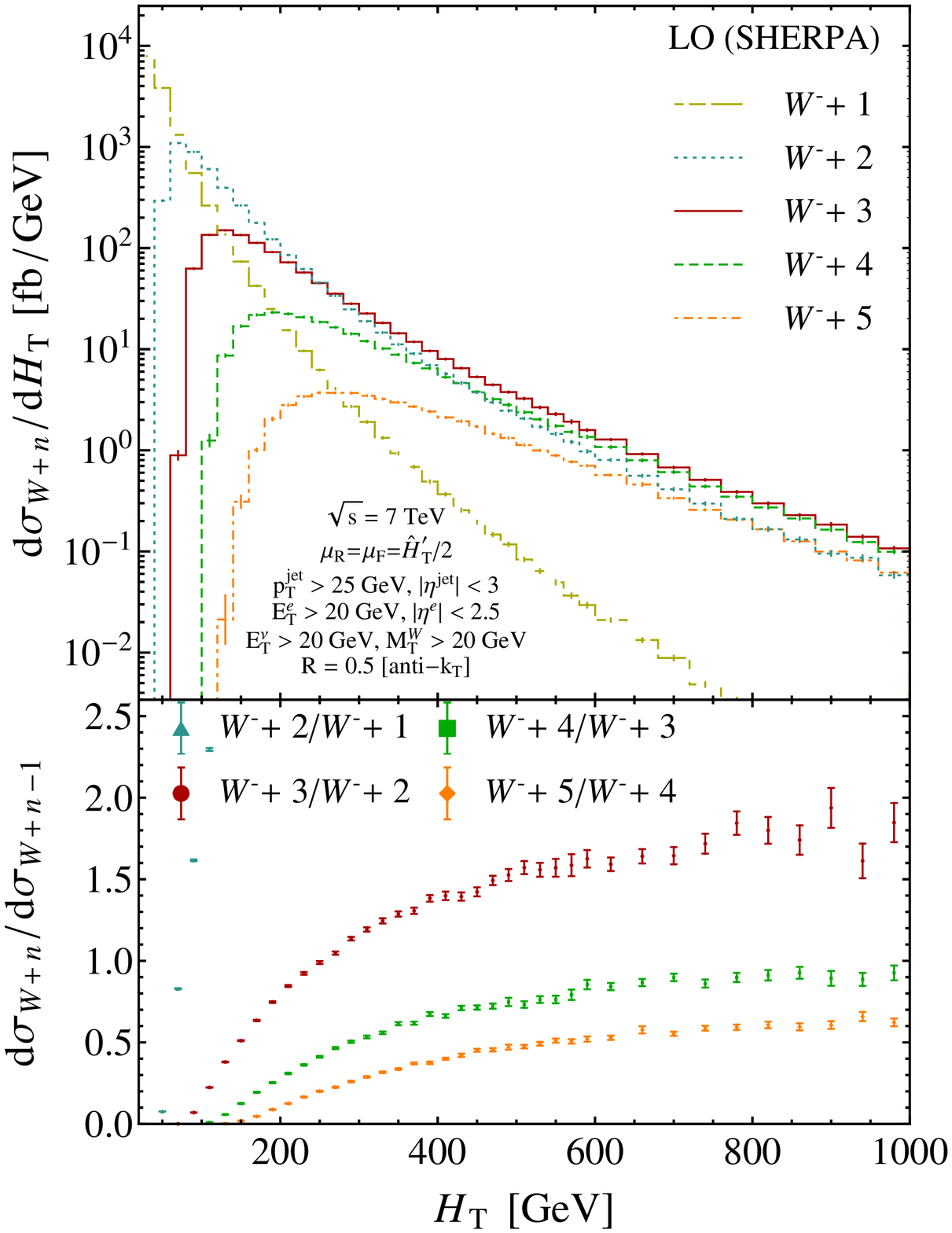}
\hspace*{-5mm}
\includegraphics[clip,scale=0.45]{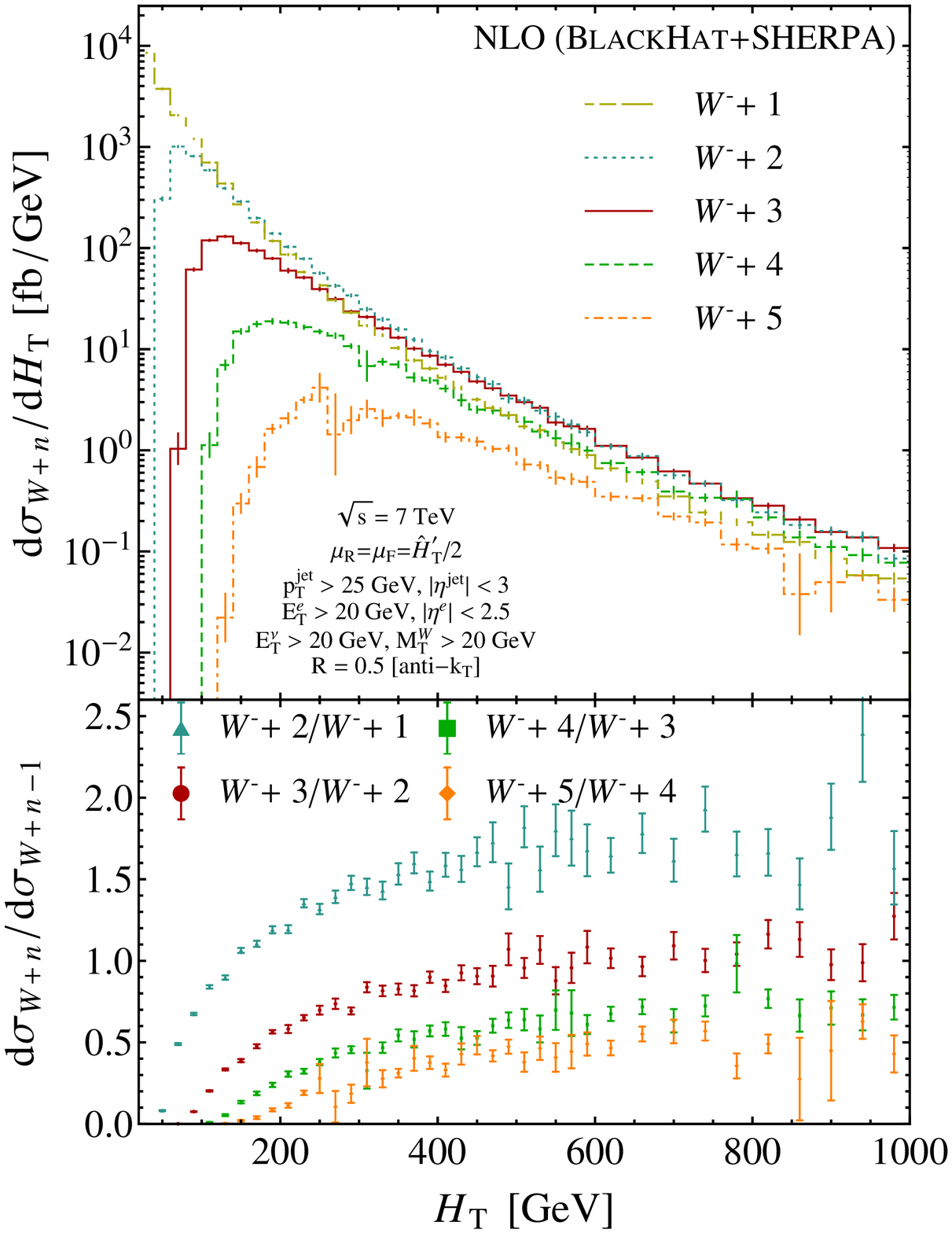}
\end{minipage}
\end{center}
\caption{The LO and NLO $\HT$ distributions
for \Wmjn-jet production at the LHC. The upper panels
show the distributions in fb/GeV.  From top to bottom, we display
results for \Wmj-jet production
through \Wmjjjjj-jet production.  The lower panels show the
jet-production ratios.
The left plot shows the distributions and ratios at LO,
the right plot at NLO.
}
\label{WmHTJetProductionRatioFigure}
\end{figure}

\begin{figure}[tb]
\begin{center}
\begin{minipage}[b]{1.03\linewidth}
\null\hskip -8mm%
\includegraphics[clip,scale=0.45]{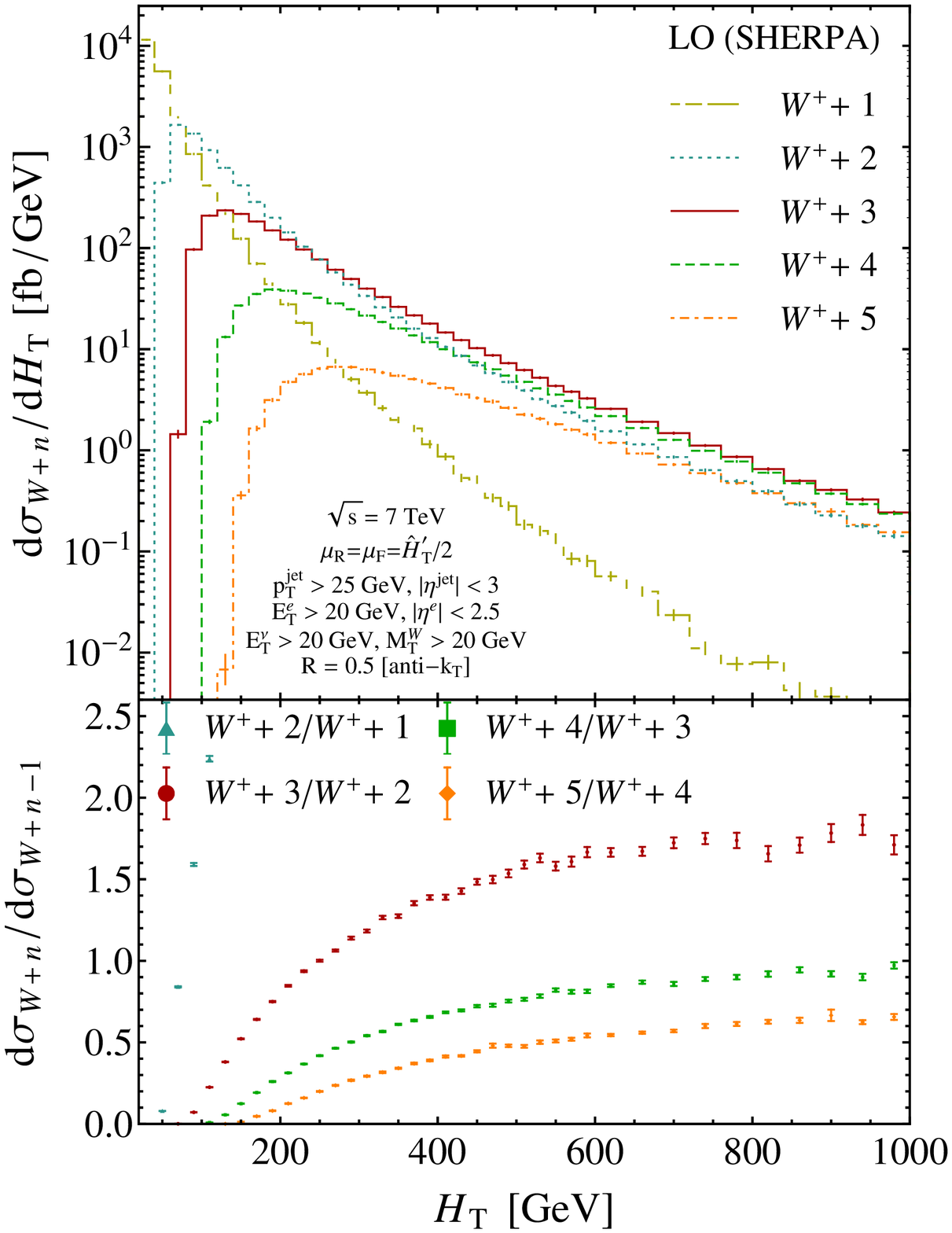}
\hspace*{-5mm}
\includegraphics[clip,scale=0.45]{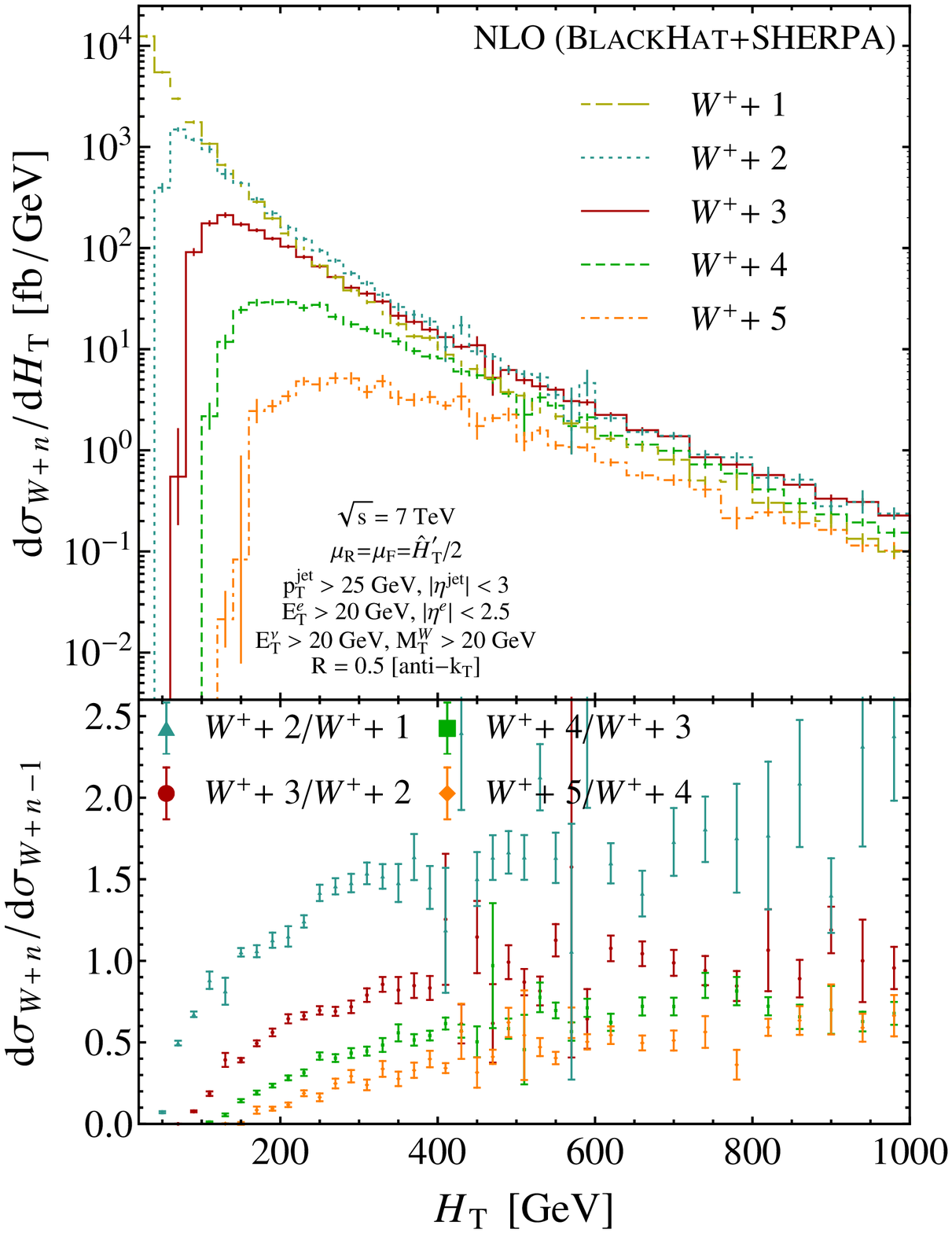}
\end{minipage}
\end{center}
\caption{The LO and NLO $\HT$ distributions
for \Wpjn-jet production at the LHC. The upper panels
show the distributions in fb/GeV.  From top to bottom, we display
\Wpj-jet production through \Wpjjjjj-jet production.  The lower panels show the
jet-production ratios.
The left plot shows the distributions and ratios at LO,
the right plot at NLO.
}
\label{WpHTJetProductionRatioFigure}
\end{figure}

\subsection{Dependence on the Total Jet Transverse Energy}
\label{WHTSection}

\def\qbar{{\overline q}}
In this section, we study the dependence of the jet-production ratio
on the total transverse energy in jets, $\HTjets$, defined in
\eqn{HTjetsdef}.
In this section we denote it simply by $\HT$.  
We provide tables of the LO and NLO differential
cross sections as a function of $\HT$ in \Wj-jet through \Wjjjjj-jet
production in appendix~\ref{WHTAppendix}:
the results for the LO differential cross section for \Wmjn-jet production in 
\tab{Wm-HTLODistributionTable}, and the results for the
NLO differential cross section in \tab{Wm-HTNLODistributionTable}.
The corresponding LO and NLO differential cross sections for \Wpjn-jet
production are shown in
\tabs{Wp-HTLODistributionTable}{Wp-HTNLODistributionTable}.  These
differential cross sections are also shown in the upper panels of
\fig{WmHTJetProductionRatioFigure} for the $W^-$, and of
\fig{WpHTJetProductionRatioFigure} for the $W^+$.

Each of these distributions rises from a threshold, passes through a
peak, and then falls off at larger $\HT$.  The peak does not, of
course, reflect any resonant behavior; it is a consequence of the
small phase space available at the
threshold of this distribution.  As the number of jets grows, the
minimum possible value of $\HT$ also rises, and the peak of the
distribution also shifts to higher values.  Beyond roughly 300~GeV,
all LO \Wjn-jet distributions for $n>1$ fall at similar rates.  The LO
distribution for \Wj-jet production falls much faster.  This rapid
fall may seem odd, but has been understood by Rubin, Salam, and
Sapeta~\cite{RubinSalamSapeta}.  At LO, the $W$ boson is forced to
have a large $\pT$ at large $\HT$, in order to balance the lone jet's
$\pT$. At NLO, because we are studying the inclusive distribution, the
leading jet's $\pT$ can be balanced by a second jet, and the $W$ boson
can be much softer, leading to large double logarithms in $\HT/M_W$.
Thus the \Wj-jet cross section increases dramatically at NLO,
as seen in the plots on the right-hand side of
\figs{WmHTJetProductionRatioFigure}{WpHTJetProductionRatioFigure}.
This partially compensates the LO behavior, though the \Wj{} distribution
still falls faster than distributions for $n\ge 2$ at the very highest
values of $\HT$.

\def\gammaH{\gamma^H}
\def\tauH{\tau^H}
\def\wH{\omega^H}
\def\HTmax{H_{\rm T}^{\rm max}}
\def\HTmin{H_{\rm T}^{\rm min}}
The corresponding jet-production ratios are shown differentially in
$\HT$ in the lower panels of these figures.  Although a factorization
argument for this distribution is less compelling than for the $W$
$\pT$ distribution, the success of the fits to the latter distribution
down to relatively low $\pT$, only just above $M_W$, suggests that we could
try fitting to a form similar to that in \eqn{WpTModel},
\begin{equation}
\frac{d\sigma_{V+n}}{d\HT} = \big(2 a_s(\HT/2)\big)^n f^H(\HT) N^H_n 
   \bigl(\ln^{n-1}\rho_{H,n} + c^H_n \ln^{n-2}\rho_{H,n}\bigr)
                         \bigl(1-\HT/\HTmax\bigr)^{\gammaH_n}\,,
\label{WHTModel0}
\end{equation}
where $a_s$ is defined in \eqn{asDefinition},
$\rho_{H,n} = \HT/(n \pTmin)$, and $\HTmax \simeq
7$~TeV is the maximum total jet transverse energy (neglecting the
effects of the $W$ transverse energy).  Such a fit can be expected to work
quite well at $\HT$ values significantly above the peaks.

However, unlike the case of the $W$ $\pT$ distribution, the 
thresholds for the distributions depend on $n$.  This fact
prevents the fit from being taken down to values just above the peak
or at the peak, which would limit our ability to perform
extrapolations using the fit forms.  

Instead, let us proceed as follows.
The threshold in the $\HT$ distribution arises from the minimum
jet $\pT$; $\HT$ cannot be less than $n \pTmin$ for $n$ jets.
The phase-space integral near the threshold has the rough form,
$$
\Big[\int \frac{dE}{E}\,g(E)\Big]^{n_*}\,,
$$
where $n_* < n$, because not all jets can be soft, and where
$g(E)$ is a slowly-varying function of the jet energy $E$.  This suggests
the following form for the threshold factor,
\begin{equation}
\ln^{\tau_n} \! \rho_{H,n}\,,
\end{equation}
where $\tau_n$ is determined by a fit.  The form of the phase-space constraints
and the form of the parton distributions at large $x$ both suggest a
large-$\HT$ fall-off factor similar to that used earlier for fitting
ratios of the $W$ $\pT$ distribution,
\begin{equation}
\bigl(1-\HT/\HTmax\bigr)^{\gamma_n^H} \,.
\end{equation}
This leads us to use the following form for fits,
\begin{equation}
\frac{d\sigma_{V+n}}{d\HT} = \big(2 a_s(\HT/2)\big)^n f^H(\HT) N^H_n 
\ln^{\tauH_n}\!\rho_{H,n}\,
             \bigl(1-\HT/\HTmax\bigr)^{\gammaH_n}\,,
\label{WHTModel}
\end{equation}
instead of the form in \eqn{WHTModel0}.  The remaining factor of
$f^H(\HT)$ can be assumed to be $n$-independent for $n\ge 2$.  We omit
terms with additional subleading logarithms, as our results do not
have the statistical precision to incisively determine their coefficients,
and allowing them can lead some fits into unphysical regions for some
parameters.  The parameters $N^H_n$, $\gammaH_n$, and $\tauH_n$ are all
dimensionless.

\def\fitWmHTLOa{~\hfil---~& ~$0.75 \pm 0.07$~& ~\hfil---~}
\def\fitWmHTNLOa{~\hfil---~& ~$1.1 \pm 0.2$~& ~\hfil---~}
\def\fitWmHTLOb{~$8.5 \pm 0.2$~& ~$1.93 \pm 0.04$~& ~$6.3 \pm 0.3$~}
\def\fitWmHTNLOb{~$7.7 \pm 0.4$~& ~$1.9 \pm 0.1$~& ~$5.0 \pm 0.7$~}
\def\fitWmHTLOc{~$6.46 \pm 0.08$~& ~$2.93 \pm 0.02$~& ~$8.0 \pm 0.2$~}
\def\fitWmHTNLOc{~$6.7 \pm 0.3$~& ~$2.74 \pm 0.05$~& ~$7.2 \pm 0.8$~}
\def\fitWmHTLOd{~$5.3 \pm 0.1$~& ~$3.85 \pm 0.02$~& ~$9.5 \pm 0.3$~}
\def\fitWmHTNLOd{~$6.2 \pm 0.7$~& ~$3.7 \pm 0.1$~& ~$11 \pm 2$~}

\begin{table}
\begin{tabular}{||cc|l|l|l||}
\hline
\multicolumn{2}{||c}{\multirow{2}{*}{Process}} & \multicolumn{3}{|c||}{Fit Values}\\
\cline{3-5}
&& \multicolumn{1}{|c|}{$N^H_n/N^H_{n-1}$} & \multicolumn{1}{|c|}{$\tauH_n$}
& \multicolumn{1}{|c||}{$\gammaH_n$} \\
\hline
\multirow{2}{*}{$\displaystyle\quad{W^-+ 2}\quad$}
& \multicolumn{1}{|c|}{LO} &\fitWmHTLOa  \\
\cline{2-5}
& \multicolumn{1}{|c|}{~NLO~~} &\fitWmHTNLOa \\
\hline
\multirow{2}{*}{$\displaystyle\quad{W^-+ 3}\quad$}
& \multicolumn{1}{|c|}{LO} &\fitWmHTLOb \\
\cline{2-5}
& \multicolumn{1}{|c|}{~NLO~~} &\fitWmHTNLOb \\
\hline
\multirow{2}{*}{$\displaystyle{W^-+ 4}$}
& \multicolumn{1}{|c|}{LO} &\fitWmHTLOc \\
\cline{2-5}
& \multicolumn{1}{|c|}{~NLO~~} &\fitWmHTNLOc \\
\hline
\multirow{2}{*}{$\displaystyle{W^-+ 5}$}
& \multicolumn{1}{|c|}{LO} &\fitWmHTLOd \\
\cline{2-5}
& \multicolumn{1}{|c|}{~NLO~~} &\fitWmHTNLOd \\
\hline
\end{tabular}
\caption{Fit parameters for the jet-production ratio in \Wmjn{} jets as a function of the jet $\HT$, using the form in \eqn{WHTModel}.  Dashes indicate parameters that are fixed as described in the text rather than fitted, or 
where we have
not carried out a fit.
}
\label{WmHTFit}
\end{table}

\def\fitWpHTLOa{~\hfil---~& ~$0.70 \pm 0.06$~& ~\hfil---~}
\def\fitWpHTNLOa{~\hfil---~& ~$0.8 \pm 0.5$~& ~\hfil---~}
\def\fitWpHTLOb{~$8.5 \pm 0.1$~& ~$1.90 \pm 0.03$~& ~$6.4 \pm 0.2$~}
\def\fitWpHTNLOb{~$7 \pm 1$~& ~$1.7 \pm 0.3$~& ~$7 \pm 1$~}
\def\fitWpHTLOc{~$6.51 \pm 0.04$~& ~$2.921 \pm 0.006$~& ~$8.3 \pm 0.1$~}
\def\fitWpHTNLOc{~$6.5 \pm 0.3$~& ~$2.64 \pm 0.06$~& ~$9.4 \pm 0.8$~}
\def\fitWpHTLOd{~$5.41 \pm 0.07$~& ~$3.89 \pm 0.02$~& ~$10.0 \pm 0.2$~}
\def\fitWpHTNLOd{~$5.0 \pm 0.5$~& ~$3.5 \pm 0.1$~& ~$10 \pm 2$~}

\begin{table}
\begin{tabular}{||cc|l|l|l||}
\hline
\multicolumn{2}{||c}{\multirow{2}{*}{Process}} & \multicolumn{3}{|c||}{Fit Values}\\
\cline{3-5}
&& \multicolumn{1}{|c|}{$N^H_n/N^H_{n-1}$} & \multicolumn{1}{|c|}{$\tauH_n$}
& \multicolumn{1}{|c||}{$\gammaH_n$} \\
\hline
\multirow{2}{*}{$\displaystyle\quad{W^++ 2}\quad$}
& \multicolumn{1}{|c|}{LO} &\fitWpHTLOa \\
\cline{2-5}
& \multicolumn{1}{|c|}{~NLO~~} &\fitWpHTNLOa \\
\hline
\multirow{2}{*}{$\displaystyle\quad{W^++ 3}\quad$}
& \multicolumn{1}{|c|}{LO} &\fitWpHTLOb \\
\cline{2-5}
& \multicolumn{1}{|c|}{~NLO~~} &\fitWpHTNLOb \\
\hline
\multirow{2}{*}{$\displaystyle{W^++ 4}$}
& \multicolumn{1}{|c|}{LO} &\fitWpHTLOc \\
\cline{2-5}
& \multicolumn{1}{|c|}{~NLO~~} &\fitWpHTNLOc \\
\hline
\multirow{2}{*}{$\displaystyle{W^++ 5}$}
& \multicolumn{1}{|c|}{LO} &\fitWpHTLOd \\
\cline{2-5}
& \multicolumn{1}{|c|}{~NLO~~} &\fitWpHTNLOd \\
\hline
\end{tabular}
\caption{Fit parameters for the jet-production ratio in \Wpjn{} jets as a function of the jet $\HT$, using the form in \eqn{WHTModel}.  Dashes indicate parameters that are fixed as described in the text rather than fitted, or where
we have
not carried out a fit.
}
\label{WpHTFit}
\end{table}

\def\oH{\omega^H} 

Because we have computed the $\HT$ distribution out
to larger values than the $W$ $\pT$ distribution, the effect of the
large-$\HT$ suppression factor is more noticeable.  The different
behavior of the \Wj-jet distribution at LO, as discussed above, makes
it unsuitable for the same fit.  Therefore we drop it, and do not
fit the LO \Wjj-jet to \Wj-jet ratio.  Instead, we set $N^H_2=1$.  We
start fitting with the \Wjjj-jet to \Wjj-jet ratio, where in addition
to $N^H_3$, $\tauH_3$, and $\gammaH_3$, we also fit for $\tauH_2$.  We
set $\gammaH_2=17/4$, which is approximately the value we would obtain
from an NLO fit to the \Wjj-jet to \Wj-jet ratio with $\gammaH_1=0$.
We then use these values in fitting the ratio of forms in
\eqn{WHTModel} to the \Wjjjj-jet to \Wjjj-jet ratio in order to
determine $N^H_4$, $\tauH_4$, and $\gammaH_4$, and so on.  We repeat
this procedure at NLO, starting again with the \Wjjj-jet to \Wjj-jet
ratio.  In all the fits, we drop the bin nearest
threshold\footnote{Were we to include this point, we should replace
  the threshold factor $\ln^{\tauH_n}\rho_{H,n}$ by its average over a
  bin, but the fits remain poor.}, but otherwise include {\it all\/}
bins up to $1000$~GeV.  (The bin sizes are larger for larger $\HT$, as
shown in
tables~\ref{Wm-HTLODistributionTable}--\ref{Wp-HTNLODistributionTable}.)
We perform a nonlinear fit, and obtain fit parameters directly, and
the uncertainties using the Monte-Carlo procedure described in
\sect{MinJetPTSection}.  We obtain fits that range from marginally
adequate (probabilities of 1--2\%) to adequate (probabilities of
$\gtrsim 10$\%).  The results of the fits are shown in \tab{WmHTFit}
for jet production in association with a $W^-$ boson, and in
\tab{WpHTFit} for production in association with a $W^+$ boson.

\begin{figure}[tb]
\begin{center}
\null\hskip -3mm\begin{minipage}[b]{1.\linewidth}
\begin{tabular}{cc}
\includegraphics[clip,scale=0.43]{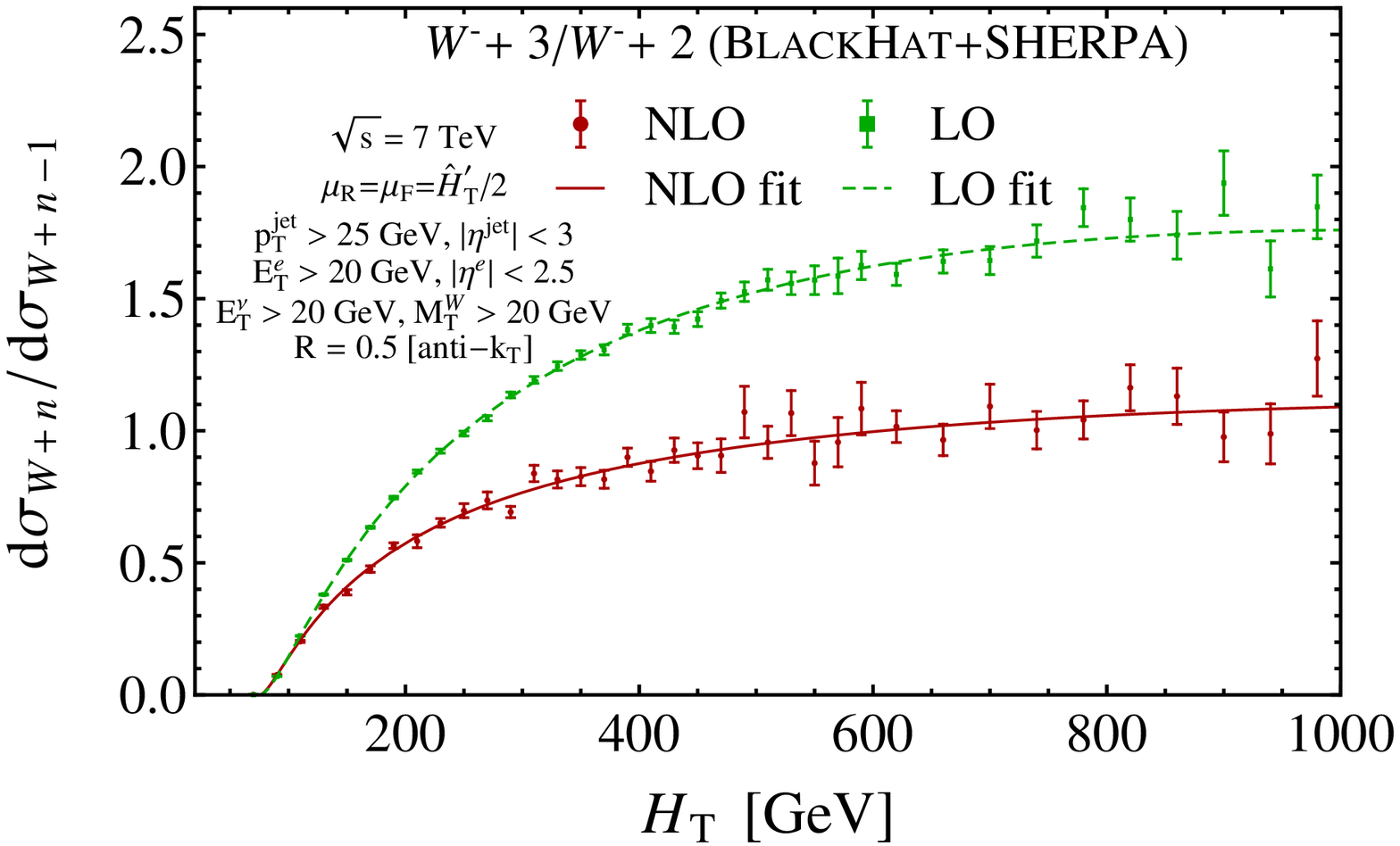}
&\includegraphics[clip,scale=0.43]{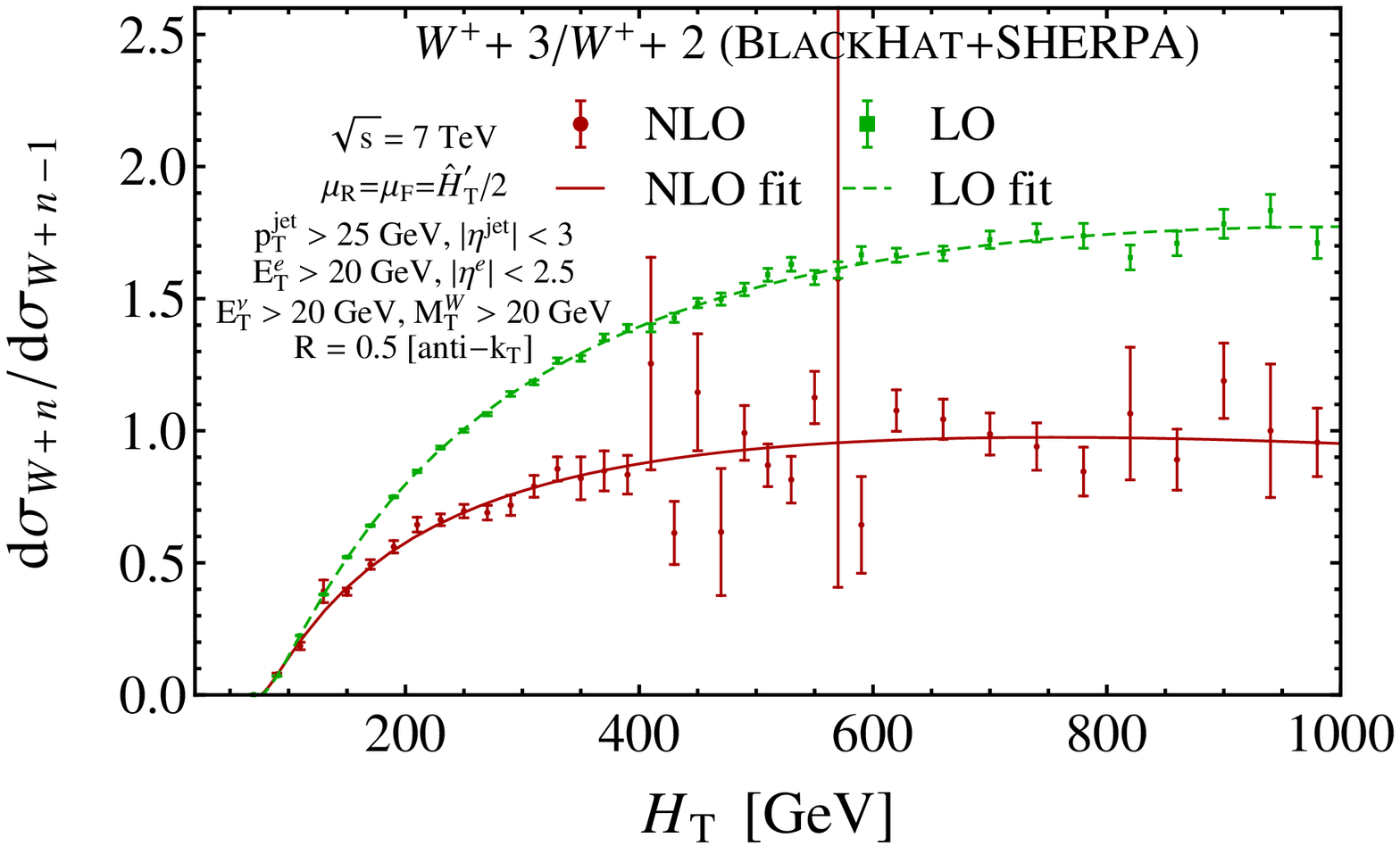}
\\
(a)&(b)\\
\includegraphics[clip,scale=0.43]{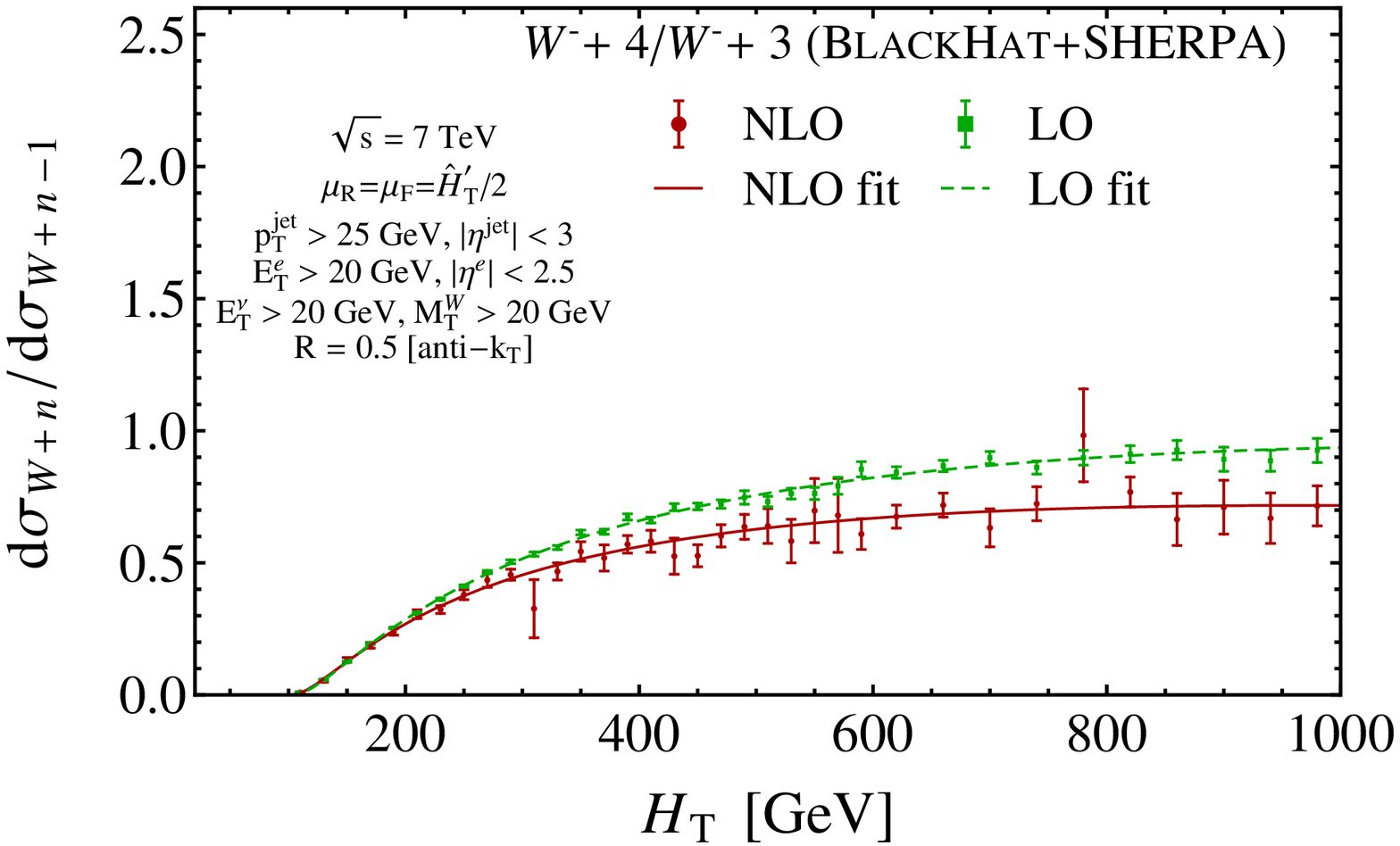}
&\includegraphics[clip,scale=0.43]{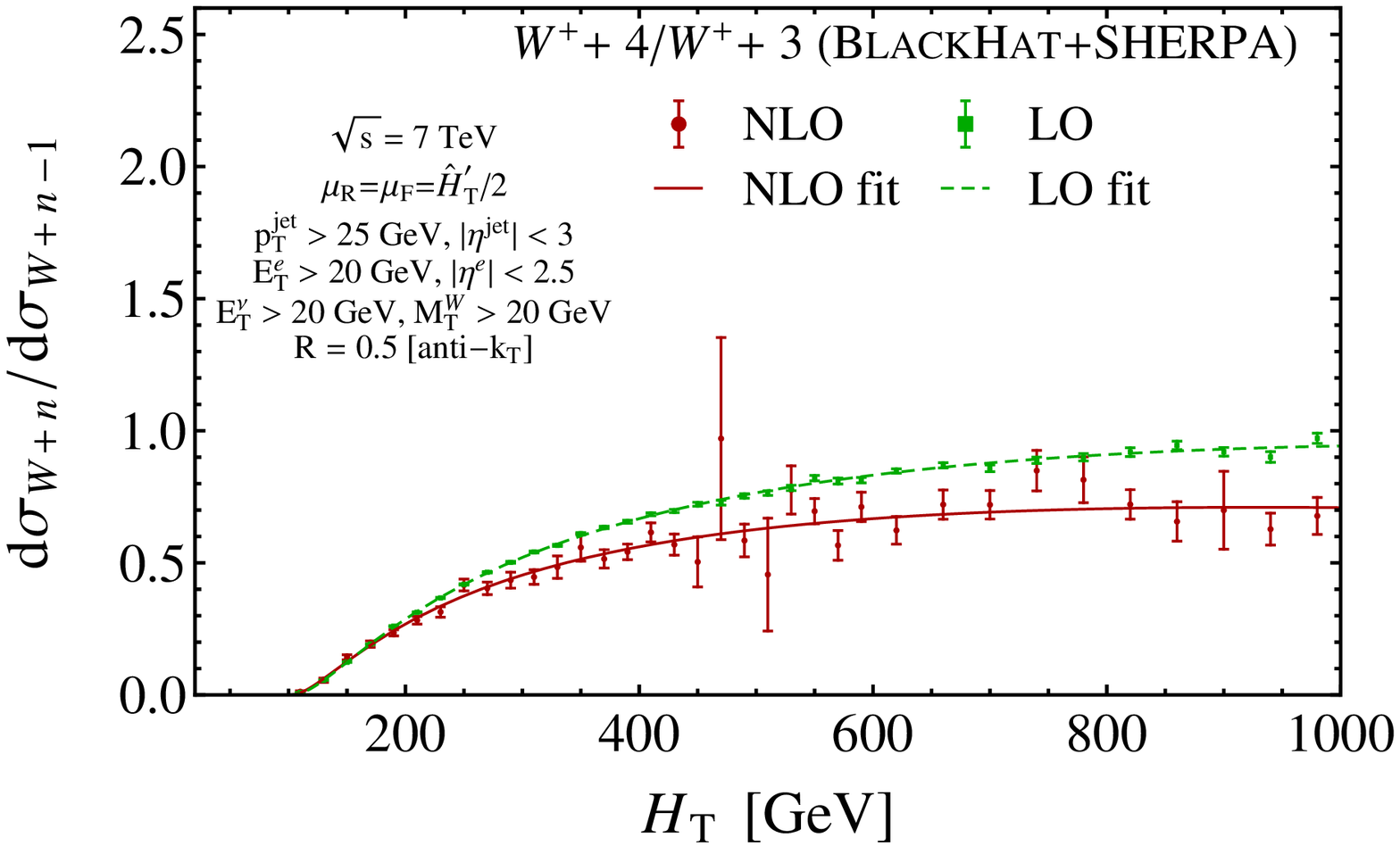}
\\
(c)&(d)\\
\includegraphics[clip,scale=0.43]{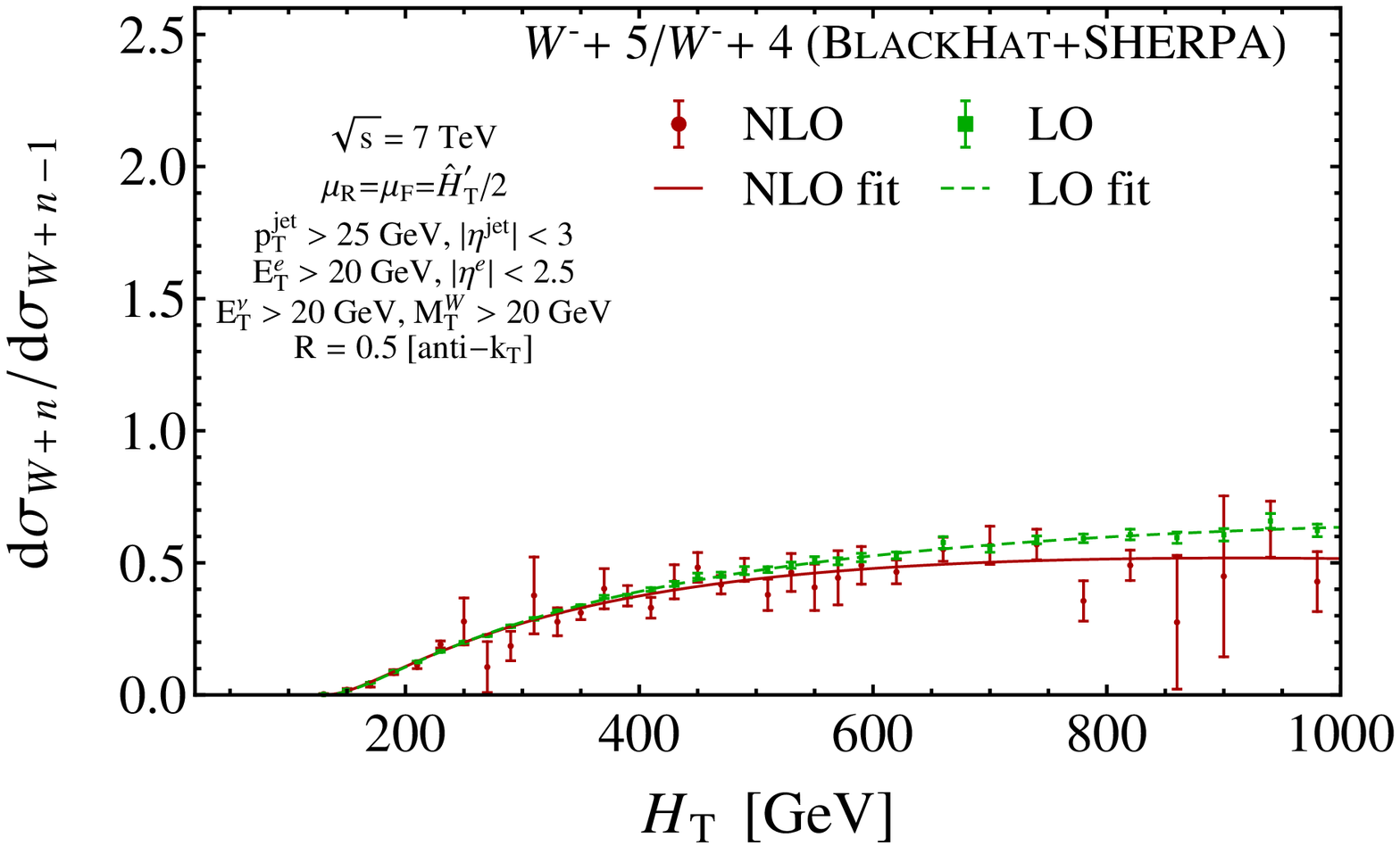}
&\includegraphics[clip,scale=0.43]{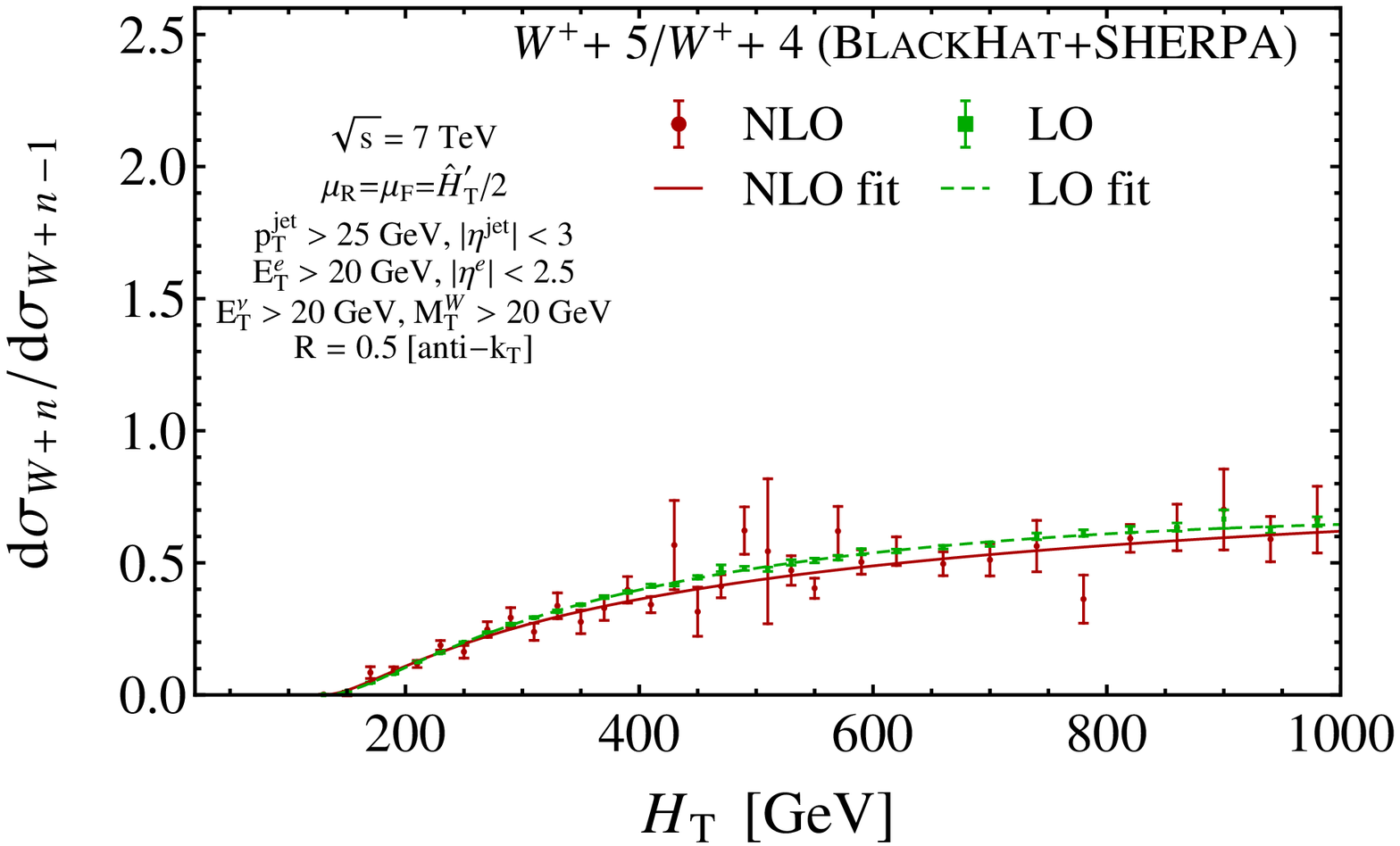}
\\
(e)&(f)
\end{tabular}
\end{minipage}
\end{center}
\caption{Fits to the ratios of the \Wjn-jet to \Wjnm-jet differential
  cross sections as a function of $\HT$.  Each plot shows the computed
  LO and NLO ratios, as well as fits with the parameters in
  \tabs{WmHTFit}{WpHTFit}.  
In the left column we show the ratios for $W^-$ and in the right column 
for $W^+$.
From top to bottom, $n$ goes from 3 to 5.
}
\label{WHTJetProductionRatioFitFigure}
\end{figure}

We display these fits to the jet production ratios in
\fig{WHTJetProductionRatioFitFigure}.
The same remarks apply here as in fits to the total cross sections as
functions of the minimum jet $\pT$, and to the differential cross
sections in the $W$ boson $\pT$: the functional forms are only
approximate, and the adequacy of the fits reflects the limited
statistical precision of our Monte-Carlo integrations.  Nonetheless,
it is remarkable that the results can be fit with so few parameters,
and fits to experimental data should be expected to be quite good,
given anticipated experimental uncertainties.

In our study of \Wjjjjj-jet production~\cite{W5j}, we used the results
for the NLO total cross sections for \Wjjj-jet, \Wjjjj-jet, and
\Wjjjjj-jet production to extrapolate and obtain predictions for the
total cross sections at NLO for \Wmjjjjjj-jet and \Wpjjjjjj-jet
production.  Here, we will go a bit further, and extrapolate to obtain
a prediction for the $\HT$ differential cross section in \Wjjjjjj-jet
production at NLO.

We will use the form in \eqn{WHTModel}, and extrapolate the NLO
parameters in \tabs{WmHTFit}{WpHTFit}.  Because of the different
behavior of the \Wjj-jet to \Wj-jet ratio, we exclude it from the fit.
We perform a linear extrapolation on the $\tauH_n$ and $\gammaH_n$,
and determine $N^H_6$ by matching to the extrapolated total cross
section rather than by direct extrapolation of the ratios in
\tabs{WmHTFit}{WpHTFit}.  This gives us a prediction for the
ratio of $\HT$
distributions in \Wjjjjjj-jet and \Wjjjjj-jet production.  In order to obtain
an estimate of an uncertainty band for this prediction, we
again use a Monte Carlo approach.  Because the uncertainties on 
the $\tauH$s and $\gammaH$s are highly correlated, we estimate the
uncertainties in the extrapolation as follows. 

We generate an ensemble of 1,000 synthetic data sets, where each bin
of the different $\HT$ distributions is chosen from a
normally-distributed ensemble with mean given by the computed value,
and width given by the estimated statistical uncertainty in our NLO
computation.  For each collection of distributions from the ensemble,
we form the \Wjn- to \Wjnm-jet ratios, and fit to ratios of the
corresponding forms from \eqn{WHTModel}.  We also compute the
corresponding total cross section by summing all bins.  For each
collection, we then extrapolate the total cross section as well as the
$\tauH_n$ and $\gammaH_n$ parameters linearly.  That is, we perform
the entire fitting and extrapolation procedure independently for each
synthetic data set.  This provides us with an element of an ensemble
of curves around our central prediction.  For our uncertainty band, we
retain all curves whose parameters are within the correlated 68\%
confidence-level ellipsoid of the central values of the fits.  In the
present case, the parameters are highly correlated;
ignoring the correlations would yield a much wider uncertainty band.

In order to extrapolate the distributions (and not just ratios of
distributions), we need to have an estimate of the $f^H$ function
appearing in \eqn{WHTModel}.  To obtain one, we
make use of the \Wjj-jet differential distribution,
\begin{equation}
f^H(H) \stackrel{\rm est}{=}
\frac{d\sigma_{W+2}}{d\HT}(H)  (2 a_s(H/2))^{-2} (N_2^H)^{-1} 
\ln^{-\tauH_2}\!\rho_H \, \bigl(1 - H/\HTmax\bigr)^{-\gammaH_2}\,.
\end{equation}
One could imagine using this equation to obtain values for $f^H(H)$
point-by-point, but it turns out to be more convenient and more stable to
have an analytic form for it.  In order to obtain such a form, we
need a fit to the \Wjj-jet NLO differential cross section.  It turns
out that we can use the following form to fit $f^H(H)$,
\begin{equation}
g_2 \ln^r(H/10) \biggl(\frac{H}{2\pTmin}\biggr)^{\omega_2} e^{-h_* H}\,,
\end{equation}
and obtain an adequate fit.  The parameters are,
\begin{equation}
g_2 = 100 {}^{+339}_{-77}\cdot 10^3\,,\quad \omega_2 = -4.2 \pm 0.7\,,\quad h_* = 0.0025 \pm 0.0004\,,\quad r = 3 \pm 2
\label{WmjjHTFitEqn}
\,,\end{equation}
for $W^-$ and,
\begin{equation}
g_2 = 17 {}^{+328}_{-16}\cdot 10^3\,,\quad \omega_2 = -5 \pm 1\,,\quad h_* = 0.0016 \pm 0.0008\,,\quad r = 6 \pm 4
\,,\label{WpjjHTFitEqn}
\end{equation}
for $W^+$.  (The distribution for $g_2$ is log-Gaussian rather than Gaussian.  The uncertainties on it are much larger than
the statistical uncertainty on the total cross section, because it is strongly correlated with the other parameters.  Two of the parameters, $\omega_2$ and $r$, are dimensionless; $g_2$ has units of fb/GeV, and $h_*$ of GeV${}^{-1}$.)

\begin{figure}[tb]
\begin{center}
\begin{minipage}[b]{1.03\linewidth}
\null\hskip -8mm%
\includegraphics[clip,scale=0.43]{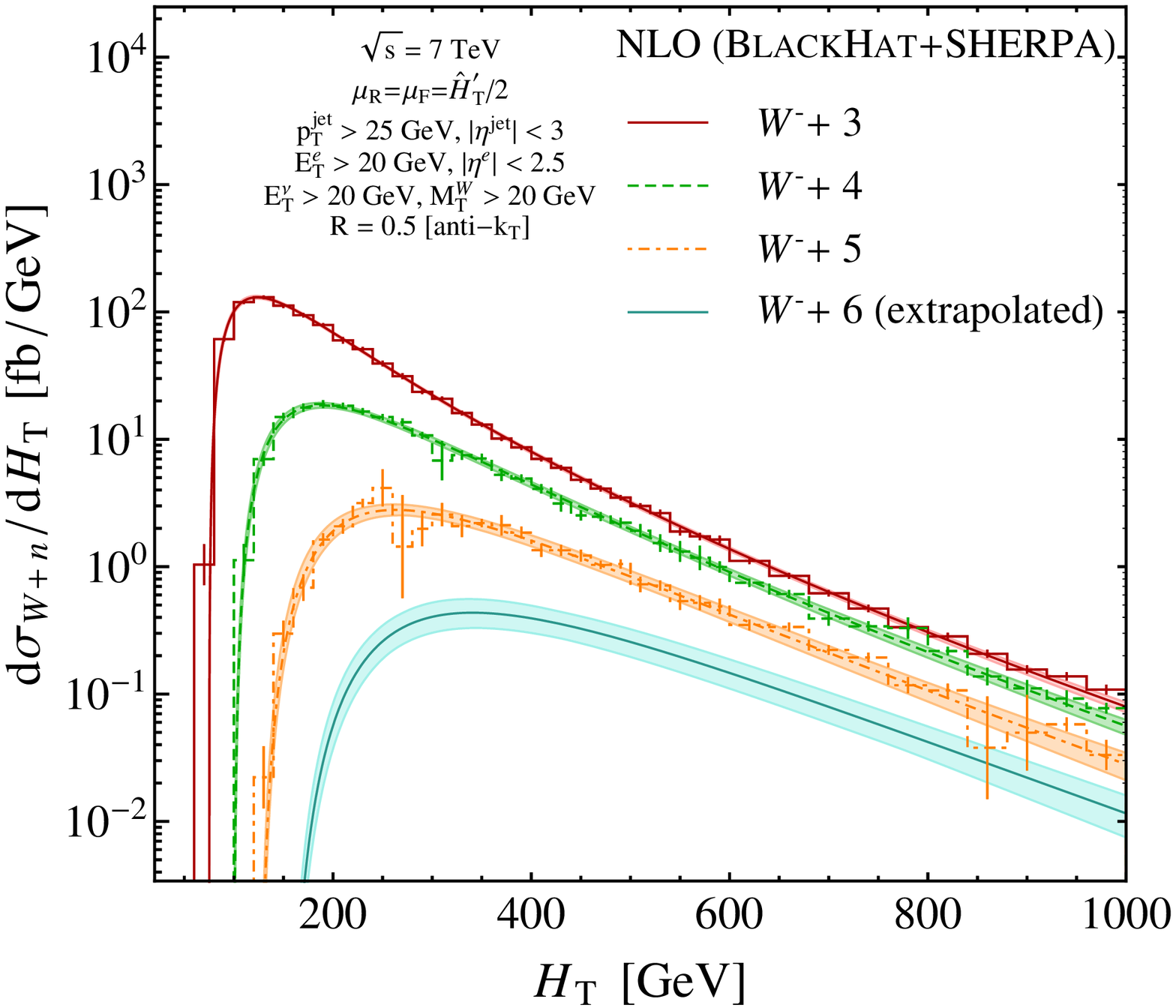}
\includegraphics[clip,scale=0.43]{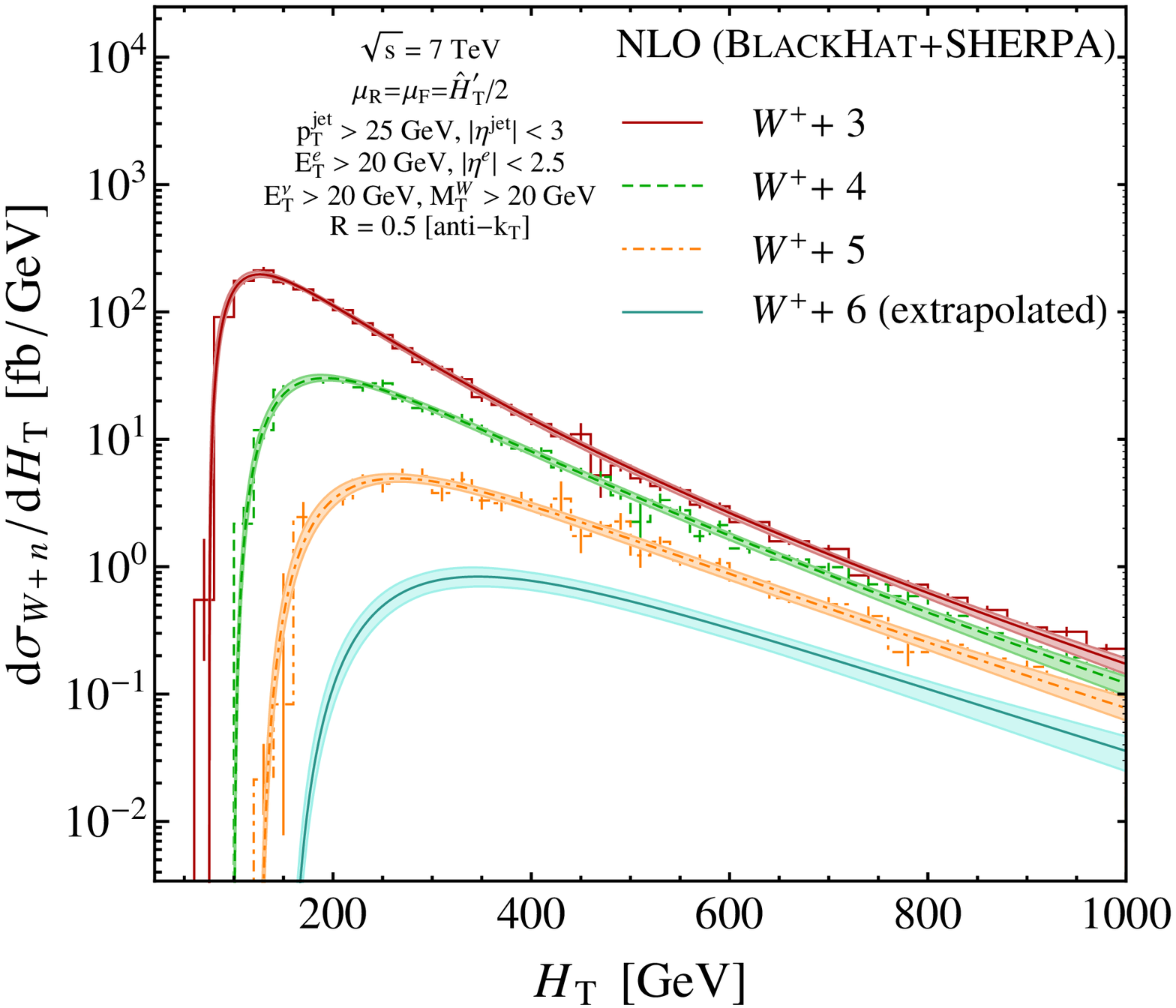}
\end{minipage}
\end{center}
\caption{NLO predictions for $\HT$ distributions
for \Wjn-jet production at the LHC, based on fits to ratios of
distributions, along with an extrapolation for $n=6$.  The 
computed numerical distributions for $n=3,4,5$ are shown against
the predictions
based on direct fits to ratios of distributions.  From the top down,
the curves (dark red, green, and orange) correspond to $n=3,4,5$.
The extrapolation-based prediction for $n=6$ is the bottom distribution,
shown in a solid (turquoise) line, with a shaded (light turquoise) band
showing the statistical uncertainty in our prediction. 
The normalizations of the curves
are adjusted to the total cross sections given in
\tab{CrossSectionAnti-kt-R5Table} for \Wjjj{} through \Wjjjjj-jet~\cite{W5j},
and to the extrapolated total cross section for \Wjjjjjj-jet.
The left plot is for $W^-$, and the right plot for $W^+$.
}
\label{WHTJetPredictionFigure}
\end{figure}

We can verify the consistency of this procedure, by comparing the
total cross sections for $n=3,4,5$ obtained by integrating the
`predicted' differential cross sections over the entire range from
threshold to the maximum $\HT$ with the corresponding NLO total cross
sections computed by summing the $\HT$ histograms\footnote{
  The statistical
  uncertainties in the summed histograms are typically substantially
  larger than in the total cross section, presumably because 
  real-emission configurations and corresponding
  counter-configurations can fall into different bins.  The summed
  histogram is probably a more suitable reference, because the
  analytic forms are effectively fit to the distribution.}.  We find
that they are in agreement.  After extrapolating the exponents in
\eqn{WHTModel}, and fixing the normalization $N^H_6$ using the
extrapolated value of the total cross section~\cite{W5j},
\begin{eqnarray}
&&W^- + 6 \hbox{ jets}:\quad 0.15 \pm 0.01\ \hbox{pb}\,,\nonumber\\
&&W^+ + 6 \hbox{ jets}:\quad 0.30 \pm 0.03\ \hbox{pb}\,,
\end{eqnarray}
we find for $W^-$,
\begin{equation}
N^H_6 = 2.0\cdot 10^3\,, \quad \tauH_6 = 4.6 \pm 0.1\,,\quad \gammaH_6 = 13 \pm 2
\,,\end{equation}
and for $W^+$,
\begin{equation}
N^H_6 = 1.0\cdot 10^3\,,\quad \tauH_6 = 4.3 \pm 0.2\,,\quad \gammaH_6 = 11 \pm 3
\,.
\end{equation}
We do not quote an error for the normalization $N^H_6$ because its value is
tightly correlated with the values of the exponents.
The predictions for the \Wjjjjjj-jet $\HT$ distributions are
shown in \fig{WHTJetPredictionFigure}.

\begin{figure}[tb]
\begin{center}
\begin{minipage}[b]{1.03\linewidth}
\null\hskip -8mm%
\includegraphics[clip,scale=0.43]{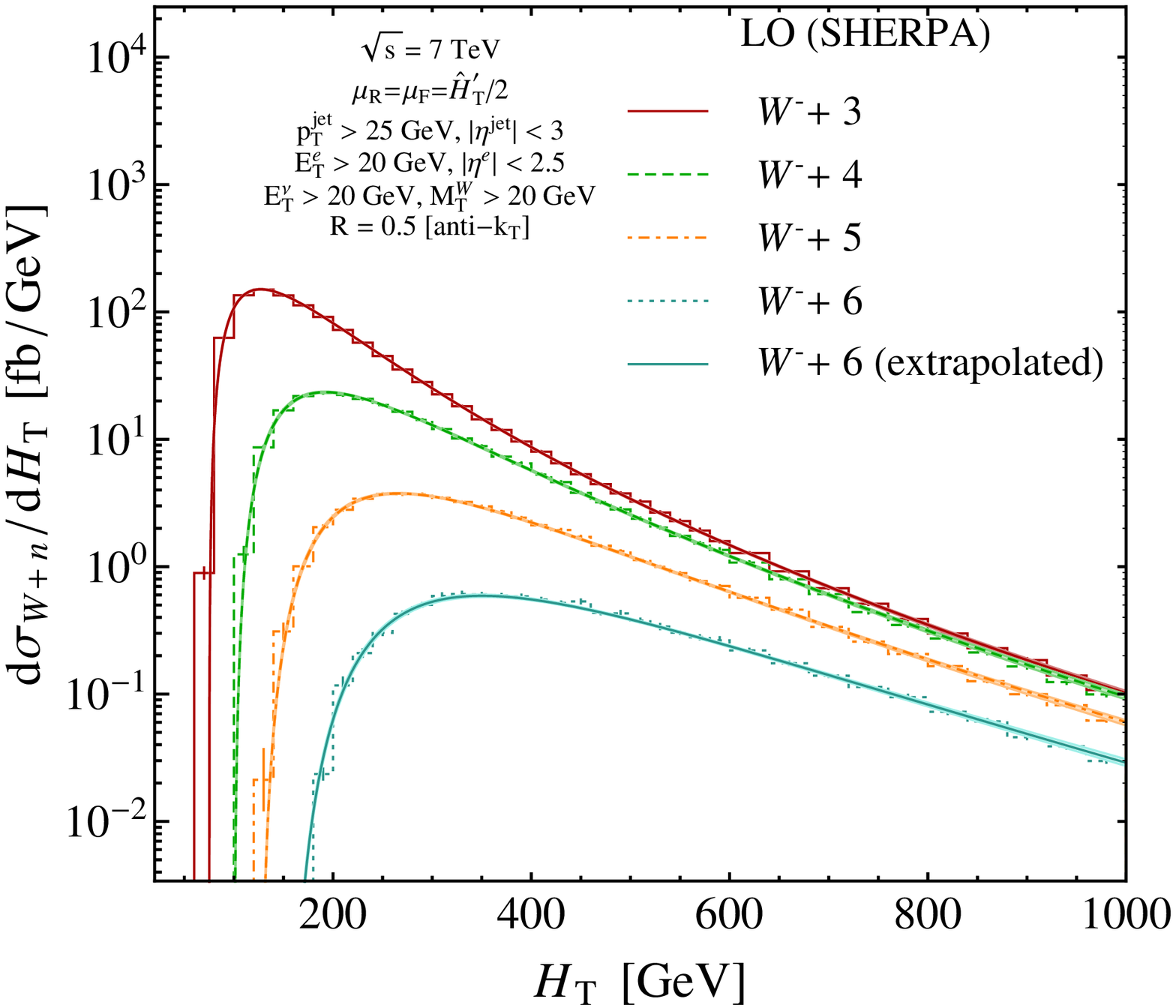}
\includegraphics[clip,scale=0.43]{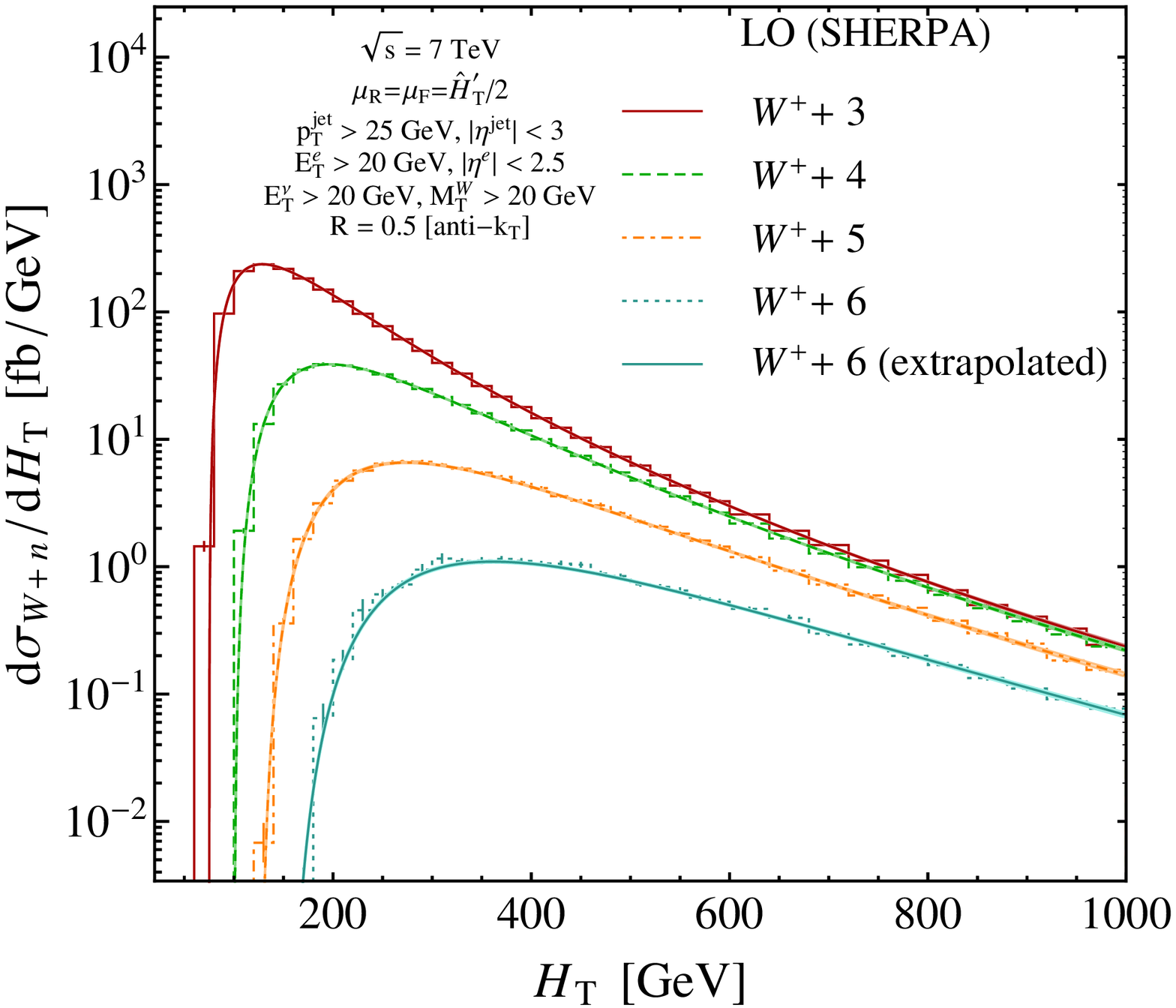}
\end{minipage}
\end{center}
\caption{LO test of an extrapolation to \Wjjjjjj{} jets.
The $\HT$ distributions for \Wjn-jet production at the LHC
are computed at LO for $n=3,4,5,6$ (histograms).
The predictions based on fits to ratios of LO distributions
are the curves, from the top down (dark red, green, and orange)
corresponding to $n=3,4,5$.
The extrapolation-based prediction for $n=6$ is the bottom curve, shown as a
solid (turquoise) line, with a shaded (light turquoise) band showing the
statistical uncertainty in our prediction.  The histogram shows a direct
computation of the same distribution.  The normalizations of the curves
are adjusted, respectively, to the total cross sections given in
\tab{CrossSectionAnti-kt-R5Table} for \Wjjj{} through \Wjjjjj-jet,
and to the extrapolated total cross section for \Wjjjjjj-jet.
The left plot is for $W^-$, and the right plot for $W^+$.
}
\label{WHTJetLOPredictionFigure}
\end{figure}

\begin{figure}[tb]
\begin{center}
\begin{minipage}[b]{1.03\linewidth}
\null\hskip -8mm%
\includegraphics[clip,scale=0.42]{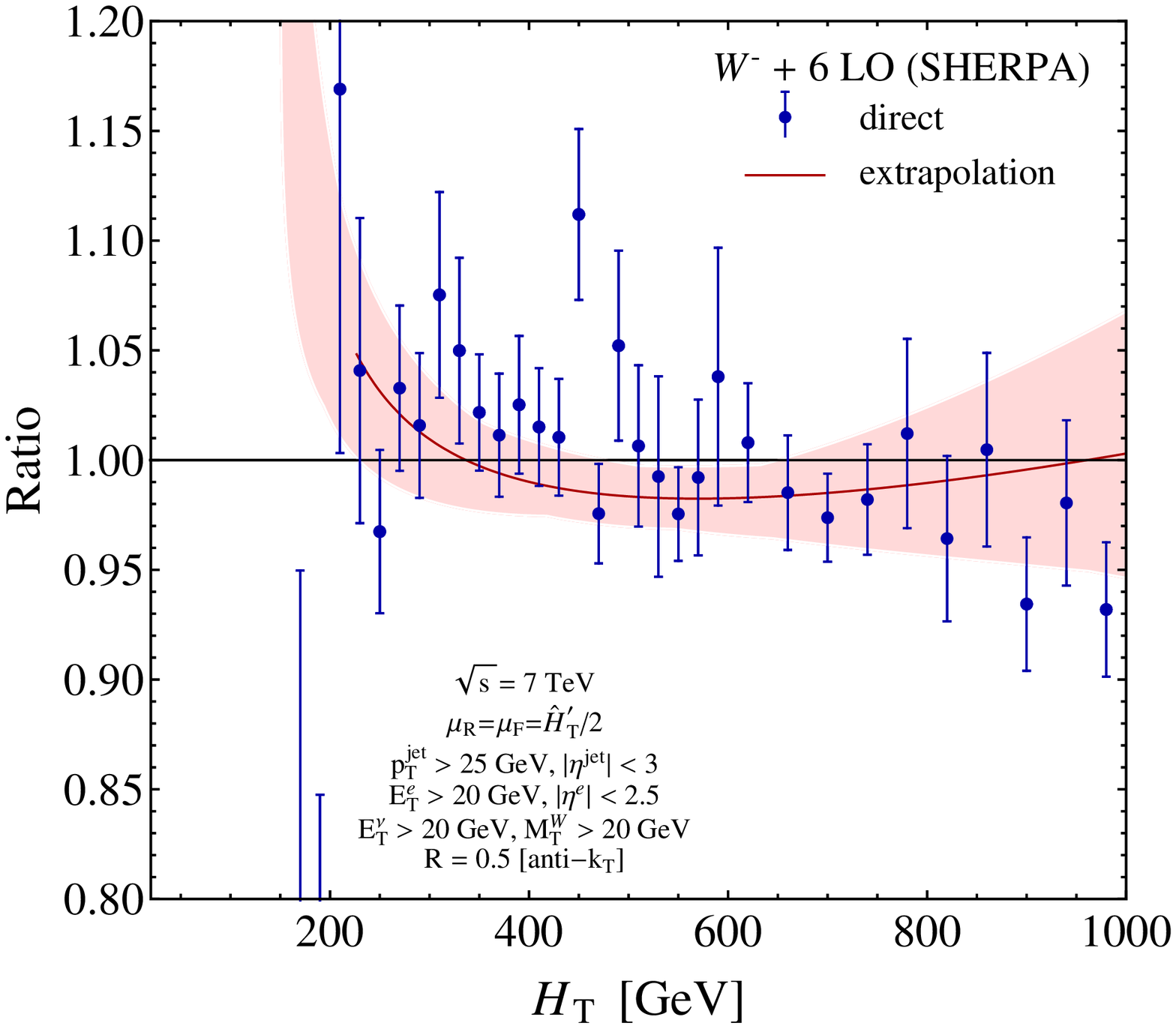}
\includegraphics[clip,scale=0.42]{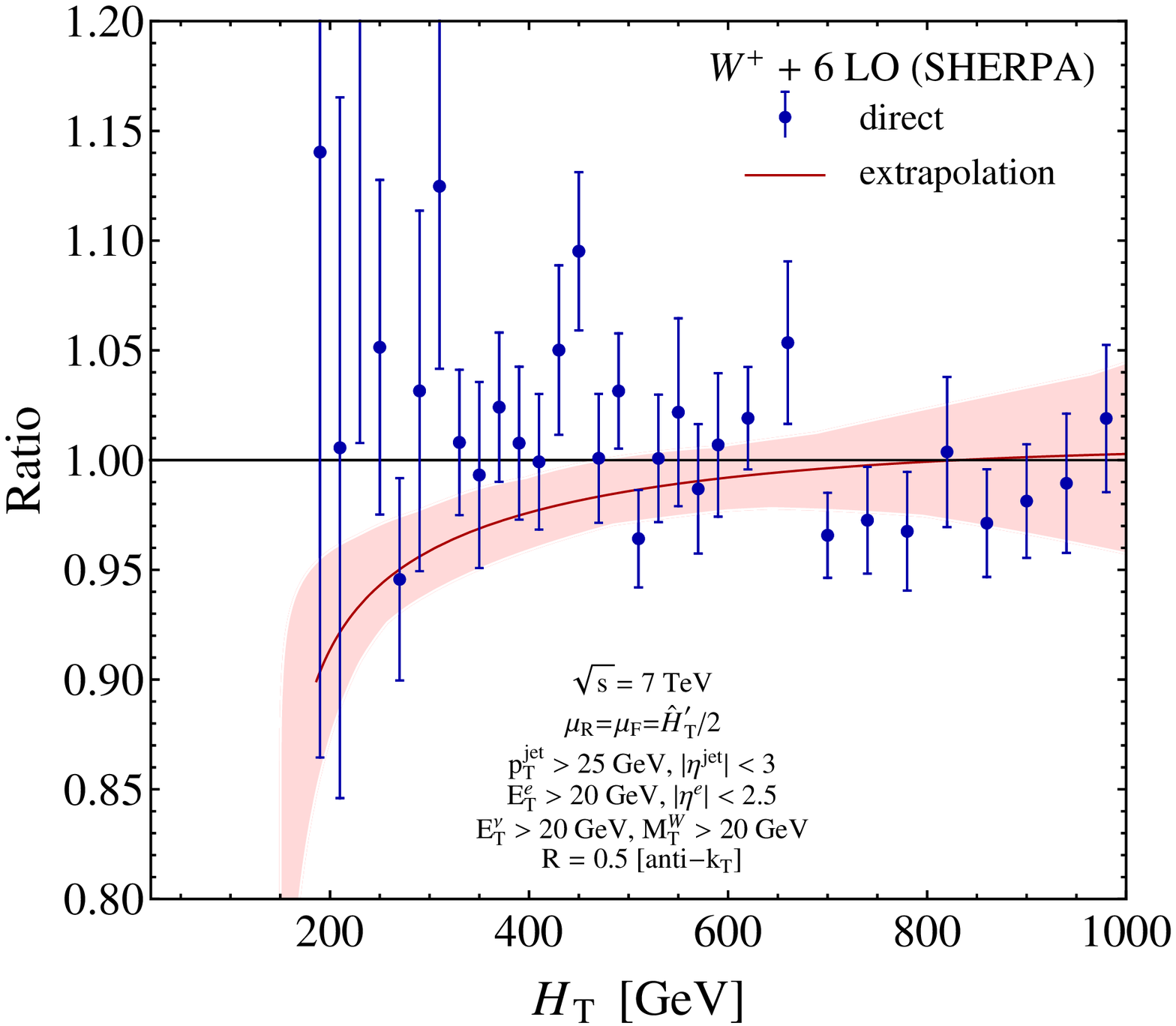}
\end{minipage}
\end{center}
\vskip -.5 cm 
\caption{\baselineskip 19 pt
The ratio of the extrapolation-based LO prediction for the
  $\HT$ distribution for \Wjjjjjj-jet production at the LHC to a fit
  to the direct computation (solid dark red line), along with a 68\%
  uncertainty band for the extrapolation.  The (blue) points show the
  ratio of the direct computation to a fit to that computation, with
  statistical uncertainties alone.  The left plot is for $W^-$, and
  the right plot for $W^+$.  }
\label{WHTJetLOComparisonFigure}
\end{figure}

We can also check the extrapolation procedure described above by
comparing an extrapolation against a direct calculation at LO.  We
have done this for the LO \Wjjjjjj-jet $\HT$ distribution, where the
direct calculation is feasible.  The results are shown in
\fig{WHTJetLOPredictionFigure}.  The direct fit is again made by
fitting ratios of the model in \eqn{WHTModel} to the computed
\Wjjjjjj-jet to \Wjjjjj-jet ratio, and using the parameters obtained,
along with those in \eqns{WmjjHTFitEqn}{WpjjHTFitEqn}, to obtain a
curve for the \Wjjjjjj-jet $\HT$ distribution itself.  The agreement
of the extrapolation with the direct calculation is excellent.  On a
logarithmic scale, any differences are hard to see, so in
\fig{WHTJetLOComparisonFigure} we show the ratios of the extrapolated
distribution and of the computed distributions to the direct fit to
the computation.  The direct fit in this figure is thus represented by
the horizontal axis; the computation by itself by the points with
associated statistical uncertainties and the extrapolated distribution
by the colored curve. The uncertainty in the extrapolation is given by 
the shaded band.  For $W^{-}$ production, the extrapolation,
direct fit, and calculation all agree within 5\% except right above
threshold.  In the region contributing the bulk of the cross section,
the extrapolation and direct fit agree within 3\%.  For $W^+$
production, the agreement is not as good, but the extrapolation,
direct fit, and calculation again agree to within 5\% over most of the
range.

\begin{figure}[tb]
\begin{center}
\begin{minipage}[b]{1.03\linewidth}
\null\hskip -5mm
\includegraphics[clip,scale=0.42]{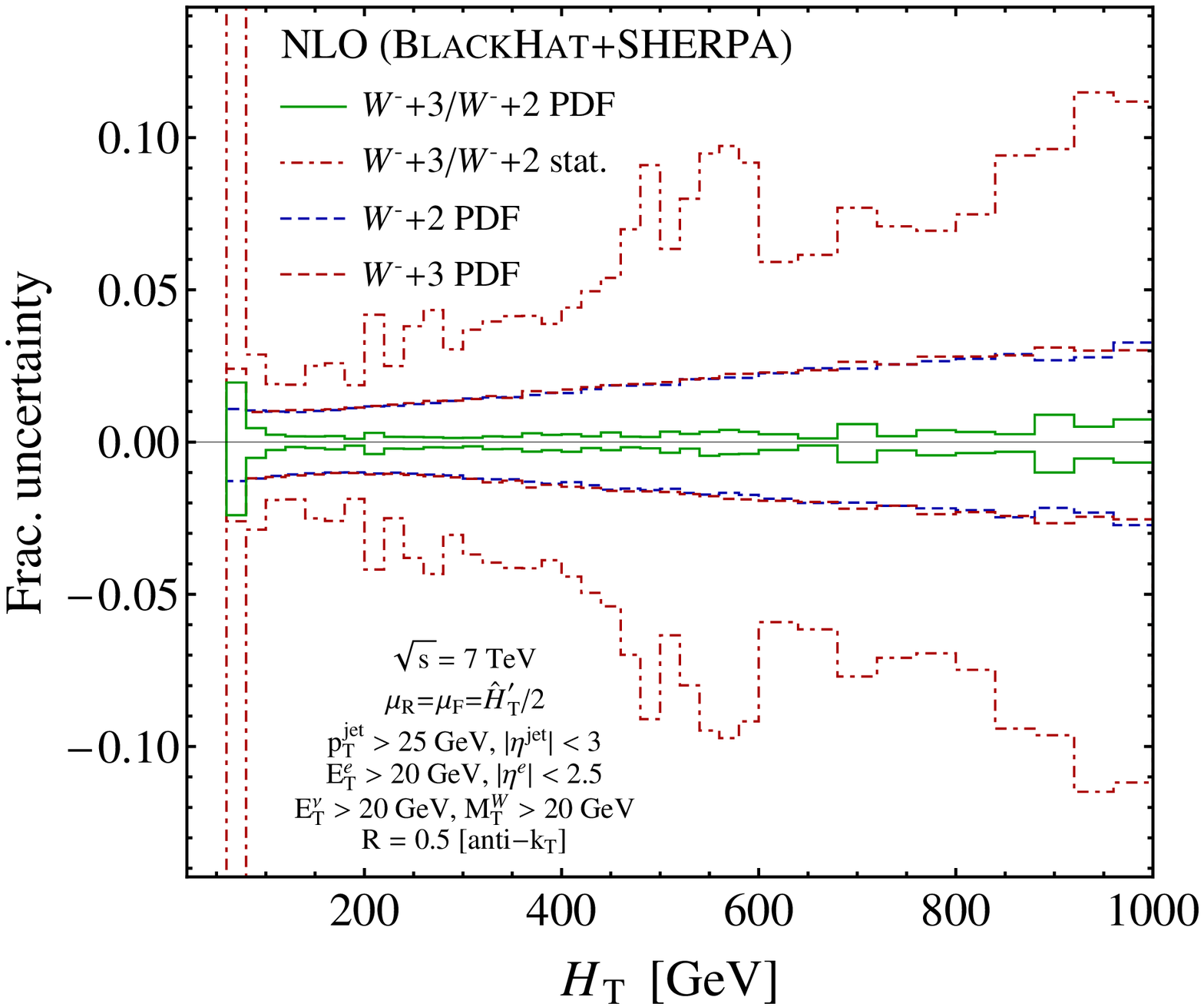}
\includegraphics[clip,scale=0.42]{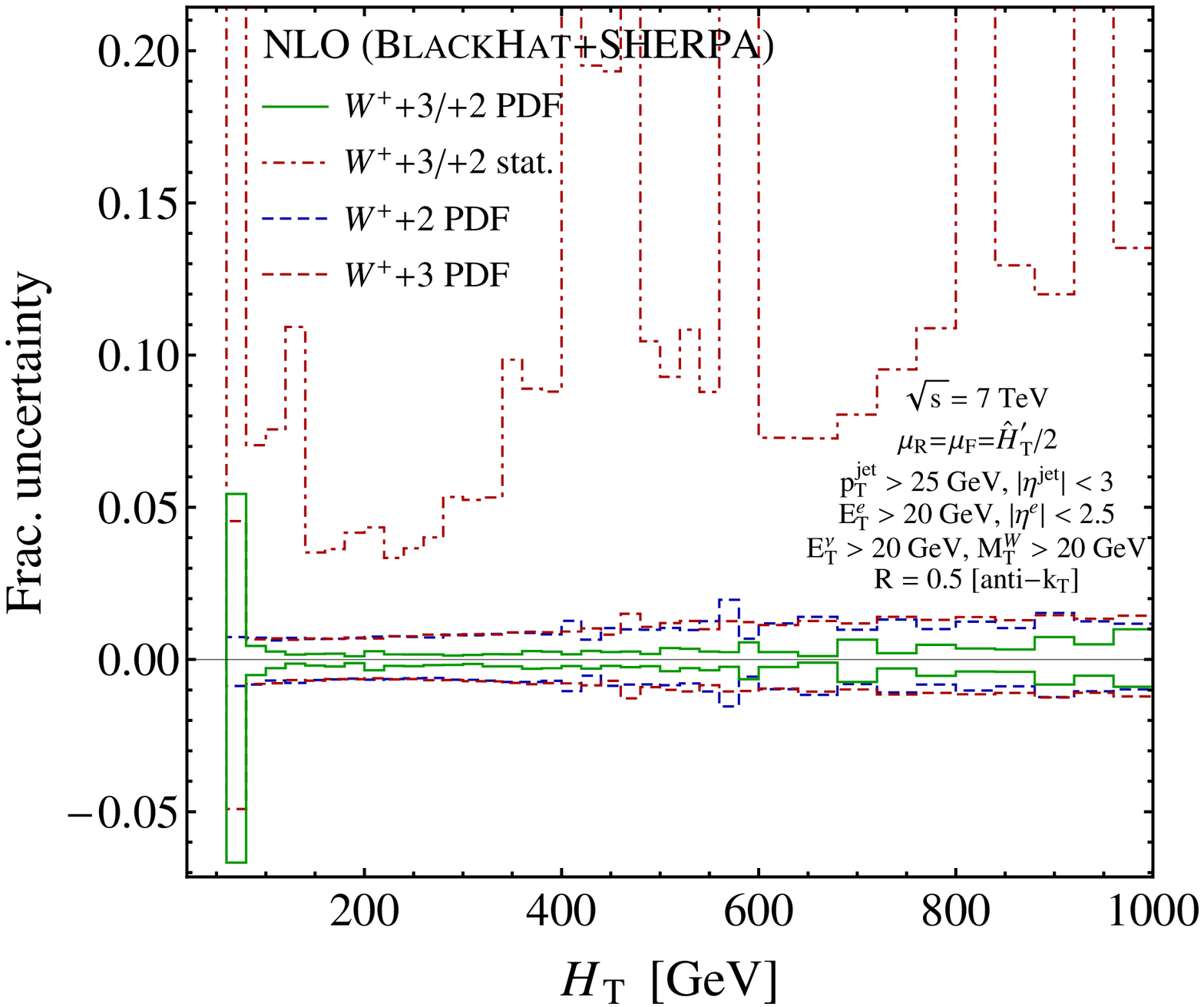}
\end{minipage}
\end{center}
\vskip -0.5 cm 
\caption{\baselineskip 19 pt
PDF uncertainties in the ratio of the \Wjjj-jet to \Wjj-jet
  $\HT$ distribution at NLO.  The plots show, from the smallest
  absolute values to the largest: the PDF uncertainties on the
  \Wjjj/\Wjj-jet ratio (solid green); the separate uncertainties on
  the \Wjjj- (dashed dark red) and \Wjj-jet (dashed blue)
  distributions; and the statistical uncertainty on the \Wjjj/\Wjj-jet
  ratio (dot-dashed dark red) shown for comparison.  The left plot
  shows the $W^{-}$ differential cross sections, and the right plot
  the $W^+$ ones.  In the right plot, only the positive statistical
  uncertainty is shown.  }
\label{HTPDFUncertaintyFigure}
\end{figure}

We have also evaluated the PDF uncertainties for the $\HT$
distribution, using the same approach as for the total cross sections
studied in \sect{MinJetPTSection}.  We display the results in
\fig{HTPDFUncertaintyFigure}.  The left figure shows that the PDF
uncertainties in the separate \Wmjj-jet and \Wmjjj-jet distributions
are small, ranging from 1\% at smaller values of $\HT$ above the
\Wjjj-jet threshold to just below 3\% at the highest values.  The PDF
uncertainties in the ratio are considerably smaller, ranging from
0.5\%{} at smaller $\HT$ values to less than 1\% even at the
highest $\HT$ values.  In \Wpjn-jet production, the PDF uncertainties
on the separate distributions are smaller than in the \Wmjn-jet case,
but the uncertainties on the ratios remain comparably smaller.  With
the number of events we have collected, these uncertainties are
considerably smaller than the statistical uncertainties in the ratio,
especially in \Wpjn-jet production.

\FloatBarrier

\subsection{Dependence on Jet Transverse Momenta}
\label{JetPTSection}

\begin{figure}[tb]
\begin{center}
\null\hskip -3mm\begin{minipage}[b]{1.\linewidth}
\begin{tabular}{cc}
\includegraphics[clip,scale=0.43]{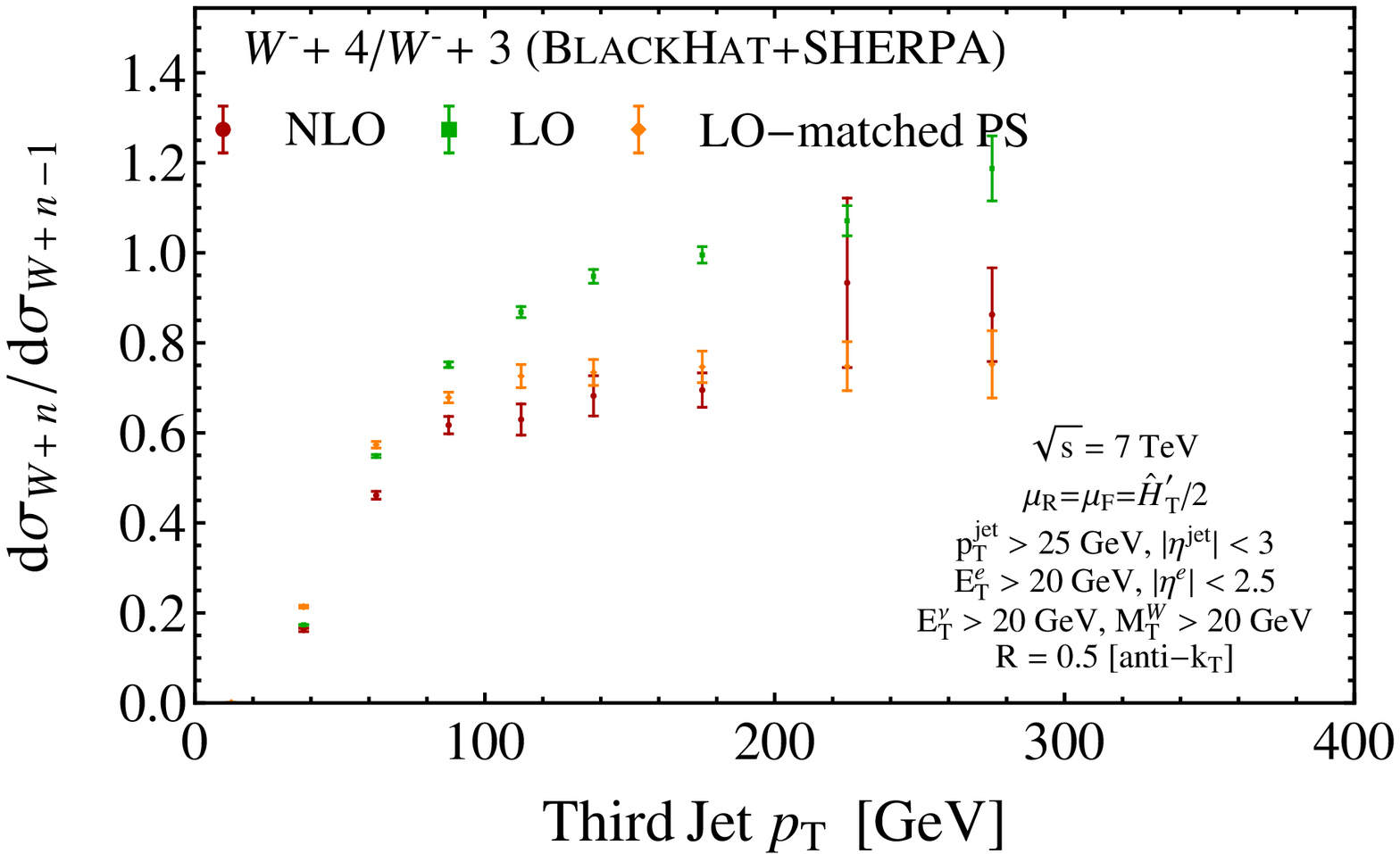}
&\includegraphics[clip,scale=0.43]{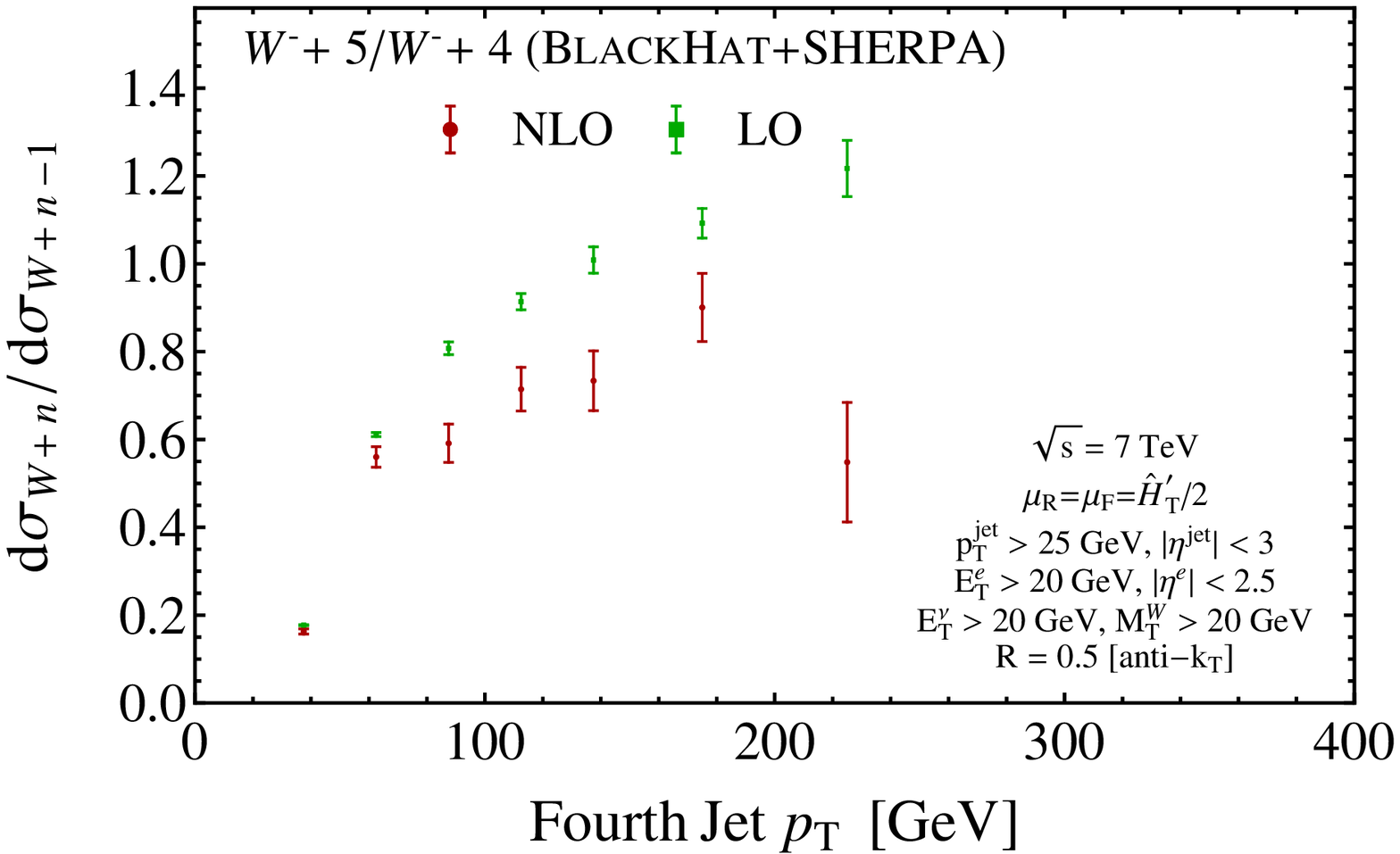}
\\
(a)&(b)
\end{tabular}
\end{minipage}
\end{center}
\caption{The ratios of \Wmjn-jet to \Wmjnm-jet cross sections as a
  function of the softest-jet $\pT$, at LO, at NLO,
  and for the \Wmjjjj-jet to \Wmjjj-jet ratio, in a parton-shower calculation
  matched to LO.  In (a) we show
  the \Wmjjjj-jet to \Wmjjj-jet ratio as a function of the third-jet
  $\pT$; and in (b) the \Wmjjjjj-jet to \Wmjjjj-jet ratio as a
  function of the fourth-jet $\pT$.  }
\label{WmSJetProductionRatioFigure}
\end{figure}

\begin{figure}[tb]
\begin{center}
\null\hskip -3mm\begin{minipage}[b]{1.\linewidth}
\begin{tabular}{cc}
\includegraphics[clip,scale=0.43]{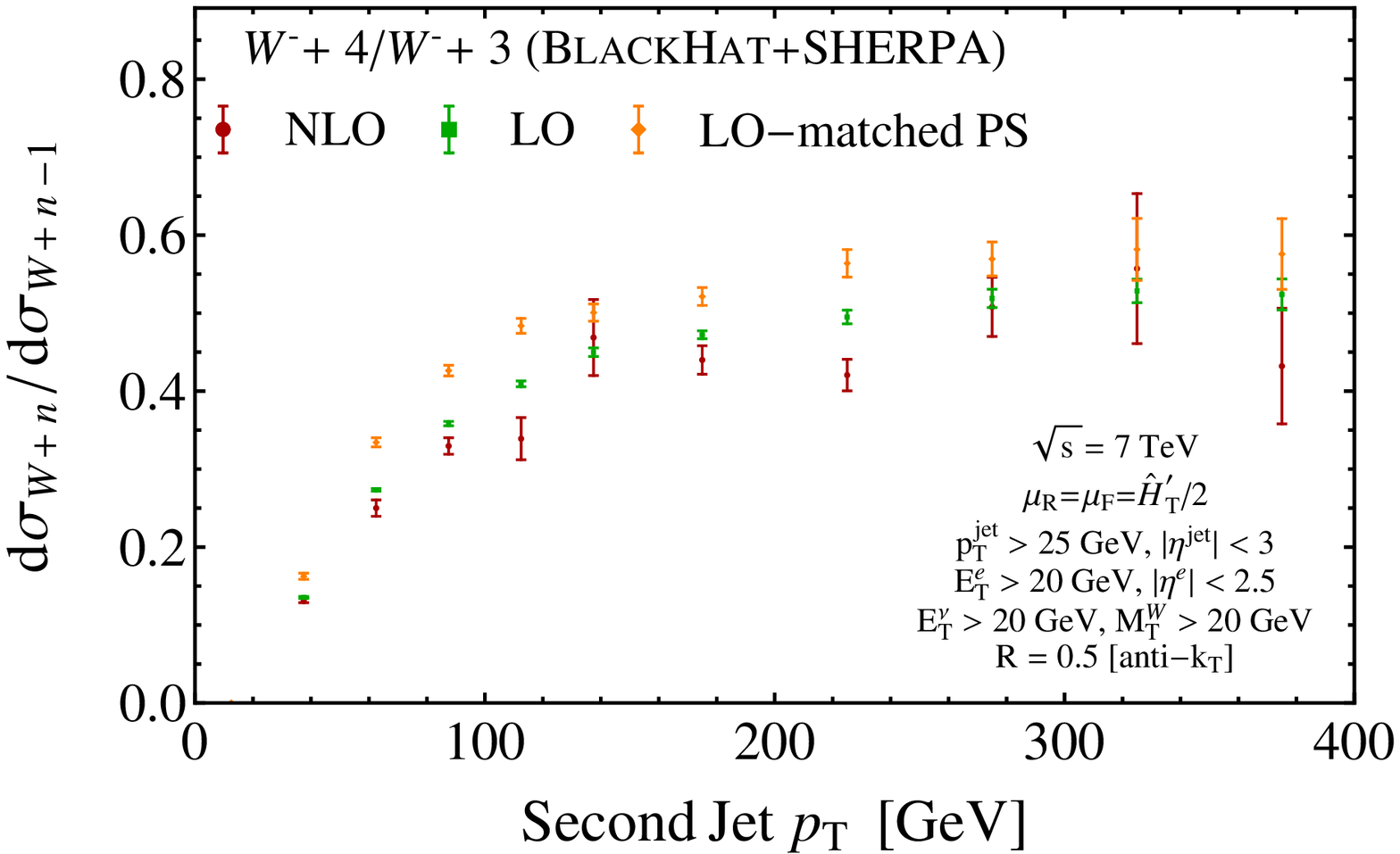}
&\includegraphics[clip,scale=0.43]{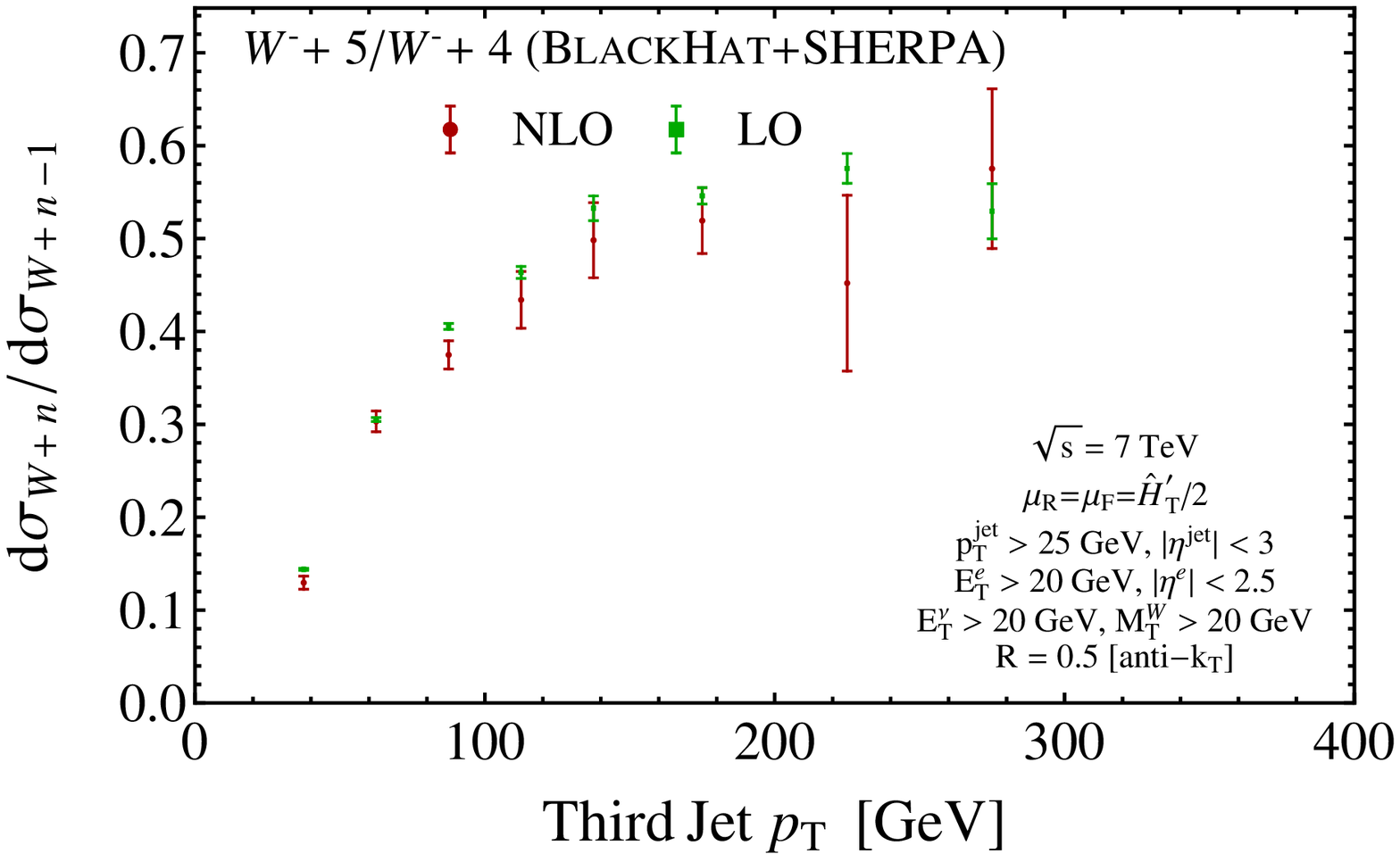}
\\
(a)&(b)
\end{tabular}
\end{minipage}
\end{center}
\caption{The ratios of \Wmjn-jet to \Wmjnm-jet cross sections as a
  function of the next-to-softest-jet $\pT$, at LO, at NLO,
  and for the \Wmjjjj-jet to \Wmjjj-jet ratio, in a parton-shower calculation
  matched to LO.  In (a)
  we show the \Wmjjjj-jet to \Wmjjj-jet ratio as a function of the
  second-jet $\pT$; and in (b) the \Wmjjjjj-jet to \Wmjjjj-jet ratio
  as a function of the third-jet $\pT$.  }
\label{WmNSJetProductionRatioFigure}
\end{figure}

In this subsection, we study jet-production ratios as a function
of various jet transverse momenta.  We give detailed results for the LO
and NLO 
\Wmjn-jet differential cross section as a function of the
second-, third-, and fourth-hardest jets' transverse momenta in
appendix~\ref{JetpTAppendix}.
  We display jet-production ratios as functions
of these jet transverse momenta 
in~\figs{WmSJetProductionRatioFigure}{WmNSJetProductionRatioFigure}, 
organized not by the
ordinal jet number (as in appendix~\ref{JetpTAppendix}),
but rather according to whether the jet is the softest 
(ordered $n-1$) in the
\Wmjnm-jet process, or the next-to-softest (ordered $n-2$).
That is, we consider the ratios,
\begin{equation}
\frac{d\sigma^{W+n}/d\pTn{,n-1}}
     {d\sigma^{W+n-1}/d\pTn{,n-1}}
{\hskip .3cm \rm\ and\ \hskip .3cm  }
\frac{d\sigma^{W+n}/d\pTn{,n-2}}
     {d\sigma^{W+n-1}/d\pTn{,n-2}}\,,
\end{equation}
in \figs{WmSJetProductionRatioFigure}{WmNSJetProductionRatioFigure},
respectively.

We see that the jet-production ratio as a function of the softest
jet's $\pT$ (\fig{WmSJetProductionRatioFigure})
suffers large NLO corrections, whereas the corrections as
a function of harder jet $\pT$s (\fig{WmNSJetProductionRatioFigure})
are much more modest. This is consistent with previous results~\cite{W4j}.
For comparison, we also show
results for the \Wjjjj-jet to \Wjjj-jet ratio obtained using the \SHERPA{}
parton shower matched to LO.  The parton-shower result
is somewhat above the LO and NLO results for the next-to-softest
jet, but in rough agreement with both.  It is closer to the NLO result
for the softest-jet distribution, suggesting that corrections beyond
NLO are not large.
\FloatBarrier

\section{Conclusions}
\label{ConclusionSection}

The first years of data and analyses from experiments at the Large
Hadron Collider (LHC) emphasize the need for reliable theoretical
calculations in searches for new physics beyond the Standard Model.
In many channels, new-physics signals can hide in broad distributions
underneath Standard Model backgrounds.  Extraction of signals requires
accurate predictions for the background processes, for which
next-to-leading order (NLO) cross sections in perturbative QCD are an
important first step.

In this paper, we computed jet-production ratios to NLO in QCD for $W$
production in association with up to five jets.  Ratios of cross
sections and of distributions are expected to have reduced theoretical
and experimental uncertainties, most notably a reduced uncertainty in
the jet-energy scale.  The \Wjn- to \Wjnm-jet ratios are formally of
$\Ord(\alpha_s)$, and accordingly are expected to have a much smaller
scale dependence than the scale dependence of the underlying cross
sections.  We used a leading-color approximation for the virtual terms
in \Wjjjj-{} and \Wjjjjj-jet production. We had previously validated
this approximation for \Wjjj-{} and \Wjjjj-jet
production~\cite{W3jDistributions,ItaOzeren}.  We expect the
subleading-color terms to contribute less than 3\% of the total cross
section, though it would be useful to verify this for \Wjjjjj-jet
production.  We expect other approximations we have used, such as
neglecting real and virtual top-quarks, as well as real-emission
contributions with four quark pairs, to have even smaller effects.

We have studied a number of quantities constructed from ratios of
cross sections.  In \sect{MinJetPTSection}, we studied the ratio of
cross sections as a function of the minimum jet transverse momentum
$\pTmin$.  We saw that for ratios with $n\ge 3$, the dependence is
quite similar for different values of $n$ even at LO, and very similar
at NLO.  This suggests a universal behavior that is in accord with the
(weak) linear dependence on $n$ of ratios of total cross sections at
fixed $\pTmin$~\cite{W5j}.  The dependence on $\pTmin$ can be
described very well by a three-parameter functional form.  Its
behavior is dominated by a power-law behavior with an exponent
consistent with being independent of the number of jets, both at LO
and at NLO.  We have studied the uncertainty on the ratios due to
imprecise knowledge of the parton distribution functions, and find
that as expected, the ratio is less sensitive than the underlying
cross sections.  For the range of $\pTmin$ values we have studied (25
to 120 GeV), the fractional uncertainty in the ratio due to PDF
uncertainty varies from 0.6\%{} to 1.5\%.

We have also studied more differential quantities.  In
\sect{WpTSection}, we studied ratios of differential distributions
with respect to the transverse momentum of the $W$ boson $\pTV$.  The
ratios of distributions can be described by a simple three-parameter
fit form, a simpler form than needed for the distributions themselves.
The \Wjj-{} to \Wj-jet ratio again behaves differently than ratios for
$n\ge 3$; but the latter ratios are similar in shape, and depend
primarily on powers of a single log of $\pTV/\pTmin$.  The simplicity
of the ratios suggests that at large $\pTV$, the production process
can be understood as the production of a $W$ boson recoiling against a
jet system, with a universal function describing the ``fragmentation''
(perturbative splitting) of the jet system into individual jets.  The
NLO corrections are significant for $n=2$, but small for larger $n$.

In \sect{WHTSection}, we studied the ratios of differential
distributions with respect to the total transverse energy in jets,
$\HTjets$.  We again find that the ratios can be described in terms of
a simple three-parameter fit function, with a universal function
describing the production of the $W$-boson plus jet system dropping
out of ratios.  The parameters of these fit function for $n\ge 3$ are
in turn very well described by a linear fit form.  We have used this
observation to extrapolate the $\HTjets$ distribution to \Wjjjjjj-jet
production at NLO.  A similar extrapolation at LO agrees very nicely
with a direct calculation, suggesting the procedure adds only 5\%
uncertainty over much of the range to the scale uncertainty.  We
expect that other distributions can be extrapolated using a similar
approach.

Finally, in \sect{JetPTSection}, we studied the ratios of differential
distributions in jet transverse momenta.  These ratios have large NLO
corrections for the softest identified jet, but only modest
corrections for harder jets.  A comparison to parton shower results
suggests that corrections beyond NLO are modest.

The study of extrapolations in \sect{WHTSection} is motivated by the
increasing difficulty of precision QCD calculations as the number of
jets increases.  The availability of the \Wjjjjj-jet calculation was
critical in allowing us to extrapolate to
\Wjjjjjj-jet production, and in principle, beyond it.  We look forward
to comparing the quantities studied here, and extrapolations of this
type, to LHC data.

\section*{Acknowledgments}
We are grateful to Kemal Ozeren for important contributions to early stages of
this project. We also thank Joey Huston, David Saltzberg, Maria
Spiropulu, Eric Takasugi, and Matthias Webber for helpful discussions.
This research was supported by the US Department of Energy under
contracts DE--AC02--76SF00515 and DE-{S}C0009937.  DAK’s research is
supported by the European Research Council under Advanced Investigator
Grant ERC--AdG--228301.  DM's work was supported by the Research
Executive Agency (REA) of the European Union under the Grant Agreement
number PITN--GA--2010--264564 (LHCPhenoNet). SH's work was partly
supported by a grant from the US LHC Theory Initiative through NSF
contract PHY--0705682.  The work of FFC is supported by the Alexander
von Humboldt Foundation, in the framework of the Sofja Kovalevskaja
Award 2014, endowed by the German Federal Ministry of Education and
Research. The work of HI is supported by the Juniorprofessor Program
of Ministry of Science, Research and the Arts of the state of
Baden-W\"urttemberg, Germany.  This research used resources of
Academic Technology Services at UCLA, and of the National Energy
Research Scientific Computing Center, which is supported by the Office
of Science of the U.S. Department of Energy under Contract
No.~DE--AC02--05CH11231.

\appendix
\section{Cross Sections as a Function of the Minimum Jet Transverse Momentum}
\label{MinJetpTAppendix}

\begin{table}[ht]
\vskip .4 cm
\renewcommand{\arraystretch}{0.86}
\centering
\begin{tabular}{||c|l|l|l|l|}
\hline
Min jet $p_{T}$ &  \multicolumn{1}{|c|}{$\displaystyle\vphantom{\sum_{i=1}^W}\frac{W^-+2}{W^-+1}$} &  
\multicolumn{1}{|c|}{$\displaystyle\frac{W^-+3}{W^-+2}$} &  
\multicolumn{1}{|c|}{$\displaystyle\frac{W^-+4}{W^-+3}$} & 
\multicolumn{1}{|c|}{$\displaystyle\frac{W^-+5}{W^-+4}$}  \\
\hline
25 &~$0.2949(0.0004)$~&~$0.2511(0.0005)$~&~$0.2345(0.0009)$~&~$0.218(0.001)$~\\
30 &~$0.2751(0.0003)$~&~$0.2222(0.0005)$~&~$0.2101(0.0009)$~&~$0.194(0.001)$~\\
35 &~$0.2627(0.0004)$~&~$0.1988(0.0005)$~&~$0.190(0.001)$~&~$0.174(0.001)$~\\
40 &~$0.2560(0.0004)$~&~$0.1795(0.0005)$~&~$0.175(0.001)$~&~$0.157(0.001)$~\\
45 &~$0.2529(0.0004)$~&~$0.1635(0.0005)$~&~$0.159(0.001)$~&~$0.146(0.001)$~\\
50 &~$0.2527(0.0005)$~&~$0.1498(0.0005)$~&~$0.1453(0.0009)$~&~$0.136(0.001)$~\\
55 &~$0.2545(0.0005)$~&~$0.1379(0.0005)$~&~$0.134(0.001)$~&~$0.125(0.001)$~\\
60 &~$0.2572(0.0005)$~&~$0.1266(0.0005)$~&~$0.126(0.001)$~&~$0.116(0.001)$~\\
65 &~$0.2599(0.0006)$~&~$0.1166(0.0004)$~&~$0.118(0.001)$~&~$0.109(0.001)$~\\
70 &~$0.2620(0.0006)$~&~$0.1078(0.0004)$~&~$0.112(0.001)$~&~$0.101(0.001)$~\\
75 &~$0.2643(0.0007)$~&~$0.1008(0.0005)$~&~$0.106(0.001)$~&~$0.094(0.002)$~\\
80 &~$0.2669(0.0007)$~&~$0.0938(0.0004)$~&~$0.100(0.002)$~&~$0.088(0.002)$~\\
85 &~$0.2702(0.0008)$~&~$0.0875(0.0004)$~&~$0.096(0.002)$~&~$0.082(0.002)$~\\
90 &~$0.2730(0.0008)$~&~$0.0823(0.0004)$~&~$0.091(0.002)$~&~$0.078(0.003)$~\\
95 &~$0.2764(0.0009)$~&~$0.0773(0.0004)$~&~$0.087(0.003)$~&~$0.072(0.003)$~\\
100 &~$0.280(0.001)$~&~$0.0731(0.0004)$~&~$0.080(0.001)$~&~$0.069(0.001)$~\\
105 &~$0.284(0.001)$~&~$0.0691(0.0004)$~&~$0.076(0.001)$~&~$0.066(0.001)$~\\
110 &~$0.288(0.001)$~&~$0.0651(0.0005)$~&~$0.071(0.001)$~&~$0.064(0.001)$~\\
115 &~$0.292(0.001)$~&~$0.0614(0.0005)$~&~$0.068(0.001)$~&~$0.061(0.002)$~\\
120 &~$0.296(0.001)$~&~$0.0581(0.0005)$~&~$0.064(0.001)$~&~$0.059(0.002)$~\\

\hline
\end{tabular}
\caption{The jet-production ratio in \Wmjn{} jets at LO as a function of the minimum jet $\pT$ in GeV.  These values
are shown in the left plot in \fig{WmMinJetPtJetProductionRatioFigure}.
}
\label{WmJetProductionRatioJetPtLOTable}
\end{table}

\begin{table}[ht]
\vskip .4 cm
\renewcommand{\arraystretch}{0.86}
\centering
\begin{tabular}{||c|l|l|l|l|}
\hline
Min jet $p_{T}$ & \multicolumn{1}{|c|}{ $\displaystyle\vphantom{\sum_{i=1}^W}\frac{W^++2}{W^++1}$} &  
\multicolumn{1}{|c|}{$\displaystyle\frac{W^++3}{W^++2}$} &  
\multicolumn{1}{|c|}{$\displaystyle\frac{W^++4}{W^++3}$} & 
\multicolumn{1}{|c|}{$\displaystyle\frac{W^++5}{W^++4}$}  \\
\hline
25 &~$0.3119(0.0006)$~&~$0.2671(0.0005)$~&~$0.2490(0.0005)$~&~$0.2319(0.0008)$~\\
30 &~$0.2930(0.0006)$~&~$0.2388(0.0004)$~&~$0.2227(0.0004)$~&~$0.2078(0.0007)$~\\
35 &~$0.2810(0.0006)$~&~$0.2147(0.0004)$~&~$0.2023(0.0004)$~&~$0.1895(0.0008)$~\\
40 &~$0.2747(0.0006)$~&~$0.1945(0.0004)$~&~$0.1854(0.0004)$~&~$0.173(0.001)$~\\
45 &~$0.2720(0.0007)$~&~$0.1780(0.0004)$~&~$0.1704(0.0004)$~&~$0.160(0.001)$~\\
50 &~$0.2722(0.0007)$~&~$0.1637(0.0005)$~&~$0.1577(0.0005)$~&~$0.146(0.001)$~\\
55 &~$0.2734(0.0008)$~&~$0.1509(0.0004)$~&~$0.1473(0.0004)$~&~$0.136(0.001)$~\\
60 &~$0.2765(0.0008)$~&~$0.1396(0.0004)$~&~$0.1380(0.0005)$~&~$0.126(0.002)$~\\
65 &~$0.2784(0.0009)$~&~$0.1299(0.0004)$~&~$0.1293(0.0004)$~&~$0.1174(0.0009)$~\\
70 &~$0.281(0.001)$~&~$0.1214(0.0004)$~&~$0.1212(0.0004)$~&~$0.112(0.001)$~\\
75 &~$0.282(0.001)$~&~$0.1136(0.0004)$~&~$0.1139(0.0005)$~&~$0.105(0.001)$~\\
80 &~$0.284(0.001)$~&~$0.1064(0.0004)$~&~$0.1076(0.0005)$~&~$0.099(0.001)$~\\
85 &~$0.287(0.001)$~&~$0.0999(0.0004)$~&~$0.1016(0.0004)$~&~$0.094(0.001)$~\\
90 &~$0.291(0.001)$~&~$0.0941(0.0004)$~&~$0.0959(0.0004)$~&~$0.089(0.002)$~\\
95 &~$0.295(0.001)$~&~$0.0889(0.0004)$~&~$0.0911(0.0005)$~&~$0.083(0.001)$~\\
100 &~$0.297(0.001)$~&~$0.0835(0.0004)$~&~$0.0872(0.0005)$~&~$0.079(0.001)$~\\
105 &~$0.301(0.001)$~&~$0.0790(0.0004)$~&~$0.0832(0.0005)$~&~$0.075(0.001)$~\\
110 &~$0.305(0.002)$~&~$0.0744(0.0004)$~&~$0.0795(0.0005)$~&~$0.072(0.001)$~\\
115 &~$0.308(0.002)$~&~$0.0703(0.0004)$~&~$0.0760(0.0005)$~&~$0.067(0.001)$~\\
120 &~$0.314(0.002)$~&~$0.0668(0.0004)$~&~$0.0725(0.0005)$~&~$0.063(0.002)$~\\

\hline
\end{tabular}
\caption{The jet-production ratio in \Wpjn{} jets at LO, as a function of the minimum jet $\pT$ in GeV. These values
are shown in the left plot in \fig{WpMinJetPtJetProductionRatioFigure}.
}
\label{WpJetProductionRatioJetPtLOTable}
\end{table}

\begin{table}[ht]
\vskip .4 cm
\renewcommand{\arraystretch}{0.86}
\centering
\begin{tabular}{||c|l|l|l|l|}
\hline
Min jet $p_{T}$ &  \multicolumn{1}{|c|}{$\displaystyle\vphantom{\sum_{i=1}^W}\frac{W^-+2}{W^-+1}$} &  
\multicolumn{1}{|c|}{$\displaystyle\frac{W^-+3}{W^-+2}$} &  
\multicolumn{1}{|c|}{$\displaystyle\frac{W^-+4}{W^-+3}$} & 
\multicolumn{1}{|c|}{$\displaystyle\frac{W^-+5}{W^-+4}$}  \\
\hline
25 &~$0.238(0.001)$~&~$0.219(0.001)$~&~$0.211(0.003)$~&~$0.200(0.006)$~\\
30 &~$0.2163(0.0009)$~&~$0.195(0.001)$~&~$0.195(0.002)$~&~$0.181(0.003)$~\\
35 &~$0.203(0.001)$~&~$0.176(0.002)$~&~$0.178(0.006)$~&~$0.158(0.007)$~\\
40 &~$0.196(0.001)$~&~$0.157(0.002)$~&~$0.156(0.006)$~&~$0.151(0.006)$~\\
45 &~$0.187(0.001)$~&~$0.146(0.001)$~&~$0.143(0.002)$~&~$0.140(0.003)$~\\
50 &~$0.1805(0.0008)$~&~$0.134(0.001)$~&~$0.132(0.004)$~&~$0.131(0.005)$~\\
55 &~$0.1767(0.0009)$~&~$0.123(0.001)$~&~$0.125(0.003)$~&~$0.124(0.004)$~\\
60 &~$0.172(0.001)$~&~$0.1155(0.0009)$~&~$0.115(0.004)$~&~$0.117(0.004)$~\\
65 &~$0.169(0.002)$~&~$0.107(0.001)$~&~$0.105(0.004)$~&~$0.114(0.005)$~\\
70 &~$0.167(0.001)$~&~$0.0982(0.0009)$~&~$0.104(0.002)$~&~$0.101(0.004)$~\\
75 &~$0.167(0.001)$~&~$0.0909(0.0009)$~&~$0.101(0.004)$~&~$0.087(0.004)$~\\
80 &~$0.165(0.001)$~&~$0.0853(0.0009)$~&~$0.092(0.002)$~&~$0.086(0.004)$~\\
85 &~$0.165(0.001)$~&~$0.0797(0.0009)$~&~$0.089(0.003)$~&~$0.077(0.005)$~\\
90 &~$0.165(0.001)$~&~$0.0756(0.0009)$~&~$0.083(0.003)$~&~$0.076(0.004)$~\\
95 &~$0.165(0.001)$~&~$0.0709(0.0008)$~&~$0.080(0.003)$~&~$0.073(0.004)$~\\
100 &~$0.163(0.001)$~&~$0.068(0.001)$~&~$0.073(0.002)$~&~$0.069(0.004)$~\\
105 &~$0.163(0.001)$~&~$0.0637(0.0008)$~&~$0.069(0.002)$~&~$0.063(0.004)$~\\
110 &~$0.163(0.001)$~&~$0.0595(0.0009)$~&~$0.064(0.002)$~&~$0.057(0.004)$~\\
115 &~$0.157(0.002)$~&~$0.059(0.001)$~&~$0.061(0.002)$~&~$0.057(0.004)$~\\
120 &~$0.159(0.002)$~&~$0.054(0.001)$~&~$0.058(0.003)$~&~$0.055(0.004)$~\\

\hline
\end{tabular}
\caption{The jet-production ratio in \Wmjn{} jets at NLO, as a function of the minimum jet $\pT$ in GeV.  These values
are shown in the right plot in \fig{WmMinJetPtJetProductionRatioFigure}.
}
\label{WmJetProductionRatioJetPtNLOTable}
\end{table}

\begin{table}[ht]
\vskip .4 cm
\renewcommand{\arraystretch}{0.86}
\centering
\begin{tabular}{||c|c|c|c|c|}
\hline
Min jet $p_{T}$ & \multicolumn{1}{|c|}{ $\displaystyle\vphantom{\sum_{i=1}^W}\frac{W^++2}{W^++1}$} &  
\multicolumn{1}{|c|}{$\displaystyle\frac{W^++3}{W^++2}$} &  
\multicolumn{1}{|c|}{$\displaystyle\frac{W^++4}{W^++3}$} & 
\multicolumn{1}{|c|}{$\displaystyle\frac{W^++5}{W^++4}$}  \\
\hline
25 &~$0.242(0.002)$~&~$0.235(0.002)$~&~$0.225(0.003)$~&~$0.218(0.006)$~\\
30 &~$0.226(0.002)$~&~$0.210(0.002)$~&~$0.201(0.003)$~&~$0.187(0.006)$~\\
35 &~$0.215(0.001)$~&~$0.185(0.002)$~&~$0.187(0.003)$~&~$0.177(0.005)$~\\
40 &~$0.204(0.002)$~&~$0.172(0.002)$~&~$0.170(0.002)$~&~$0.155(0.006)$~\\
45 &~$0.198(0.002)$~&~$0.157(0.002)$~&~$0.156(0.003)$~&~$0.151(0.004)$~\\
50 &~$0.195(0.002)$~&~$0.141(0.002)$~&~$0.149(0.003)$~&~$0.131(0.004)$~\\
55 &~$0.192(0.002)$~&~$0.131(0.002)$~&~$0.138(0.003)$~&~$0.129(0.003)$~\\
60 &~$0.188(0.002)$~&~$0.123(0.002)$~&~$0.129(0.002)$~&~$0.114(0.004)$~\\
65 &~$0.184(0.002)$~&~$0.116(0.001)$~&~$0.119(0.002)$~&~$0.110(0.003)$~\\
70 &~$0.179(0.002)$~&~$0.109(0.001)$~&~$0.113(0.002)$~&~$0.102(0.003)$~\\
75 &~$0.181(0.002)$~&~$0.100(0.002)$~&~$0.107(0.003)$~&~$0.103(0.007)$~\\
80 &~$0.178(0.002)$~&~$0.096(0.001)$~&~$0.100(0.002)$~&~$0.085(0.006)$~\\
85 &~$0.177(0.002)$~&~$0.091(0.001)$~&~$0.092(0.002)$~&~$0.081(0.004)$~\\
90 &~$0.175(0.002)$~&~$0.087(0.001)$~&~$0.086(0.003)$~&~$0.081(0.004)$~\\
95 &~$0.174(0.002)$~&~$0.082(0.001)$~&~$0.078(0.006)$~&~$0.083(0.007)$~\\
100 &~$0.175(0.003)$~&~$0.076(0.002)$~&~$0.080(0.002)$~&~$0.071(0.004)$~\\
105 &~$0.177(0.003)$~&~$0.071(0.001)$~&~$0.077(0.002)$~&~$0.068(0.005)$~\\
110 &~$0.173(0.003)$~&~$0.067(0.001)$~&~$0.074(0.002)$~&~$0.070(0.004)$~\\
115 &~$0.177(0.003)$~&~$0.064(0.001)$~&~$0.070(0.002)$~&~$0.063(0.004)$~\\
120 &~$0.177(0.003)$~&~$0.061(0.001)$~&~$0.067(0.003)$~&~$0.063(0.004)$~\\

\hline
\end{tabular}
\caption{The jet-production ratio in \Wpjn{} jets at NLO, as a function of the minimum jet $\pT$ in GeV.  These values
are shown in the right plot in \fig{WpMinJetPtJetProductionRatioFigure}.
}
\label{WpJetProductionRatioJetPtNLOTable}
\end{table}

\FloatBarrier In this appendix, we present the detailed results at LO
and at NLO, for the ratio of \Wjn-jet to \Wjnm-jet production when
varying the cut on the minimum jet transverse energy across a range of
values from 25 to 120~GeV.  Our results for the ratio \Wmjn-jet to
\Wmjnm-jet production are given in
\tabs{WmJetProductionRatioJetPtLOTable}{WmJetProductionRatioJetPtNLOTable},
and the results for the ratio of \Wpjn-jet to \Wpjnm-jet production in
\tabs{WpJetProductionRatioJetPtLOTable}{WpJetProductionRatioJetPtNLOTable}.
We show corresponding numerical integration errors in parentheses.

\vfill\eject

\section{Differential Cross Sections as a Function of the $W$ 
Transverse Momentum}
\label{WpTAppendix}

\begin{table}[th]
\vskip .4 cm \renewcommand{\arraystretch}{0.76} \centering
\begin{tabular}{||c|l|l|l|l|l|}
\hline
$W^- \pT$ & \hfil\Wmj & \hfil\Wmjj & \hfil\Wmjjj & \hfil\Wmjjjj & \hfil\Wmjjjjj \\
\hline
30 &~$7919(7)$~&~$1110(3)$~&~$247(1)$~&~$48.1(0.4)$~&~$9.29(0.08)$~\\
50 &~$3827(4)$~&~$919(2)$~&~$227.5(0.9)$~&~$49.5(0.4)$~&~$9.82(0.07)$~\\
70 &~$1319(2)$~&~$612(2)$~&~$155.4(0.7)$~&~$36.8(0.3)$~&~$7.81(0.07)$~\\
90 &~$552(1)$~&~$338(1)$~&~$104.6(0.5)$~&~$27.1(0.3)$~&~$6.01(0.05)$~\\
110 &~$263.0(0.8)$~&~$189.4(0.7)$~&~$69.5(0.4)$~&~$18.8(0.2)$~&~$4.56(0.06)$~\\
130 &~$136.4(0.5)$~&~$108.7(0.4)$~&~$45.0(0.3)$~&~$13.4(0.1)$~&~$3.29(0.03)$~\\
150 &~$73.6(0.3)$~&~$64.7(0.3)$~&~$29.1(0.2)$~&~$9.3(0.1)$~&~$2.43(0.03)$~\\
170 &~$42.3(0.3)$~&~$39.4(0.2)$~&~$18.9(0.2)$~&~$6.34(0.08)$~&~$1.71(0.02)$~\\
190 &~$25.0(0.2)$~&~$24.5(0.2)$~&~$12.7(0.1)$~&~$4.51(0.06)$~&~$1.30(0.02)$~\\
220 &~$12.51(0.08)$~&~$12.94(0.07)$~&~$6.88(0.05)$~&~$2.62(0.04)$~&~$0.781(0.009)$~\\
260 &~$5.15(0.04)$~&~$5.78(0.04)$~&~$3.27(0.03)$~&~$1.29(0.02)$~&~$0.417(0.006)$~\\
300 &~$2.31(0.03)$~&~$2.71(0.02)$~&~$1.58(0.02)$~&~$0.66(0.01)$~&~$0.218(0.005)$~\\
340 &~$1.13(0.02)$~&~$1.39(0.02)$~&~$0.82(0.01)$~&~$0.365(0.008)$~&~$0.116(0.002)$~\\
380 &~$0.59(0.01)$~&~$0.73(0.01)$~&~$0.45(0.01)$~&~$0.202(0.005)$~&~$0.067(0.002)$~\\
420 &~$0.312(0.007)$~&~$0.398(0.006)$~&~$0.246(0.006)$~&~$0.114(0.004)$~&~$0.040(0.001)$~\\
460 &~$0.171(0.004)$~&~$0.219(0.004)$~&~$0.144(0.004)$~&~$0.067(0.003)$~&~$0.023(0.001)$~\\
500 &~$0.101(0.003)$~&~$0.131(0.003)$~&~$0.080(0.002)$~&~$0.037(0.002)$~&~$1.42(0.05) \cdot 10^{-2}$~\\

\hline
\end{tabular}
\caption{The LO \Wmjn{}-jet cross section taken differentially in the
  $W^-$ $\pT$, in fb/GeV.  These values are shown in the upper panel of the left figure in
\fig{WmPtJetProductionRatioFigure}.}
\label{Wm-pTLODistributionTable}
\end{table}

\begin{table}
\vskip .4 cm
\renewcommand{\arraystretch}{0.76}
\centering
\begin{tabular}{||c|l|l|l|l|l|}
\hline
$W^- \pT$ & \hfil\Wmj & \hfil\Wmjj & \hfil\Wmjjj & \hfil\Wmjjjj & \hfil\Wmjjjjj \\
\hline
30 &~$7517(40)$~&~$1132(10)$~&~$211(3)$~&~$38.0(0.9)$~&~$5.9(0.6)$~\\
50 &~$4825(24)$~&~$915(7)$~&~$200(2)$~&~$40(1)$~&~$7.3(0.4)$~\\
70 &~$1793(9)$~&~$612(5)$~&~$137(1)$~&~$29.6(0.8)$~&~$5.6(0.3)$~\\
90 &~$767(5)$~&~$342(4)$~&~$91(2)$~&~$20.7(0.7)$~&~$4.5(0.3)$~\\
110 &~$368(3)$~&~$195(2)$~&~$60.6(0.8)$~&~$15.2(0.5)$~&~$3.7(0.3)$~\\
130 &~$194(2)$~&~$112(1)$~&~$39.9(0.6)$~&~$8(2)$~&~$2.4(0.1)$~\\
150 &~$107(1)$~&~$67.5(0.9)$~&~$25.2(0.5)$~&~$7.1(0.3)$~&~$1.83(0.09)$~\\
170 &~$61(1)$~&~$40.1(0.6)$~&~$16.8(0.4)$~&~$4.9(0.3)$~&~$1.35(0.08)$~\\
190 &~$38(1)$~&~$24.8(0.6)$~&~$10.9(0.2)$~&~$3.6(0.1)$~&~$1.00(0.08)$~\\
220 &~$18.5(0.4)$~&~$12.9(0.2)$~&~$6.1(0.1)$~&~$2.0(0.1)$~&~$0.46(0.06)$~\\
260 &~$7.7(0.1)$~&~$5.9(0.1)$~&~$2.84(0.05)$~&~$1.09(0.05)$~&~$0.32(0.02)$~\\
300 &~$3.57(0.08)$~&~$2.67(0.05)$~&~$1.38(0.05)$~&~$0.54(0.03)$~&~$0.13(0.04)$~\\
340 &~$1.73(0.04)$~&~$1.39(0.03)$~&~$0.74(0.02)$~&~$0.27(0.02)$~&~$0.04(0.04)$~\\
380 &~$0.90(0.03)$~&~$0.74(0.02)$~&~$0.38(0.02)$~&~$0.16(0.02)$~&~$0.049(0.007)$~\\
420 &~$0.49(0.02)$~&~$0.40(0.01)$~&~$0.21(0.01)$~&~$0.089(0.006)$~&~$0.032(0.004)$~\\
460 &~$0.25(0.02)$~&~$0.220(0.009)$~&~$0.129(0.005)$~&~$0.061(0.004)$~&~$0.018(0.003)$~\\
500 &~$0.05(0.09)$~&~$0.137(0.005)$~&~$0.073(0.004)$~&~$0.034(0.004)$~&~$0.013(0.003)$~\\

\hline
\end{tabular}
\caption{The NLO \Wmjn{}-jet cross section taken differentially in the
  $W^-$ $\pT$, in fb/GeV.  These values are shown in the upper panel of the right figure in
\fig{WmPtJetProductionRatioFigure}.}
\label{Wm-pTNLODistributionTable}
\end{table}

\begin{table}[ht]
\vskip .4 cm
\renewcommand{\arraystretch}{0.76}
\centering
\begin{tabular}{||c|l|l|l|l|l|}
\hline
$W^+ \pT$ & \hfil\Wpj & \hfil\Wpjj & \hfil\Wpjjj & \hfil\Wpjjjj & \hfil\Wpjjjjj \\
\hline
30 &~$11480(27)$~&~$1691(3)$~&~$397(1)$~&~$82.8(0.2)$~&~$16.6(0.1)$~\\
50 &~$5584(16)$~&~$1406(3)$~&~$373(1)$~&~$83.7(0.2)$~&~$18.1(0.2)$~\\
70 &~$1965(8)$~&~$938(2)$~&~$254.2(0.8)$~&~$64.3(0.4)$~&~$14.3(0.1)$~\\
90 &~$851(5)$~&~$531(1)$~&~$174.0(0.8)$~&~$46.6(0.1)$~&~$11.2(0.1)$~\\
110 &~$415(3)$~&~$304.3(0.7)$~&~$116.3(0.4)$~&~$33.66(0.09)$~&~$8.51(0.07)$~\\
130 &~$220(2)$~&~$179.9(0.5)$~&~$76.1(0.3)$~&~$23.91(0.08)$~&~$6.40(0.08)$~\\
150 &~$124(1)$~&~$108.2(0.3)$~&~$50.4(0.3)$~&~$16.87(0.07)$~&~$4.55(0.04)$~\\
170 &~$70.1(0.9)$~&~$67.8(0.3)$~&~$33.5(0.2)$~&~$11.89(0.05)$~&~$3.41(0.04)$~\\
190 &~$43.9(0.7)$~&~$43.2(0.2)$~&~$22.3(0.1)$~&~$8.31(0.04)$~&~$2.50(0.04)$~\\
220 &~$23.0(0.3)$~&~$23.69(0.08)$~&~$13.1(0.1)$~&~$5.07(0.02)$~&~$1.63(0.02)$~\\
260 &~$9.6(0.2)$~&~$10.91(0.04)$~&~$6.43(0.05)$~&~$2.65(0.02)$~&~$0.89(0.03)$~\\
300 &~$4.4(0.1)$~&~$5.38(0.03)$~&~$3.32(0.03)$~&~$1.44(0.01)$~&~$0.486(0.007)$~\\
340 &~$2.30(0.08)$~&~$2.84(0.02)$~&~$1.78(0.01)$~&~$0.794(0.007)$~&~$0.280(0.005)$~\\
380 &~$1.40(0.06)$~&~$1.53(0.01)$~&~$0.99(0.01)$~&~$0.446(0.004)$~&~$0.166(0.005)$~\\
420 &~$0.70(0.03)$~&~$0.866(0.008)$~&~$0.566(0.007)$~&~$0.263(0.003)$~&~$0.092(0.002)$~\\
460 &~$0.43(0.02)$~&~$0.522(0.007)$~&~$0.339(0.005)$~&~$0.158(0.002)$~&~$0.057(0.001)$~\\
500 &~$0.23(0.01)$~&~$0.303(0.004)$~&~$0.21(0.01)$~&~$0.096(0.002)$~&~$0.038(0.001)$~\\

\hline
\end{tabular}
\caption{The LO \Wpjn{}-jet cross section taken differentially in the
  $W^+$ $\pT$, in fb/GeV.  These values are shown in the upper panel of the left figure in
\fig{WpPtJetProductionRatioFigure}.}
\label{Wp-pTLODistributionTable}
\end{table}

\begin{table}[ht]
\vskip .4 cm
\renewcommand{\arraystretch}{0.76}
\centering
\begin{tabular}{||c|l|l|l|l|l|}
\hline
$W^+ \pT$ & \hfil\Wpj & \hfil\Wpjj & \hfil\Wpjjj & \hfil\Wpjjjj & \hfil\Wpjjjjj \\
\hline
30 &~$10990(151)$~&~$1649(26)$~&~$333(5)$~&~$65(2)$~&~$12.9(0.7)$~\\
50 &~$6998(40)$~&~$1375(17)$~&~$326(6)$~&~$64(1)$~&~$15(1)$~\\
70 &~$2644(20)$~&~$904(11)$~&~$212(3)$~&~$49(1)$~&~$9.3(0.6)$~\\
90 &~$1125(12)$~&~$516(7)$~&~$149(2)$~&~$35.3(0.8)$~&~$7.6(0.3)$~\\
110 &~$557(6)$~&~$299(4)$~&~$99(1)$~&~$25.6(0.9)$~&~$6.0(0.2)$~\\
130 &~$301(4)$~&~$183(3)$~&~$65(1)$~&~$18.3(0.4)$~&~$4.4(0.6)$~\\
150 &~$171(3)$~&~$107(2)$~&~$44(1)$~&~$12(1)$~&~$2.9(0.2)$~\\
170 &~$103(2)$~&~$68(1)$~&~$28.5(0.5)$~&~$9.5(0.3)$~&~$2.3(0.1)$~\\
190 &~$60(1)$~&~$42.4(0.9)$~&~$18.9(0.4)$~&~$6.8(0.9)$~&~$1.91(0.08)$~\\
220 &~$33.2(0.7)$~&~$23(1)$~&~$10.6(0.3)$~&~$4.1(0.2)$~&~$1.01(0.07)$~\\
260 &~$14.3(0.4)$~&~$10.5(0.3)$~&~$5.2(0.1)$~&~$1.80(0.08)$~&~$0.5(0.2)$~\\
300 &~$6.4(0.2)$~&~$5.2(0.1)$~&~$2.65(0.08)$~&~$1.09(0.06)$~&~$0.32(0.03)$~\\
340 &~$3.4(0.1)$~&~$2.86(0.09)$~&~$1.50(0.05)$~&~$0.59(0.03)$~&~$0.20(0.02)$~\\
380 &~$1.9(0.1)$~&~$1.41(0.04)$~&~$0.81(0.05)$~&~$0.33(0.02)$~&~$0.13(0.02)$~\\
420 &~$0.99(0.06)$~&~$0.81(0.02)$~&~$0.47(0.02)$~&~$0.22(0.01)$~&~$0.062(0.009)$~\\
460 &~$0.61(0.04)$~&~$0.46(0.02)$~&~$0.28(0.02)$~&~$0.12(0.01)$~&~$0.035(0.007)$~\\
500 &~$0.30(0.04)$~&~$0.29(0.01)$~&~$0.19(0.01)$~&~$0.080(0.007)$~&~$0.021(0.005)$~\\

\hline
\end{tabular}
\caption{The NLO \Wpjn{}-jet cross section taken differentially in the
  $W^+$ $\pT$, in fb/GeV.  These values are shown in the upper panel of the right figure in
\fig{WpPtJetProductionRatioFigure}.}
\label{Wp-pTNLODistributionTable}
\end{table}

\FloatBarrier In this appendix, we present the detailed results at LO
and at NLO for the cross section for \Wjn-jet production taken
differentially in the $W$'s transverse momentum.  We give the LO
differential cross sections for \Wmj-jet through \Wmjjjjj-jet
production in \tab{Wm-pTLODistributionTable}, and the corresponding
results for the NLO differential cross section in
\tab{Wm-pTNLODistributionTable}.  We give the LO and NLO differential
cross sections for \Wpj-jet through \Wpjjjjj-jet production in
\tabs{Wp-pTLODistributionTable}{Wp-pTNLODistributionTable}
respectively.  We show corresponding numerical integration errors in
parentheses.

\vfill\eject

\section{Differential Cross Sections as a Function of the Total Jet Transverse Energy}
\label{WHTAppendix}

\begin{table}[ht]
\renewcommand{\arraystretch}{0.5}
\centering
\scalebox{0.96}{
\begin{tabular}{||c|l|l|l|l|l||}
\hline
$\HT$ & \hfil\Wmj & \hfil\Wmjj & \hfil\Wmjjj & \hfil\Wmjjjj & \hfil\Wmjjjjj \\
\hline
30 &~$7919(7)$~&~\hfil---~&~\hfil---~&~\hfil---~&~\hfil---~\\
50 &~$3827(4)$~&~$294(2)$~&~\hfil---~&~\hfil---~&~\hfil---~\\
70 &~$1319(2)$~&~$1093(3)$~&~$0.89(0.09)$~&~\hfil---~&~\hfil---~\\
90 &~$552(1)$~&~$893(2)$~&~$62.6(0.7)$~&~\hfil---~&~\hfil---~\\
110 &~$263.0(0.8)$~&~$604(2)$~&~$135.1(0.8)$~&~$1.2(0.2)$~&~\hfil---~\\
130 &~$136.4(0.5)$~&~$395(1)$~&~$149.9(0.8)$~&~$8.6(0.2)$~&~$0.02(0.01)$~\\
150 &~$73.6(0.3)$~&~$263.9(0.9)$~&~$134.7(0.6)$~&~$16.9(0.3)$~&~$0.31(0.04)$~\\
170 &~$42.3(0.3)$~&~$177.5(0.7)$~&~$112.6(0.5)$~&~$21.8(0.3)$~&~$1.01(0.03)$~\\
190 &~$25.0(0.2)$~&~$122.1(0.5)$~&~$91.3(0.5)$~&~$23.1(0.3)$~&~$2.04(0.07)$~\\
210 &~$15.4(0.1)$~&~$85.4(0.4)$~&~$72.2(0.4)$~&~$22.3(0.3)$~&~$2.80(0.04)$~\\
230 &~$9.61(0.09)$~&~$62.1(0.4)$~&~$57.3(0.4)$~&~$20.7(0.2)$~&~$3.42(0.05)$~\\
250 &~$6.23(0.07)$~&~$45.6(0.3)$~&~$45.1(0.3)$~&~$18.6(0.2)$~&~$3.71(0.05)$~\\
270 &~$4.07(0.05)$~&~$33.7(0.2)$~&~$35.3(0.2)$~&~$16.4(0.2)$~&~$3.70(0.04)$~\\
290 &~$2.71(0.04)$~&~$24.8(0.2)$~&~$28.2(0.2)$~&~$14.2(0.2)$~&~$3.69(0.05)$~\\
310 &~$1.91(0.04)$~&~$18.9(0.2)$~&~$22.6(0.2)$~&~$12.0(0.2)$~&~$3.46(0.04)$~\\
330 &~$1.33(0.03)$~&~$14.6(0.1)$~&~$18.2(0.2)$~&~$10.1(0.1)$~&~$3.22(0.03)$~\\
350 &~$0.93(0.02)$~&~$11.2(0.1)$~&~$14.4(0.1)$~&~$8.8(0.1)$~&~$2.97(0.03)$~\\
370 &~$0.69(0.02)$~&~$9.1(0.1)$~&~$11.8(0.1)$~&~$7.31(0.09)$~&~$2.71(0.03)$~\\
390 &~$0.49(0.01)$~&~$6.95(0.08)$~&~$9.61(0.09)$~&~$6.5(0.1)$~&~$2.42(0.03)$~\\
410 &~$0.37(0.01)$~&~$5.71(0.09)$~&~$7.99(0.09)$~&~$5.29(0.07)$~&~$2.11(0.02)$~\\
430 &~$0.256(0.008)$~&~$4.65(0.07)$~&~$6.49(0.07)$~&~$4.61(0.07)$~&~$1.94(0.03)$~\\
450 &~$0.195(0.006)$~&~$3.74(0.06)$~&~$5.32(0.06)$~&~$3.80(0.05)$~&~$1.72(0.03)$~\\
470 &~$0.147(0.005)$~&~$2.98(0.05)$~&~$4.44(0.05)$~&~$3.21(0.06)$~&~$1.46(0.02)$~\\
490 &~$0.118(0.004)$~&~$2.47(0.04)$~&~$3.76(0.06)$~&~$2.81(0.08)$~&~$1.32(0.02)$~\\
510 &~$0.083(0.003)$~&~$2.07(0.04)$~&~$3.25(0.06)$~&~$2.38(0.05)$~&~$1.13(0.01)$~\\
530 &~$0.064(0.003)$~&~$1.71(0.03)$~&~$2.66(0.05)$~&~$2.03(0.04)$~&~$1.00(0.01)$~\\
550 &~$0.049(0.002)$~&~$1.45(0.04)$~&~$2.28(0.05)$~&~$1.74(0.03)$~&~$0.89(0.01)$~\\
570 &~$0.037(0.002)$~&~$1.21(0.03)$~&~$1.92(0.07)$~&~$1.52(0.03)$~&~$0.77(0.01)$~\\
590 &~$0.030(0.002)$~&~$0.98(0.03)$~&~$1.59(0.03)$~&~$1.35(0.04)$~&~$0.70(0.01)$~\\
620 &~$2.10(0.08) \cdot 10^{-2}$~&~$0.80(0.02)$~&~$1.28(0.02)$~&~$1.08(0.02)$~&~$0.569(0.007)$~\\
660 &~$1.33(0.06) \cdot 10^{-2}$~&~$0.56(0.01)$~&~$0.92(0.01)$~&~$0.80(0.02)$~&~$0.46(0.01)$~\\
700 &~$8.8(0.5) \cdot 10^{-3}$~&~$0.41(0.01)$~&~$0.68(0.01)$~&~$0.61(0.01)$~&~$0.336(0.004)$~\\
740 &~$4.8(0.3) \cdot 10^{-3}$~&~$0.297(0.009)$~&~$0.51(0.01)$~&~$0.440(0.009)$~&~$0.258(0.003)$~\\
780 &~$3.4(0.2) \cdot 10^{-3}$~&~$0.210(0.007)$~&~$0.388(0.008)$~&~$0.348(0.008)$~&~$0.206(0.003)$~\\
820 &~$2.3(0.1) \cdot 10^{-3}$~&~$0.166(0.007)$~&~$0.299(0.007)$~&~$0.273(0.007)$~&~$0.166(0.003)$~\\
860 &~$1.5(0.1) \cdot 10^{-3}$~&~$0.131(0.006)$~&~$0.229(0.005)$~&~$0.212(0.007)$~&~$0.126(0.002)$~\\
900 &~$1.02(0.08) \cdot 10^{-3}$~&~$0.095(0.005)$~&~$0.184(0.007)$~&~$0.165(0.006)$~&~$0.100(0.002)$~\\
940 &~$6.7(0.6) \cdot 10^{-4}$~&~$0.087(0.005)$~&~$0.140(0.004)$~&~$0.124(0.004)$~&~$0.082(0.002)$~\\
980 &~$3.3(0.3) \cdot 10^{-4}$~&~$0.058(0.003)$~&~$0.107(0.004)$~&~$0.099(0.003)$~&~$0.062(0.001)$~\\

\hline
\end{tabular}}
\caption{The LO \Wmjn{}-jet cross section taken differentially in the
  total jet transverse energy $\HT$, in fb/GeV.  These values are shown in the upper panel of the left plot in
\fig{WmHTJetProductionRatioFigure}.}
\label{Wm-HTLODistributionTable}
\end{table}

\begin{table}[ht]
\renewcommand{\arraystretch}{0.7}
\centering
\scalebox{0.97}{
\begin{tabular}{||c|l|l|l|l|l|}
\hline
$\HT$ & \hfil\Wmj & \hfil\Wmjj & \hfil\Wmjjj & \hfil\Wmjjjj & \hfil\Wmjjjjj \\
\hline
30 &~$8514(51)$~&~\hfil---~&~\hfil---~&~\hfil---~&~\hfil---~\\
50 &~$3754(39)$~&~$307(10)$~&~\hfil---~&~\hfil---~&~\hfil---~\\
70 &~$2061(10)$~&~$1009(11)$~&~$1.0(0.4)$~&~\hfil---~&~\hfil---~\\
90 &~$1203(6)$~&~$811(7)$~&~$61(2)$~&~\hfil---~&~\hfil---~\\
110 &~$701(5)$~&~$589(6)$~&~$119(2)$~&~$1.1(0.3)$~&~\hfil---~\\
130 &~$435(3)$~&~$390(5)$~&~$130(2)$~&~$7.0(0.7)$~&~$0.02(0.01)$~\\
150 &~$271(2)$~&~$288(3)$~&~$112(2)$~&~$15.0(0.8)$~&~$0.30(0.06)$~\\
170 &~$179(2)$~&~$198(3)$~&~$94(2)$~&~$17.7(0.9)$~&~$0.7(0.2)$~\\
190 &~$117(2)$~&~$140(2)$~&~$79(1)$~&~$19(1)$~&~$1.6(0.2)$~\\
210 &~$86(1)$~&~$103(1)$~&~$60(2)$~&~$18.3(0.7)$~&~$2.1(0.2)$~\\
230 &~$58.0(0.6)$~&~$78(1)$~&~$51(1)$~&~$16.5(0.7)$~&~$3.2(0.2)$~\\
250 &~$42.8(0.6)$~&~$56(1)$~&~$39(1)$~&~$14.9(0.6)$~&~$4(1)$~\\
270 &~$30.5(0.4)$~&~$42(1)$~&~$31(1)$~&~$13.6(0.7)$~&~$1(1)$~\\
290 &~$23.0(0.4)$~&~$34.0(0.8)$~&~$23.6(0.4)$~&~$10.7(0.5)$~&~$2.0(0.6)$~\\
310 &~$17.1(0.3)$~&~$24.9(0.8)$~&~$20.8(0.4)$~&~$7(2)$~&~$2.6(0.5)$~\\
330 &~$13.8(0.3)$~&~$19.7(0.6)$~&~$16.1(0.4)$~&~$7.5(0.5)$~&~$2.1(0.4)$~\\
350 &~$10.3(0.2)$~&~$15.8(0.6)$~&~$13.0(0.3)$~&~$7.1(0.5)$~&~$2.2(0.1)$~\\
370 &~$7.8(0.2)$~&~$12.4(0.4)$~&~$10.1(0.3)$~&~$5.3(0.5)$~&~$2.1(0.4)$~\\
390 &~$6.4(0.2)$~&~$9.6(0.3)$~&~$8.6(0.2)$~&~$4.9(0.3)$~&~$1.8(0.2)$~\\
410 &~$5.2(0.2)$~&~$8.3(0.3)$~&~$7.0(0.2)$~&~$4.1(0.3)$~&~$1.3(0.1)$~\\
430 &~$4.1(0.1)$~&~$6.4(0.3)$~&~$6.0(0.2)$~&~$3.1(0.4)$~&~$1.3(0.1)$~\\
450 &~$3.18(0.09)$~&~$5.3(0.2)$~&~$4.8(0.1)$~&~$2.5(0.2)$~&~$1.2(0.1)$~\\
470 &~$2.62(0.09)$~&~$4.5(0.3)$~&~$4.1(0.1)$~&~$2.5(0.2)$~&~$1.03(0.06)$~\\
490 &~$2.2(0.1)$~&~$3.2(0.3)$~&~$3.5(0.1)$~&~$2.2(0.2)$~&~$1.05(0.06)$~\\
510 &~$1.72(0.07)$~&~$3.1(0.2)$~&~$3.00(0.09)$~&~$1.9(0.2)$~&~$0.73(0.09)$~\\
530 &~$1.58(0.09)$~&~$2.5(0.2)$~&~$2.6(0.1)$~&~$1.5(0.2)$~&~$0.71(0.05)$~\\
550 &~$1.19(0.05)$~&~$2.2(0.2)$~&~$1.9(0.1)$~&~$1.3(0.2)$~&~$0.54(0.07)$~\\
570 &~$1.03(0.04)$~&~$1.8(0.2)$~&~$1.73(0.08)$~&~$1.2(0.2)$~&~$0.52(0.06)$~\\
590 &~$0.90(0.05)$~&~$1.5(0.1)$~&~$1.63(0.08)$~&~$0.99(0.08)$~&~$0.49(0.06)$~\\
620 &~$0.66(0.03)$~&~$1.09(0.05)$~&~$1.11(0.04)$~&~$0.75(0.04)$~&~$0.35(0.03)$~\\
660 &~$0.49(0.02)$~&~$0.88(0.05)$~&~$0.85(0.03)$~&~$0.61(0.03)$~&~$0.34(0.02)$~\\
700 &~$0.35(0.02)$~&~$0.57(0.04)$~&~$0.62(0.02)$~&~$0.39(0.04)$~&~$0.22(0.02)$~\\
740 &~$0.24(0.01)$~&~$0.47(0.02)$~&~$0.47(0.02)$~&~$0.34(0.03)$~&~$0.19(0.01)$~\\
780 &~$0.19(0.01)$~&~$0.32(0.02)$~&~$0.34(0.01)$~&~$0.33(0.06)$~&~$0.12(0.01)$~\\
820 &~$0.146(0.009)$~&~$0.24(0.01)$~&~$0.28(0.01)$~&~$0.22(0.01)$~&~$0.11(0.01)$~\\
860 &~$0.124(0.009)$~&~$0.18(0.01)$~&~$0.21(0.01)$~&~$0.14(0.02)$~&~$0.04(0.03)$~\\
900 &~$0.085(0.006)$~&~$0.16(0.01)$~&~$0.156(0.009)$~&~$0.11(0.01)$~&~$0.05(0.03)$~\\
940 &~$0.058(0.004)$~&~$0.14(0.01)$~&~$0.138(0.008)$~&~$0.09(0.01)$~&~$0.058(0.006)$~\\
980 &~$0.054(0.006)$~&~$0.085(0.007)$~&~$0.108(0.008)$~&~$0.077(0.006)$~&~$0.033(0.008)$~\\

\hline
\end{tabular}}
\caption{The NLO \Wmjn{}-jet cross section taken differentially in the
  total jet transverse energy $\HT$, in fb/GeV.  These values are shown in the upper panel of the right plot in
\fig{WmHTJetProductionRatioFigure}.}
\label{Wm-HTNLODistributionTable}
\end{table}

\begin{table}
\renewcommand{\arraystretch}{0.7}
\centering
\scalebox{0.97}{
\begin{tabular}{||c|l|l|l|l|l|}
\hline
$\HT$ & \hfil\Wpj & \hfil\Wpjj & \hfil\Wpjjj & \hfil\Wpjjjj & \hfil\Wpjjjjj \\
\hline
30 &~$11480(27)$~&~\hfil---~&~\hfil---~&~\hfil---~&~\hfil---~\\
50 &~$5584(16)$~&~$443(4)$~&~\hfil---~&~\hfil---~&~\hfil---~\\
70 &~$1965(8)$~&~$1652(4)$~&~$1.4(0.1)$~&~\hfil---~&~\hfil---~\\
90 &~$851(5)$~&~$1354(3)$~&~$97(1)$~&~\hfil---~&~\hfil---~\\
110 &~$415(3)$~&~$930(2)$~&~$209(1)$~&~$1.91(0.05)$~&~\hfil---~\\
130 &~$220(2)$~&~$622(1)$~&~$236(1)$~&~$13.2(0.1)$~&~$0.007(0.002)$~\\
150 &~$124(1)$~&~$417(1)$~&~$217.6(0.9)$~&~$27.0(0.2)$~&~$0.36(0.03)$~\\
170 &~$70.1(0.9)$~&~$285.6(0.8)$~&~$183.1(0.6)$~&~$35.2(0.1)$~&~$1.65(0.06)$~\\
190 &~$43.9(0.7)$~&~$199.6(0.6)$~&~$149.7(0.6)$~&~$38.9(0.4)$~&~$3.14(0.07)$~\\
210 &~$27.8(0.5)$~&~$142.8(0.5)$~&~$121.0(0.6)$~&~$37.9(0.1)$~&~$4.74(0.07)$~\\
230 &~$18.2(0.4)$~&~$103.3(0.4)$~&~$96.7(0.5)$~&~$35.5(0.1)$~&~$5.67(0.07)$~\\
250 &~$11.5(0.3)$~&~$77.1(0.3)$~&~$77.2(0.3)$~&~$32.3(0.1)$~&~$6.43(0.08)$~\\
270 &~$7.6(0.2)$~&~$57.4(0.2)$~&~$61.0(0.3)$~&~$28.33(0.09)$~&~$6.69(0.07)$~\\
290 &~$5.0(0.2)$~&~$43.4(0.2)$~&~$49.5(0.3)$~&~$24.85(0.09)$~&~$6.7(0.1)$~\\
310 &~$3.7(0.1)$~&~$33.6(0.2)$~&~$39.7(0.2)$~&~$21.51(0.08)$~&~$6.31(0.07)$~\\
330 &~$2.6(0.1)$~&~$25.9(0.2)$~&~$32.8(0.2)$~&~$18.56(0.07)$~&~$5.88(0.09)$~\\
350 &~$2.0(0.1)$~&~$20.6(0.1)$~&~$26.2(0.1)$~&~$16.00(0.07)$~&~$5.47(0.06)$~\\
370 &~$1.65(0.09)$~&~$16.0(0.1)$~&~$21.6(0.1)$~&~$13.69(0.06)$~&~$5.07(0.07)$~\\
390 &~$1.15(0.08)$~&~$12.9(0.1)$~&~$17.9(0.1)$~&~$11.74(0.06)$~&~$4.57(0.06)$~\\
410 &~$0.87(0.05)$~&~$10.53(0.09)$~&~$14.6(0.1)$~&~$10.01(0.05)$~&~$4.14(0.06)$~\\
430 &~$0.53(0.04)$~&~$8.62(0.08)$~&~$12.30(0.09)$~&~$8.56(0.04)$~&~$3.57(0.04)$~\\
450 &~$0.51(0.04)$~&~$6.91(0.06)$~&~$10.26(0.07)$~&~$7.40(0.05)$~&~$3.30(0.04)$~\\
470 &~$0.34(0.03)$~&~$5.80(0.06)$~&~$8.7(0.1)$~&~$6.32(0.03)$~&~$3.03(0.08)$~\\
490 &~$0.28(0.02)$~&~$4.74(0.06)$~&~$7.28(0.07)$~&~$5.48(0.03)$~&~$2.63(0.04)$~\\
510 &~$0.18(0.02)$~&~$3.91(0.04)$~&~$6.21(0.07)$~&~$4.75(0.02)$~&~$2.26(0.04)$~\\
530 &~$0.16(0.02)$~&~$3.21(0.04)$~&~$5.23(0.06)$~&~$4.10(0.02)$~&~$2.05(0.04)$~\\
550 &~$0.13(0.01)$~&~$2.74(0.04)$~&~$4.34(0.05)$~&~$3.56(0.02)$~&~$1.81(0.03)$~\\
570 &~$0.08(0.01)$~&~$2.36(0.04)$~&~$3.80(0.04)$~&~$3.08(0.02)$~&~$1.60(0.02)$~\\
590 &~$0.081(0.009)$~&~$1.96(0.03)$~&~$3.26(0.04)$~&~$2.65(0.02)$~&~$1.44(0.03)$~\\
620 &~$0.057(0.005)$~&~$1.54(0.02)$~&~$2.57(0.02)$~&~$2.18(0.01)$~&~$1.19(0.01)$~\\
660 &~$0.040(0.004)$~&~$1.14(0.01)$~&~$1.91(0.02)$~&~$1.663(0.008)$~&~$0.93(0.01)$~\\
700 &~$0.023(0.003)$~&~$0.86(0.01)$~&~$1.48(0.02)$~&~$1.269(0.009)$~&~$0.724(0.008)$~\\
740 &~$0.011(0.001)$~&~$0.64(0.01)$~&~$1.12(0.01)$~&~$0.991(0.006)$~&~$0.59(0.01)$~\\
780 &~$0.008(0.001)$~&~$0.50(0.01)$~&~$0.86(0.01)$~&~$0.776(0.005)$~&~$0.476(0.009)$~\\
820 &~$0.008(0.001)$~&~$0.394(0.009)$~&~$0.65(0.01)$~&~$0.600(0.004)$~&~$0.376(0.006)$~\\
860 &~$4.2(0.7) \cdot 10^{-3}$~&~$0.292(0.007)$~&~$0.499(0.007)$~&~$0.472(0.004)$~&~$0.300(0.007)$~\\
900 &~$3.7(0.7) \cdot 10^{-3}$~&~$0.227(0.006)$~&~$0.406(0.006)$~&~$0.373(0.003)$~&~$0.25(0.01)$~\\
940 &~$1.8(0.3) \cdot 10^{-3}$~&~$0.178(0.005)$~&~$0.326(0.007)$~&~$0.294(0.003)$~&~$0.183(0.003)$~\\
980 &~$1.3(0.2) \cdot 10^{-3}$~&~$0.142(0.004)$~&~$0.243(0.004)$~&~$0.236(0.002)$~&~$0.155(0.004)$~\\

\hline
\end{tabular}}
\caption{The LO \Wpjn{}-jet cross section taken differentially in the
  total jet transverse energy $\HT$, in fb/GeV.  These values are shown in the upper panel of the left plot in
\fig{WpHTJetProductionRatioFigure}.}
\label{Wp-HTLODistributionTable}
\end{table}

\begin{table}
\renewcommand{\arraystretch}{0.7}
\centering
\scalebox{0.97}{
\begin{tabular}{||c|l|l|l|l|l|}
\hline
$\HT$ & \hfil\Wpj & \hfil\Wpjj & \hfil\Wpjjj & \hfil\Wpjjjj & \hfil\Wpjjjjj \\
\hline
30 &~$12404(152)$~&~\hfil---~&~\hfil---~&~\hfil---~&~\hfil---~\\
50 &~$5475(45)$~&~$394(34)$~&~\hfil---~&~\hfil---~&~\hfil---~\\
70 &~$3000(29)$~&~$1487(42)$~&~$0.5(0.6)$~&~\hfil---~&~\hfil---~\\
90 &~$1756(31)$~&~$1180(22)$~&~$91(6)$~&~\hfil---~&~\hfil---~\\
110 &~$1075(13)$~&~$947(56)$~&~$176(8)$~&~$2.2(0.6)$~&~\hfil---~\\
130 &~$664(8)$~&~$540(55)$~&~$212(9)$~&~$12(2)$~&~$0.02(0.01)$~\\
150 &~$417(6)$~&~$439(9)$~&~$172(5)$~&~$24(2)$~&~$0.1(0.2)$~\\
170 &~$287(5)$~&~$304(9)$~&~$150(3)$~&~$29(2)$~&~$2.4(0.6)$~\\
190 &~$196(4)$~&~$221(8)$~&~$124(2)$~&~$29(1)$~&~$2.7(0.3)$~\\
210 &~$139(5)$~&~$160(6)$~&~$103(2)$~&~$29(1)$~&~$3.4(0.4)$~\\
230 &~$99(2)$~&~$123(3)$~&~$82(2)$~&~$26(2)$~&~$4.8(0.4)$~\\
250 &~$67(1)$~&~$95(3)$~&~$66(2)$~&~$27(1)$~&~$4.5(0.6)$~\\
270 &~$52(1)$~&~$75(2)$~&~$52(1)$~&~$21(1)$~&~$5.2(0.6)$~\\
290 &~$38.2(0.7)$~&~$56(2)$~&~$40(2)$~&~$18(1)$~&~$5.1(0.6)$~\\
310 &~$29.2(0.8)$~&~$45(1)$~&~$35(1)$~&~$15.8(0.7)$~&~$3.8(0.5)$~\\
330 &~$22.8(0.7)$~&~$35(1)$~&~$30(1)$~&~$14(1)$~&~$4.8(0.6)$~\\
350 &~$17.7(0.5)$~&~$26(2)$~&~$21(1)$~&~$12.0(0.8)$~&~$3.3(0.5)$~\\
370 &~$13.4(0.4)$~&~$22(2)$~&~$18.6(0.6)$~&~$9.6(0.6)$~&~$3.2(0.4)$~\\
390 &~$12.9(0.5)$~&~$19(2)$~&~$15.6(0.5)$~&~$8.5(0.4)$~&~$3.4(0.4)$~\\
410 &~$8.8(0.5)$~&~$10(3)$~&~$13.2(0.5)$~&~$8.1(0.4)$~&~$2.8(0.2)$~\\
430 &~$7.2(0.3)$~&~$17(3)$~&~$10.6(0.4)$~&~$6.0(0.4)$~&~$3(1)$~\\
450 &~$6.4(0.5)$~&~$9.6(0.7)$~&~$11(2)$~&~$5.5(0.3)$~&~$1.7(0.5)$~\\
470 &~$5.2(0.3)$~&~$8.5(0.5)$~&~$5(2)$~&~$5.1(0.4)$~&~$2.1(0.1)$~\\
490 &~$3.8(0.2)$~&~$6.3(0.4)$~&~$6.2(0.5)$~&~$3.6(0.2)$~&~$2.3(0.3)$~\\
510 &~$3.5(0.2)$~&~$5.7(0.4)$~&~$4.9(0.3)$~&~$2(1)$~&~$1.2(0.2)$~\\
530 &~$2.5(0.1)$~&~$5.3(0.4)$~&~$4.3(0.3)$~&~$3.3(0.3)$~&~$1.6(0.1)$~\\
550 &~$2.2(0.1)$~&~$3.5(0.3)$~&~$4.0(0.2)$~&~$2.8(0.2)$~&~$1.12(0.09)$~\\
570 &~$1.8(0.1)$~&~$2(1)$~&~$3.1(0.1)$~&~$1.7(0.2)$~&~$1.1(0.1)$~\\
590 &~$1.7(0.2)$~&~$5(1)$~&~$3.0(0.1)$~&~$2.1(0.1)$~&~$1.07(0.07)$~\\
620 &~$1.30(0.06)$~&~$2.1(0.1)$~&~$2.24(0.09)$~&~$1.4(0.1)$~&~$0.76(0.05)$~\\
660 &~$1.07(0.09)$~&~$1.52(0.09)$~&~$1.58(0.07)$~&~$1.14(0.07)$~&~$0.57(0.04)$~\\
700 &~$0.81(0.08)$~&~$1.39(0.09)$~&~$1.38(0.07)$~&~$0.99(0.06)$~&~$0.51(0.05)$~\\
740 &~$0.50(0.04)$~&~$0.91(0.07)$~&~$0.85(0.04)$~&~$0.73(0.05)$~&~$0.41(0.06)$~\\
780 &~$0.49(0.08)$~&~$0.85(0.08)$~&~$0.72(0.04)$~&~$0.59(0.05)$~&~$0.21(0.05)$~\\
820 &~$0.30(0.04)$~&~$0.5(0.1)$~&~$0.57(0.03)$~&~$0.41(0.02)$~&~$0.24(0.02)$~\\
860 &~$0.25(0.04)$~&~$0.51(0.06)$~&~$0.46(0.03)$~&~$0.30(0.03)$~&~$0.19(0.02)$~\\
900 &~$0.20(0.03)$~&~$0.28(0.03)$~&~$0.33(0.02)$~&~$0.23(0.05)$~&~$0.16(0.02)$~\\
940 &~$0.13(0.01)$~&~$0.31(0.08)$~&~$0.31(0.02)$~&~$0.19(0.01)$~&~$0.11(0.01)$~\\
980 &~$0.10(0.01)$~&~$0.24(0.03)$~&~$0.23(0.02)$~&~$0.15(0.01)$~&~$0.10(0.02)$~\\

\hline
\end{tabular}}
\caption{The NLO \Wpjn{}-jet cross section taken differentially in the
  total jet transverse energy $\HT$, in fb/GeV.  These values are shown in the upper panel of the right plot in
\fig{WpHTJetProductionRatioFigure}.}
\label{Wp-HTNLODistributionTable}
\end{table}

\FloatBarrier In this appendix, we present the detailed results at LO
and at NLO for the cross section for \Wjn-jet production taken
differentially in the total jet transverse energy.  We give the LO
differential cross sections for \Wmj-jet through \Wmjjjjj-jet
production in \tab{Wm-HTLODistributionTable}, and the corresponding
results for the NLO differential cross section in
\tab{Wm-HTNLODistributionTable}.  We give the LO and NLO differential
cross sections for \Wpj-jet through \Wpjjjjj-jet production in
\tabs{Wp-HTLODistributionTable}{Wp-HTNLODistributionTable}
respectively.  We show corresponding numerical integration errors in
parentheses.

\vfill\eject

\section{Cross Sections as a Function of Jet Transverse Momenta}
\label{JetpTAppendix}

\begin{table}[tb]
\renewcommand{\arraystretch}{0.86}
\centering
\begin{tabular}{||c|l|l|l||}
\hline
$\pT$ & \hfil\Wmjjj & \hfil\Wmjjjj & \hfil\Wmjjjjj \\
\hline
37.5 &~$418(1)$~&~$56.6(0.5)$~&~$6.57(0.08)$~\\
62.5 &~$228.8(0.6)$~&~$62.5(0.3)$~&~$12.84(0.07)$~\\
87.5 &~$95.9(0.4)$~&~$34.4(0.2)$~&~$9.11(0.05)$~\\
112.5 &~$44.5(0.2)$~&~$18.2(0.1)$~&~$5.56(0.03)$~\\
137.5 &~$22.6(0.1)$~&~$10.1(0.1)$~&~$3.32(0.03)$~\\
175.0 &~$9.63(0.06)$~&~$4.55(0.04)$~&~$1.62(0.01)$~\\
225.0 &~$3.54(0.04)$~&~$1.75(0.02)$~&~$0.649(0.007)$~\\
275.0 &~$1.43(0.02)$~&~$0.74(0.01)$~&~$0.280(0.005)$~\\
325.0 &~$0.63(0.01)$~&~$0.335(0.008)$~&~$0.125(0.002)$~\\
375.0 &~$0.310(0.008)$~&~$0.163(0.005)$~&~$0.065(0.002)$~\\

\hline
\end{tabular}
\caption{The LO \Wmjn{}-jet cross section taken differentially in the
  second jet transverse momentum, in fb/GeV.}
\label{Wm-Jet2pTLODistributionTable}
\end{table}

\begin{table}
\centering
\begin{tabular}{||c|l|l|l||}
\hline
$\pT$ & \hfil\Wmjjj & \hfil\Wmjjjj & \hfil\Wmjjjjj \\
\hline
37.5 &~$725(1)$~&~$125.8(0.6)$~&~$18.1(0.1)$~\\
62.5 &~$87.8(0.3)$~&~$48.1(0.2)$~&~$14.69(0.06)$~\\
87.5 &~$19.4(0.1)$~&~$14.58(0.09)$~&~$5.91(0.03)$~\\
112.5 &~$5.90(0.04)$~&~$5.12(0.06)$~&~$2.37(0.02)$~\\
137.5 &~$2.08(0.02)$~&~$1.97(0.02)$~&~$1.05(0.02)$~\\
175.0 &~$0.617(0.007)$~&~$0.614(0.009)$~&~$0.335(0.003)$~\\
225.0 &~$0.135(0.003)$~&~$0.144(0.004)$~&~$0.083(0.001)$~\\
275.0 &~$0.035(0.001)$~&~$0.042(0.002)$~&~$2.21(0.04) \cdot 10^{-2}$~\\

\hline
\end{tabular}
\caption{The LO \Wmjn{}-jet cross section taken differentially in the
  third jet transverse momentum, in fb/GeV.}
\label{Wm-Jet3pTLODistributionTable}
\end{table}

\begin{table}[th]
\renewcommand{\arraystretch}{0.86}
\centering
\begin{tabular}{||c|l|l||}
\hline
$\pT$ & \hfil\Wmjjjj & \hfil\Wmjjjjj \\
\hline
37.5 &~$180.3(0.6)$~&~$32.0(0.1)$~\\
62.5 &~$13.90(0.09)$~&~$8.49(0.04)$~\\
87.5 &~$2.29(0.04)$~&~$1.85(0.01)$~\\
112.5 &~$0.54(0.01)$~&~$0.492(0.004)$~\\
137.5 &~$0.150(0.004)$~&~$0.151(0.002)$~\\
175.0 &~$3.31(0.09) \cdot 10^{-2}$~&~$3.62(0.06) \cdot 10^{-2}$~\\
225.0 &~$4.5(0.2) \cdot 10^{-3}$~&~$5.5(0.1) \cdot 10^{-3}$~\\

\hline
\end{tabular}
\caption{The LO \Wmjn{}-jet cross section taken differentially in the
fourth jet transverse momentum, in fb/GeV.}
\label{Wm-Jet4pTLODistributionTable}
\end{table}

\begin{table}
\centering
\begin{tabular}{||c|l|l|l||}
\hline
$\pT$ & \hfil\Wmjjj & \hfil\Wmjjjj & \hfil\Wmjjjjj \\
\hline
37.5 &~$388(3)$~&~$51(1)$~&~$5.5(0.2)$~\\
62.5 &~$195(2)$~&~$49(2)$~&~$9.6(0.6)$~\\
87.5 &~$76(2)$~&~$25.1(0.5)$~&~$6.5(0.4)$~\\
112.5 &~$34.2(0.4)$~&~$11.6(0.9)$~&~$3.7(0.2)$~\\
137.5 &~$17.2(0.3)$~&~$8.0(0.8)$~&~$2.29(0.09)$~\\
175.0 &~$7.0(0.1)$~&~$3.1(0.1)$~&~$0.99(0.04)$~\\
225.0 &~$2.61(0.05)$~&~$1.10(0.05)$~&~$0.39(0.03)$~\\
275.0 &~$0.96(0.03)$~&~$0.49(0.03)$~&~$0.17(0.01)$~\\
325.0 &~$0.43(0.02)$~&~$0.24(0.04)$~&~$0.06(0.03)$~\\
375.0 &~$0.21(0.01)$~&~$0.09(0.02)$~&~$0.01(0.03)$~\\

\hline
\end{tabular}
\caption{The NLO \Wmjn{}-jet cross section taken differentially in the
  second jet transverse momentum, in fb/GeV.}
\label{Wm-Jet2pTNLODistributionTable}
\end{table}

\begin{table}
\centering
\begin{tabular}{||c|l|l|l||}
\hline
$\pT$ & \hfil\Wmjjj & \hfil\Wmjjjj & \hfil\Wmjjjjj \\
\hline
37.5 &~$628(4)$~&~$102(2)$~&~$13.2(0.7)$~\\
62.5 &~$79.0(0.6)$~&~$36.4(0.6)$~&~$11.0(0.4)$~\\
87.5 &~$17.2(0.2)$~&~$10.6(0.3)$~&~$4.0(0.1)$~\\
112.5 &~$5.4(0.1)$~&~$3.4(0.2)$~&~$1.48(0.08)$~\\
137.5 &~$1.88(0.05)$~&~$1.28(0.08)$~&~$0.64(0.03)$~\\
175.0 &~$0.56(0.01)$~&~$0.39(0.02)$~&~$0.20(0.01)$~\\
225.0 &~$0.125(0.005)$~&~$0.12(0.02)$~&~$0.053(0.004)$~\\
275.0 &~$0.029(0.002)$~&~$0.025(0.003)$~&~$0.014(0.002)$~\\

\hline
\end{tabular}
\caption{The NLO \Wmjn{}-jet cross section taken differentially in the
  third jet transverse momentum, in fb/GeV.}
\label{Wm-Jet3pTNLODistributionTable}
\end{table}

\begin{table}
\centering
\begin{tabular}{||c|l|l||}
\hline
$\pT$ & \Wmjjjj & \Wmjjjjj \\
\hline
37.5 &~$141(2)$~&~$22.9(0.7)$~\\
62.5 &~$11.3(0.4)$~&~$6.3(0.1)$~\\
87.5 &~$2.0(0.1)$~&~$1.17(0.07)$~\\
112.5 &~$0.45(0.02)$~&~$0.32(0.02)$~\\
137.5 &~$0.121(0.009)$~&~$0.089(0.005)$~\\
175.0 &~$0.026(0.002)$~&~$0.024(0.001)$~\\
225.0 &~$4.3(0.5) \cdot 10^{-3}$~&~$2.4(0.5) \cdot 10^{-3}$~\\

\hline
\end{tabular}
\caption{The NLO \Wmjn{}-jet cross section taken differentially in the
fourth jet transverse momentum, in fb/GeV.}
\label{Wm-Jet4pTNLODistributionTable}
\end{table}

\FloatBarrier
In this appendix, we present the detailed results at LO and at NLO
for the cross section for \Wmjn-jet production taken differentially
in the second-, third-, and fourth-hard jet transverse momenta.
  We give the LO differential cross sections
for the second-hard jet in \Wmjjj-jet through \Wmjjjjj-jet production in 
\tab{Wm-Jet2pTLODistributionTable}, for the third-hardest
jet in \tab{Wm-Jet3pTLODistributionTable},
and for the fourth-hardest jet in \Wmjjjj-jet and \Wmjjjjj-jet 
production
in \tab{Wm-Jet4pTLODistributionTable}.
We give the corresponding NLO differential cross sections
in tables~\ref{Wm-Jet2pTNLODistributionTable},
\ref{Wm-Jet3pTNLODistributionTable},
and \ref{Wm-Jet4pTNLODistributionTable}, respectively.
 We show corresponding numerical integration errors in
parentheses.


\end{document}